\documentclass[12pt]{article}
\usepackage{subeqnarray,indent,amsfonts,amssymb,amsmath}
\usepackage{eqsection}
\usepackage{bm}    
\usepackage{cite}  
\usepackage{pstricks,pst-node,pst-text,pst-3d,pst-plot}  
\usepackage{epic,eepic}
\usepackage{graphicx}

\begin{document}

\bibliographystyle{plain}

\date{October 25, 2005 \\[2mm]
      Revised Jan 18, 2006} 

\title{
       Complex-temperature phase diagram \\ of Potts and RSOS models
      }

\author{
  {\small Jesper Lykke Jacobsen${}^{1,2}$,
          Jean-Fran\c{c}ois Richard${}^{1,3}$, and
          Jes\'us Salas${}^4$}                           \\[1mm]
  {\small\it ${}^1$Laboratoire de Physique Th\'eorique
  et Mod\`eles Statistiques}                             \\[-0.2cm]
  {\small\it Universit\'e Paris-Sud}                     \\[-0.2cm]
  {\small\it B\^atiment 100}                             \\[-0.2cm]
  {\small\it 91405 Orsay, FRANCE}                        \\[-0.2cm]
  {\small\tt JACOBSEN@IPNO.IN2P3.FR}                     \\[1mm]
  {\small\it ${}^2$Service de Physique Th\'eorique}      \\[-0.2cm]
  {\small\it CEA Saclay}                                 \\[-0.2cm]
  {\small\it Orme des Merisiers}                        \\[-0.2cm]
  {\small\it 91191 Gif sur Yvette, FRANCE}               \\[1mm]
  {\small\it ${}^3$Laboratoire de Physique Th\'eorique
  et Hautes Energies}                                    \\[-0.2cm]
  {\small\it Universit\'e Paris VI}                      \\[-0.2cm]
  {\small\it Bo\^{\i}te 126, Tour 24,
  5${}^{\rm e}$ {\'e}tage}                           \\[-0.2cm]
  {\small\it 4 place Jussieu}                            \\[-0.2cm]
  {\small\it 75252 Paris cedex 05, FRANCE}               \\[-0.2cm]
  {\small\tt JRICHARD@LPTHE.JUSSIEU.FR}                  \\[1mm]
  {\small\it ${}^4$Grupo de Modelizaci\'on, Simulaci\'on Num\'erica 
               y Matem\'atica Industrial}\\[-0.2cm]
  {\small\it Escuela Polit\'ecnica Superior}   \\[-0.2cm] 
  {\small\it Universidad Carlos III de Madrid} \\[-0.2cm]
  {\small\it Avda.\  de la Universidad, 30}    \\[-0.2cm]
  {\small\it 28911 Legan\'es, SPAIN}           \\[-0.2cm]
  {\small\tt JSALAS@MATH.UC3M.ES}              \\[-0.2cm]
  {\protect\makebox[5in]{\quad}}  
  \\
}

\maketitle
\thispagestyle{empty}   

\begin{abstract}

We study the phase diagram of $Q$-state Potts models, for $Q=4
\cos^2(\pi/p)$ a Beraha number ($p>2$ integer), in the
complex-temperature plane. The models are defined on $L \times N$
strips of the square or triangular lattice, with boundary conditions
on the Potts spins that are periodic in the longitudinal ($N$)
direction and free or fixed in the transverse ($L$) direction. The
relevant partition functions can then be computed as sums over
partition functions of an $A_{p-1}$ type RSOS model, thus making
contact with the theory of quantum groups. We compute the accumulation
sets, as $N\to\infty$, of partition function zeros for
$p=4,5,6,\infty$ and $L=2,3,4$ and study selected features for $p>6$
and/or $L>4$. This information enables us to formulate several
conjectures about the thermodynamic limit, $L\to\infty$, of these
accumulation sets. The resulting phase diagrams are quite different
from those of the generic case (irrational $p$). For free transverse
boundary conditions, the partition function zeros are found to be
dense in large parts of the complex plane, even for the Ising model
($p=4$). We show how this feature is modified by taking fixed
transverse boundary conditions.

\end{abstract}


\clearpage

%
%
\newcommand{\be}{\begin{equation}}
\newcommand{\ee}{\end{equation}}
\newcommand{\<}{\langle}
\renewcommand{\>}{\rangle}
\newcommand{\widebar}{\overline}
\def\reff#1{(\protect\ref{#1})}
\def\spose#1{\hbox to 0pt{#1\hss}}
\def\ltapprox{\mathrel{\spose{\lower 3pt\hbox{$\mathchar"218$}}
 \raise 2.0pt\hbox{$\mathchar"13C$}}}
\def\gtapprox{\mathrel{\spose{\lower 3pt\hbox{$\mathchar"218$}}
 \raise 2.0pt\hbox{$\mathchar"13E$}}}
\def\textprime{${}^\prime$}
\def\proof{\par\medskip\noindent{\sc Proof.\ }}
\def\qed{\hbox{\hskip 6pt\vrule width6pt height7pt depth1pt \hskip1pt}\bigskip}
\def\proofof#1{\bigskip\noindent{\sc Proof of #1.\ }}
\def\half{ {1 \over 2} }
\def\third{ {1 \over 3} }
\def\twothird{ {2 \over 3} }
\def\smfrac#1#2{\textstyle{#1\over #2}}
\def\smhalf{ \smfrac{1}{2} }
\newcommand{\real}{\mathop{\rm Re}\nolimits}
\renewcommand{\Re}{\mathop{\rm Re}\nolimits}
\newcommand{\imag}{\mathop{\rm Im}\nolimits}
\renewcommand{\Im}{\mathop{\rm Im}\nolimits}
\newcommand{\sgn}{\mathop{\rm sgn}\nolimits}
\newcommand{\tr}{\mathop{\rm tr}\nolimits}
\newcommand{\diag}{\mathop{\rm diag}\nolimits}
\newcommand{\Gal}{\mathop{\rm Gal}\nolimits}
\newcommand{\mycup}{\mathop{\cup}}
\newcommand{\Arg}{\mathop{\rm Arg}\nolimits}
\newcommand{\sech}{\mathop{\rm sech}\nolimits}
\def\hboxscript#1{ {\hbox{\scriptsize\em #1}} }
\def\zhat{ {\widehat{Z}} }
\def\phat{ {\widehat{P}} }
\def\qtilde{ {\widetilde{q}} }
\renewcommand{\emptyset}{\varnothing}

\def\scra{\mathcal{A}}
\def\scrb{\mathcal{B}}
\def\scrc{\mathcal{C}}
\def\scrd{\mathcal{D}}
\def\scrf{\mathcal{F}}
\def\scrg{\mathcal{G}}
\def\scrl{\mathcal{L}}
\def\scro{\mathcal{O}}
\def\scrp{\mathcal{P}}
\def\scrq{\mathcal{Q}}
\def\scrr{\mathcal{R}}
\def\scrs{\mathcal{S}}
\def\scrt{\mathcal{T}}
\def\scrv{\mathcal{V}}
\def\scrz{\mathcal{Z}}

\def\q{{\sf q}}

\def\Z{{\mathbb Z}}
\def\R{{\mathbb R}}
\def\C{{\mathbb C}}
\def\Q{{\mathbb Q}}

\def\T{{\mathsf T}}
\def\H{{\mathsf H}}
\def\V{{\mathsf V}}
\def\D{{\mathsf D}}
\def\J{{\mathsf J}}
\def\P{{\mathsf P}}
\def\QQ{{\mathsf Q}}
\def\RR{{\mathsf R}}

\def\bsigma{{\boldsymbol{\sigma}}}
\def\bone{{\mathbf 1}}
\def\vv{{\bf v}}
\def\uu{{\bf u}}
\def\w{{\bf w}}

\newtheorem{theorem}{Theorem}[section]
\newtheorem{definition}[theorem]{Definition}
\newtheorem{proposition}[theorem]{Proposition}
\newtheorem{lemma}[theorem]{Lemma}
\newtheorem{corollary}[theorem]{Corollary}
\newtheorem{conjecture}[theorem]{Conjecture}


\newenvironment{sarray}{
          \textfont0=\scriptfont0
          \scriptfont0=\scriptscriptfont0
          \textfont1=\scriptfont1
          \scriptfont1=\scriptscriptfont1
          \textfont2=\scriptfont2
          \scriptfont2=\scriptscriptfont2
          \textfont3=\scriptfont3
          \scriptfont3=\scriptscriptfont3
        \renewcommand{\arraystretch}{0.7}
        \begin{array}{l}}{\end{array}}

\newenvironment{scarray}{
          \textfont0=\scriptfont0
          \scriptfont0=\scriptscriptfont0
          \textfont1=\scriptfont1
          \scriptfont1=\scriptscriptfont1
          \textfont2=\scriptfont2
          \scriptfont2=\scriptscriptfont2
          \textfont3=\scriptfont3
          \scriptfont3=\scriptscriptfont3
        \renewcommand{\arraystretch}{0.7}
        \begin{array}{c}}{\end{array}}

\def\kk {\phantom{-}}
\def\kkk{\phantom{-1}}
%
%

%
%
\section{Introduction} \label{sec_intro}

The $Q$-state Potts model \cite{Wu_82,Baxter_book}
can be defined for general $Q$ by using the
Fortuin--Kasteleyn (FK) representation \cite{Kasteleyn_69,Fortuin_72}.
The partition function $Z_G(Q;v)$ is a polynomial in the variables $Q$
and $v$. This latter variable is related to the Potts model coupling
constant $J$ as
\be
   v \;=\; e^{J} -1.
\label{def_v}
\ee
It turns out useful to define the temperature parameter $x$ as 
\be
  x \;=\; \frac{v}{\sqrt{Q}}
\label{def_x}
\ee
and to parameterize the interval $Q\in(0,4]$ as
\be
  Q \;=\; 4\cos^2 \left( \frac{\pi}{p} \right) \,,
  \qquad p \in (2,\infty].
\label{def_q_p}
\ee

For generic%
\footnote{More precisely, a ``generic'' value of $Q$ corresponds to an
irrational value of the parameter $p$ defined in Eq.~(\ref{def_q_p}).
This point will be made more precise in Section~\ref{sec_RSOS} below.}
values of $Q$, the main features of the phase diagram of the Potts
model in the real $(Q,x)$-plane have been known for many years
\cite{Baxter_book,Nienhuis_87}. It contains in particular a curve $x_{\rm
FM}(Q)> 0$ of ferromagnetic phase transitions which are second-order in the
range $0 < Q \leq 4$, the thermal operator being relevant. The analytic
continuation of the curve $x_{\rm FM}(Q)$ into the antiferromagnetic regime
yields a second critical curve $x_{\rm BK}(Q)<0$ with $0<Q<4$ along which the
thermal operator is irrelevant. Therefore, for a fixed value of $Q$, the
critical point $x_{\rm BK}(Q)$ acts as the renormalization group (RG)
attractor of a finite range of $x$ values: this is the Berker-Kadanoff (BK)
phase \cite{Saleur_90,Saleur_91}.

\begin{figure}[hbtp]
\centering
\includegraphics[width=400pt]{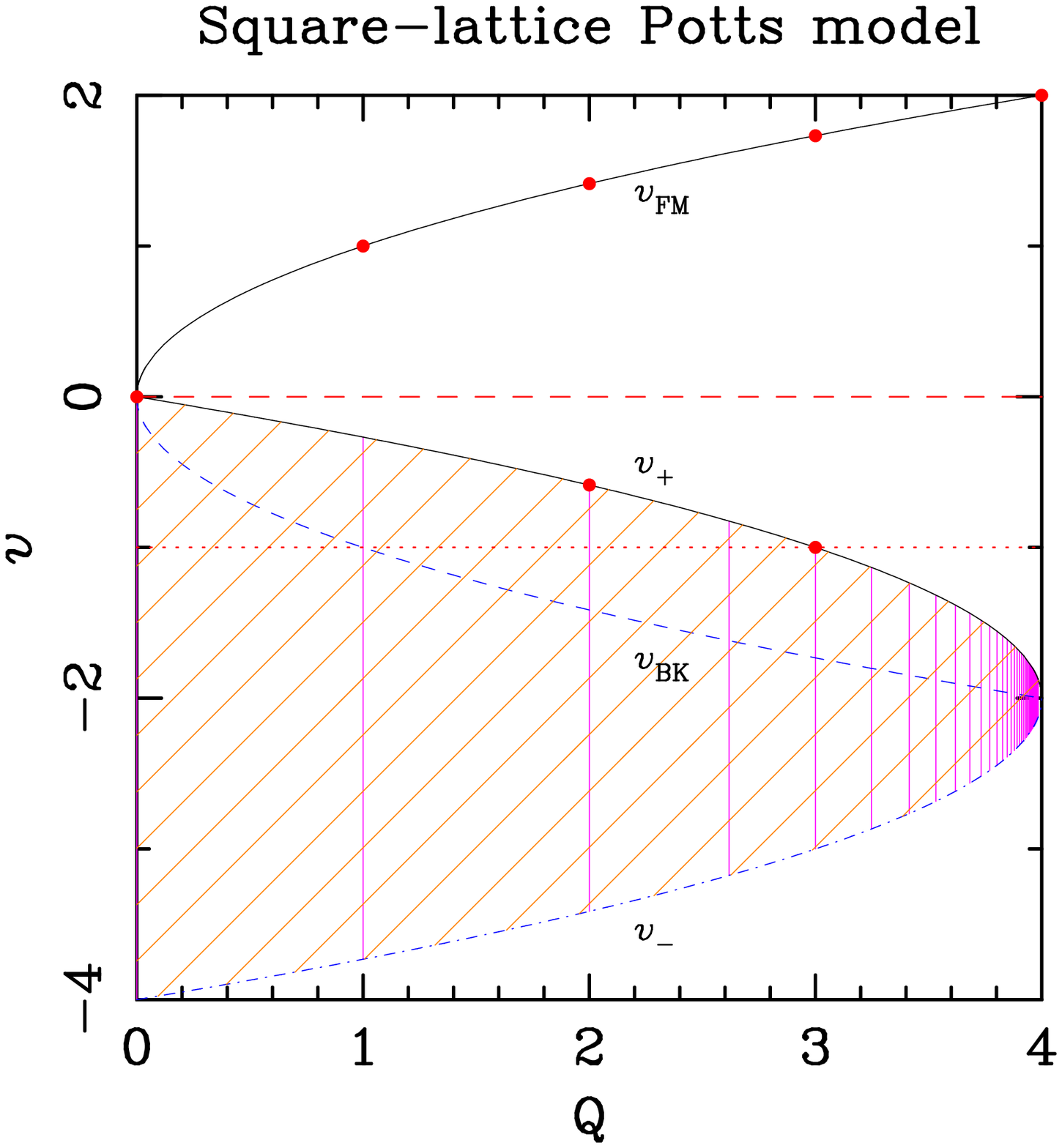}
\caption[a]{\label{figure_sq} Generic phase diagram for the two-dimensional
Potts model in the $(Q,v)$-plane. The solid black curve in the ferromagnetic
($v>0$) region shows the standard ferromagnetic phase transition curve $v_{\rm
FM}(Q)$, and the blue dashed curve is its analytic continuation $v_{\rm
BK}(Q)$ into the antiferromagnetic region. This latter curve acts as an RG
attractor for the Berker-Kadanoff phase (the orange hatched region). This is
separated from the limit of infinite temperature (red dashed curve) by the
antiferromagnetic phase-transition curve $v_+(Q)$ (solid black curve in the
$v<0$ region), and from the $v\to-\infty$ limit by its counterpart
$v_-(Q)$ (dot-dashed blue curve). The red horizontal dotted curve represents
the zero-temperature antiferromagnet ($v=-1$). The pink vertical lines show
the Beraha numbers $Q=4\cos^2(\pi/p)$ ($p=2,3,\ldots$): the phase diagram on
these lines is {\em different} from the generic one shown here and forms the
object of the present article. Note that the exact functional forms of the
curves $v_{\rm FM}(Q)$, $v_{\rm BK}(Q)$, and $v_\pm(Q)$ are lattice-dependent;
the figure shows their explicit forms for the square-lattice model.
}
\end{figure}

The generic phase diagram is shown in Fig.~\ref{figure_sq}. Since the
infinite-temperature limit ($x=0$) and the zero-temperature ferromagnet
($|x|=\infty$) are of course RG attractive, consistency of the phase
diagram requires that the BK phase be separated from these by a pair of
RG repulsive curves $x_{\pm}(Q) < 0$. 
The curve $x_+(Q)$ is  expected to correspond
to the antiferromagnetic (AF) phase transition of the model \cite{JS_05}.

The above scenario thus essentially relies on the RG attractive nature of the
curve $x_{\rm BK}(Q)$, and since this can be derived \cite{Nienhuis_87} from
very general Coulomb gas considerations, the whole picture should hold for
any two-dimensional lattice. But it remains of course of great interest
to compute the exact functional forms of the curves $x_{\rm FM}(Q)$,
$x_{\rm BK}(Q)$, and $x_{\pm}(Q)$---and the corresponding free energies---for
specific lattices.

The square-lattice Potts model is the best understood case.
Here, Baxter \cite{Baxter_book,Baxter_82} has found the exact free energy
along several curves $x=x_{\rm c}(Q)$:
\be
x_{\rm c}(Q) \;=\; \begin{cases}
   +1                                                & \qquad\text{(FM)}  \\
   -\frac{2}{\sqrt{Q}} + \sqrt{\frac{4-Q}{Q}}        & \qquad\text{(AF)} \\
   -1                                                & \qquad\text{(BK)} \\
   -\frac{2}{\sqrt{Q}} - \sqrt{\frac{4-Q}{Q}}        & \qquad\text{(AF)}
             \end{cases}
\label{values_sq_xc}
\ee
where $x_{\rm c}=1$ and $x_{\rm c}=-1$ can be identified respectively with
$x_{\rm FM}(Q)$ and $x_{\rm BK}(Q)$. The curves $x_{\pm}=-2/\sqrt{Q} \pm
\sqrt{(4-Q)/Q}$ are mutually dual (and hence equivalent) curves of AF phase
transitions, which are again second-order in the range $0<Q\le 4$. These
curves also form the boundaries of the $x$-values controlled by the BK fixed
point \cite{Saleur_91}, as outlined above. Note that the four points $x_{\rm
c}(q)$ in Eq.~\reff{values_sq_xc} correspond to the points where the circles
\begin{subeqnarray}
     |x|                             &=& 1 \\
\left| x + \frac{2}{\sqrt{Q}}\right| &=& \sqrt{\frac{4-Q}{Q}}
\label{circles_sq_x}
\end{subeqnarray}
cross the real $x$-axis. These two circles intersect at the points
\be
x \;=\; - e^{\pm i\, \pi/p}
\ee
which will be shown below to play a particular role in the phase
diagram (see Conjecture~\ref{conj_sq_1}.1).

In the case of a triangular lattice, Baxter and collaborators
\cite{Baxter_78,Baxter_86,Baxter_87} have found the free energy of the
Potts model along the curves
\begin{subeqnarray}
\sqrt{Q} x^3 + 3x^2 &=& 1 \,, \\
x                   &=& -\frac{1}{\sqrt{Q}} \,.
\label{values_tri_xc}
\end{subeqnarray}
The upper branch of Eq.~(\ref{values_tri_xc}a) is identified with the
ferromagnetic critical curve $x_\text{FM}(Q)$. We have numerical evidence that
the middle and lower branches correspond respectively to $x_\text{BK}(Q)$ and
$x_{-}(Q)$, the lower boundary of the BK phase. The position of $x_{+}(Q)$,
the upper branch of the BK phase, is at present unknown 
\cite{JSS_RGflow} (but see
Ref.~\cite{forests} for the $Q\to 0$ limit). Along the line
(\ref{values_tri_xc}b) the Potts model reduces to a coloring problem, and the
partition function is here known as the chromatic polynomial. The line
(\ref{values_tri_xc}b) belongs to the RG basin of the BK phase for
$0 < Q < 2+\sqrt{3}$ \cite{transfer3}.

The critical properties---still with $Q$ taking generic values---for these two
lattices are to a large extent universal. This is not so surprising, since the
critical exponents can largely be obtained by Coulomb gas techniques (although
the antiferromagnetic transition still reserves some challenges \cite{JS_05}).
Thus, there is numerical evidence that the exponents along the curves
$x_\text{FM}(Q)$, $x_\text{BK}(Q)$ and $x_{-}(Q)$ coincide, whereas the
evidence for the curve $x_{+}(Q)$ is non-conclusive \cite{forests}. On the
other hand, on the less-studied triangular lattice we cannot yet exclude the
possible existence of other curves of second-order phase transitions that have
no counterpart on the square lattice.

But in general we can only expect universality to hold when the
Boltzmann weights in the FK representation are non-negative
(i.e., for $Q\ge 0$, $v \ge 0$), or when the parameter $p$ takes
generic (i.e., irrational) values. The present paper aims at
studying the situation when $p$ takes non-generic values; for
simplicity we limit ourselves to the case of integer $p>2$. The
number of spin states is then equal to a so-called Beraha number $B_p$
\be
  Q \;=\; B_p \;=\; 4\cos^2 \left(\frac{\pi}{p}\right) \,,
  \qquad p=3,4,5,\ldots.
\ee
The special physics at rational values of $p$ is intimately linked to the
representation theory of the quantum group $U_q(SU(2))$, the commutant of the
Temperley-Lieb algebra, when the deformation parameter $q$ is a root of unity.
As we shall review in Section~\ref{sec_RSOS} below, the quantum group symmetry
of the Potts model at rational $p$ implies that many eigenvalues of the
transfer matrix in the FK representation have zero amplitude or cancel in
pairs because of opposite amplitudes; these eigenvalues therefore become
spurious and do not contribute to the partition function
\cite{Saleur_90,Saleur_91}. 

Remarkably, for $p$ integer and $x$ inside the BK phase, even the
leading eigenvalue acquires zero amplitude. Moreover, all the
eigenvalues which scale like the leading one in the thermodynamic
limit vanish from the partition function, and so, even the bulk free
energy $f(p;x)$ is modified \cite{JS_05}. In other words, $f(p;x)$
experiences a singularity whenever $p$ passes through an integer
value. This means in particular that for $p$ integer the critical
behavior can either disappear, or be modified, or new critical points
(and other non-critical fixed points) can emerge.

\begin{figure}[hbtp]
\centering
\includegraphics[width=400pt]{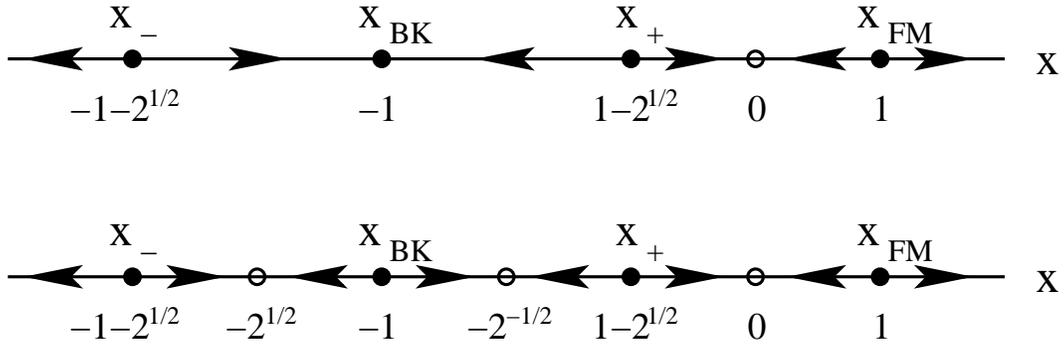}
\caption[a]{\label{fig:Ising} Phase diagram and RG flows for the
$Q\to 2$ state model (top) and the $Q=2$ Ising model (bottom), on
the real $x$-axis. Filled (resp.\ empty) circles correspond to
critical (resp.\ non-critical) fixed points.
}
\end{figure}

For the sake of clarity, we discuss the simplest example of this phenomenon.
Consider, on the square lattice, on one hand the $Q\to 2$ state model (i.e.,
with $Q$ tending to 2 through irrational values of $p$) and on the other the
$Q=2$ Ising model (i.e., with fixed integer $p=4$). For the former case, the
generic phase diagram and the associated RG flows are shown in the top part of
Fig.~\ref{fig:Ising}. The three critical points $x_{\rm FM}$ and $x_\pm$
have central charge $c=1/2$, while the fourth one $x_{\rm BK}$ has $c=-25/2$.
For the latter case, new non-critical fixed points appear (by applying
the duality and $Z_2$ gauge symmetries to the one at $x=0$), and the RG
flows become as shown in the bottom part of Fig.~\ref{fig:Ising}. One now
has $c=1/2$ for all four critical fixed points. (We shall treat the Ising
model in more detail in Section~\ref{sec:Ising} below.)

By contrast to the universality brought out for generic $Q$, the phase diagram
and critical behavior for integer $p$ is likely to have lattice dependent
features. Let us give a couple of examples of this non-universality. The
zero-temperature triangular-lattice Ising antiferromagnet, $(Q,v)=(2,-1)$, is
critical and becomes in the scaling limit a free Gaussian field with central
charge $c=1$ \cite{Stephenson_64,Blote_82b,Nienhuis_84b}, whereas the
corresponding square-lattice model is non-critical, its partition function
being trivially $Z=2$. While this observation does not in itself imply
non-universality, since the critical temperature is expected to be lattice
dependent (as is the value of $x_{\rm FM}(Q)$), the point to be noticed is
that for no value of $v$ does the $Q=2$ square-lattice model exhibit $c=1$
critical behavior. In the same vein, the square-lattice Potts model with
$(Q,v)=(3,-1)$ is equivalent to a critical six-vertex model (at $\Delta=1/2$)
\cite{Lenard_67,Baxter_70}, with again $c=1$ in the scaling limit, whereas now
the corresponding triangular-lattice model is trivial ($Z=3$). Now, the
triangular-lattice model does in fact exhibit $c=1$ behavior elsewhere (for
$x=x_-$), but the compactification radius is different from that of the
square-lattice theory and accordingly the critical exponents differ. Finally,
$(Q,v)=(4,-1)$ is a critical $c=2$ theory on the triangular lattice
\cite{Baxter_70_TRI,Kondev_96}, but is non-critical on the square lattice
\cite{Ferreira_Sokal}.

Because of the eigenvalue cancellation scenario sketched above, the FK
representation is not well suited%
\footnote{We here tacitly assume that the study relies on a transfer
matrix formulation. This is indeed so in most approaches that we know of,
whether they be analytical or numerical. An exception would be numerical
simulations of the Monte Carlo type, but in the most interesting parts of
the phase diagram this approach would probably not be possible anyway, due
to the presence of negative Boltzmann weights.}
for studying the Potts model at integer $p$. Fortunately, for $Q=B_p$ there
exists another representation of the Potts model, in terms of an RSOS model of
the $A_{p-1}$ type \cite{Pasquier_87}, in which the cancellation
phenomenon is explicitly built-in, in the sense that for generic values of $x$
all the RSOS eigenvalues contribute to the partition function. On the square
lattice, the RSOS model has been studied in great detail
\cite{ABF,Huse,Pasquier_87,Saleur_89} at the point $x=x_{\rm FM}=1$, where the
model happens to be homogeneous. Only very recently has the case of general
real $x \neq 1$ (where the RSOS model is staggered, i.e., its Boltzmann
weights are sublattice dependent) attracted some attention \cite{JS_05},
and no previous investigation of other lattices (such as the triangular
lattice included in the present study) appears to exist.

The very existence of the RSOS representation has profound links
\cite{Saleur_89,Pasquier_90} to the representation theory of the quantum group
$U_q(SU(2))$ where the deformation parameter $q$ defined by
\be
  Q = \left( q + q^{-1} \right)^2 = B_p \,, \qquad q=\exp(i\pi/p) \,,
 \label{qq-1}
\ee
is a root of unity. To ensure the quantum group invariance one needs to impose
periodic boundary conditions along the transfer direction. Further, to ensure
the exact equivalence between Potts and RSOS model partition functions the
transverse boundary conditions must be non-periodic.%
\footnote{There are however some intriguing relationships between modified
partition functions with fully periodic boundary conditions \cite{Richard_05}.
We believe that the RSOS model with such boundary conditions merits a study
similar to the one presented here, independently of its relation to the Potts
model.}
For definiteness we shall therefore study square- or triangular-lattice strips
of size $L \times N$ spins, with periodic boundary conditions in the
$N$-direction. The boundary conditions in the $L$-direction are initially
taken as free, but we shall later consider fixed transverse boundary
conditions as well. For simplicity we shall henceforth refer to these boundary
conditions as free cyclic and fixed cyclic.%
\footnote{It is convenient to introduce the notation $L_{\rm F} \times N_{\rm
P}$ (resp.\ $L_{\rm X} \times N_{\rm P}$) for a strip of size $L \times N$
spins with free (resp.\ fixed) cyclic boundary conditions.}

Using the RSOS representation we here study the phase diagram of the Potts
model at $Q=B_p$ through the loci of partition function zeros in the complex
$x$-plane. According to the Beraha-Kahane-Weiss theorem \cite{BKW_theorem},
when $N\to\infty$, the accumulation points of these zeros form either isolated
limiting points (when the amplitude of the dominant eigenvalue vanishes) or
continuous limiting curves $\scrb_L$ (when two or more dominant eigenvalues
become equimodular); we refer to Ref.~\cite{transfer1} for further details. 
In the RSOS representation only the latter scenario is possible, since all
amplitudes are strictly positive.\footnote{
 Sokal \protect\cite[Section 3]{Sokal_04} has given a slight generalization of 
 the Beraha--Kahane--Weiss theorem. In particular, when there 
 are two or more equimodular dominant eigenvalues, the set of accumulation 
 points of the partition-function zeros may include isolated limiting points 
 when {\em all}\/ the eigenvalues vanish simultaneously. See
 Section~\protect\ref{sec.sq.ising.L=2} for an example of this possibility. 
} 
As usual in such studies, branches of $\scrb_L$ that traverse the real
$x$-axis for finite $L$, or ``pinch'' it asymptotically in the thermodynamic
limit $L\to\infty$, signal the existence of a phase transition. Moreover, the
finite-size effects and the impact angles \cite{Kenna} give information about
the nature of the transition.

The limiting curves $\scrb_L$ constitute the boundaries between the
different phases of the model. Moreover, in the present set-up, each
phase can be characterized topologically by the value of the conserved
quantum group spin $S_z$, whose precise definition will be recalled in
Section~\ref{sec_RSOS} below. (A similar characterization of phases of
the chromatic polynomial was recently exploited in
Ref.~\cite{transfer4}, but in the FK representation).  One may think
of $S_z$ as a kind of ``quantum'' order parameter. A naive entropic
reasoning would seem to imply that for any real $x$ the ground state
(free energy) has $S_z=0$, since the corresponding sector of the
transfer matrix has the largest dimension. It is a most remarkable
fact that large portions of the phase diagram turn out have $S_z \neq
0$.

We have computed the limiting curves $\scrb_L$ in the complex
$x$-plane completely for $p=4,5,6,\infty$ and $L=2,3,4$ for both
lattices. Moreover, we have studied selected features thereof for
$p>6$ and/or $L>4$.  This enables us to formulate several conjectures
about the topology of $\scrb_L$ which are presumably valid for any
$L$, and therefore, provides information about the thermodynamic limit
$L\to\infty$. The resulting knowledge is a starting point for gaining
a better understanding of the fixed point structure and
renormalization group flows in these Potts models.

Our work has been motivated in particular by the following open issues:
\begin{enumerate}
 \item As outlined above, the eigenvalue cancellation phenomenon arising from
 the quantum group symmetry at integer $p$ modifies the bulk free energy in
 the Berker-Kadanoff phase. For the Ising model we have seen that this changes
 the RG nature (from attractive to repulsive) of the point $x_{\rm BK}$ as
 well as its critical exponents (from $c=-25/2$ to $c=1/2$). But for general
 integer $p$ it is not clear whether $x_{\rm BK}$ will remain a phase
 transition point, and assuming this to be the case what would be its
 properties.
 \item The chromatic line $x=-1/\sqrt{Q}$ does not appear to play any
 particular role in the generic phase diagram of the square-lattice model. By
 contrast, it is an integrable line \cite{Baxter_86,Baxter_87} for the generic
 triangular-lattice model. Qua its role as the zero-temperature
 antiferromagnet one could however expect the chromatic line to lead to
 particular (and possibly critical) behavior in the RSOS model. Even when
 critical behavior exists in the generic case (e.g., on the triangular
 lattice) the nature of the transition may change when going to the case of
 integer $p$ (e.g., from $c=-25/2$ for the $Q\to 2$ model to $c=1$ for the
 zero-temperature Ising antiferromagnet).
 \item Some features in the antiferromagnetic region might possibly exhibit an
 extreme dependence on the boundary conditions, in line with what is known,
 e.g., for the six-vertex model. It is thus of interest to study both free and
 fixed boundary conditions. To give but one example of what may be expected,
 we have discovered---rather surprisingly---that with free cyclic boundary
 conditions the partition function zeros are actually dense in substantial
 parts of the complex plane: this is true even for the simplest case of the
 square-lattice Ising model.
\item A recent numerical study \cite{JS_05} of the effective central
  charge of the RSOS model with periodic boundary conditions, as a
  function of $x$, has revealed the presence of new critical points
  inside the BK phase. In particular, strong evidence was given for a
  physical realization of the integrable flow \cite{Fateev_91} from
  parafermion to minimal models. The question arises what would be the
  location of these new points in the phase diagram.
 \item In the generic case, the spin $S_z$ of the ground state may be driven to
 arbitrary large values upon approching the point $(Q,x)=(4,-1)$ from
 within the BK phase \cite{Saleur_91,transfer4}. Is a similar mechanism
 at play for integer $p$?
\end{enumerate}

The paper is organized as follows. In Section~2 we introduce the RSOS models
and describe their precise relationship to the Potts model, largely following
Refs.~\cite{Pasquier_87,Saleur_89,Pasquier_90}. We then present, in Section~3,
the limiting curves found for the square-lattice model with free cyclic
boundary conditions, leading to the formulation of several conjectures in
Section~4. Sections~5--6 repeat this programme for the triangular-lattice
model. In Section~7 we discuss the results for free cyclic boundary
conditions, with special emphasis on the thermodynamic limit, and motivate the
need to study also fixed cyclic boundary conditions. This is then done in
Sections~8--9. Finally, Section~10 is devoted to our conclusions. An appendix
gives some technical details on the dimensions of the transfer matrices used.

%
%
\section{RSOS representation of the Potts model} \label{sec_RSOS}

The partition function of the two-dimensional Potts model can be written in
several equivalent ways, though sometimes with different domains of validity
of the relevant parameters (notably $Q$). The interplay between these
different representations is at the heart of the phenomena we wish to study.

The {\em spin representation} for $Q$ integer is well-known. Its
low-temperature expansion gives the {\em FK representation}
\cite{Kasteleyn_69,Fortuin_72} discussed in the Introduction, where $Q$ is now
an arbitrary complex number. The (interior and exterior) boundaries of the FK
clusters, which live on the medial lattice, yield the equivalent loop
representation with weight $Q^{1/2}$ per loop.

An {\em oriented loop representation} is obtained by independently assigning
an orientation to each loop, with weight $q$ (resp.\ $q^{-1}$) for
counterclockwise (resp.\ clockwise) loops, cf.~Eq.~(\ref{qq-1}). In this
representation one can define the spin $S_z$ along the transfer direction
(with parallel/antiparallel loops contributing $\pm 1/2$) which acts as a
conserved quantum number. Note that $S_z=j$ means that there are {\em at
least} $j$ non-contractible loops, i.e., loops that wind around the periodic
($N$) direction of the lattice. The weights $q^{\pm 1}$ can be further
redistributed locally, as a factor of $q^{\alpha/2\pi}$ for a counterclockwise
turn through an angle $\alpha$ \cite{Baxter_book}. While this redistribution
correctly weights contractible loops, the non-contractible loops are given
weight $2$, but this can be corrected by twisting the model, i.e., by
inserting the operator $q^{S_z}$ into the trace that defines the partition
function.

A partial resummation over the oriented-loop splittings at vertices which are
compatible with a given orientation of the edges incident to that vertex now
gives a {\em six-vertex model representation} \cite{Baxter_Kelland_Wu}. Each
edge of the medial lattice then carries an arrow, and these arrows are
conserved at the vertices: the net arrow flux defines $S_z$ as before. The
six-vertex model again needs twisting by the operator $q^{S_z}$ to ensure the
correct weighing in the $S_z \neq 0$ sectors. The Hamiltonian of the
corresponding spin chain can be extracted by taking the anisotropic limit, and
is useful for studying the model with the Bethe Ansatz technique
\cite{Baxter_book}. The fact that this Hamiltonian commutes with the
generators of the quantum group $U_q(SU(2))$ links up with the nice results of
Saleur and coworkers \cite{Saleur_89,Pasquier_90,Saleur_90,Saleur_91}.

Finally, the {\em RSOS representation}
\cite{Pasquier_87,Saleur_89,Pasquier_90} emerges from a certain simplification
of the above representations when $q=\exp(i\pi/p)$ is a root of unity (see
below).

All these formulations of the Potts model can be conveniently studied through
the corresponding transfer matrix spectra: these give access to the limiting
curves $\scrb_L$, correlation functions, critical exponents, etc.

In the FK representation the transfer matrix $\T_\text{FK}^{(2)}(L)$
is written in a basis of connectivities (set partitions) between {\em
two} time slices of the lattice (see Ref.~\cite{transfer4} for
details), and the transfer matrix propagates just one of the time
slices. Each independent connection between the two slices is called a
bridge; the number of bridges $j$ is a semi-conserved quantum number
in the sense that it cannot increase upon action of the transfer
matrix. The bridges serve to correctly weight the clusters that are
non-contractible with respect to the cyclic boundary conditions.%
\footnote{In particular, the restriction of $\T_\text{FK}^{(2)}(L)$ to
the zero-bridge sector is just the usual transfer matrix
$\T_\text{FK}$ in the FK representation, i.e., the matrix used in
Ref.~\cite{transfer1} to study the case of fully free boundary
conditions. \label{fn_FK}}
This is accomplished by writing the partition function as
\be
  Z_\text{FK} = \langle f | \T_\text{FK}^{(2)}(L)^N | i \rangle
                   = \sum_{i\geq 1} \alpha_i \, \lambda^N_i
  \label{sandwich}
\ee
for suitable initial and final vectors $| i \rangle$ and $\langle f |$. The
vector $| i \rangle$ identifies the two time slices, while $\langle f |$
imposes the periodic boundary conditions (it ``reglues'' the time slices) and
weighs the resulting non-contractible clusters. Note that these vectors
conspire to multiply the contribution of each eigenvalue $\lambda_i$ by an
amplitude $\alpha_i=\alpha_i(Q)$: this amplitude may vanish for certain values
of $Q$.

On the other hand, in the six-vertex representation the transfer matrix is
written in the purely {\em local} basis of arrows, whence the partition
function can be obtained as a trace (which however has to be twisted by
inserting $q^{S_z}$ as described above). But even without the twist
the eigenvalues are still associated with non-trivial amplitudes, as we
now review.

Let us consider first a generic value of $q$, i.e., an irrational value of
$p$. The $U_q(SU(2))$ symmetry of the spin chain Hamiltonian implies that one
can classify eigenvalues according to their value $j$ of $S_z$, and consider
only highest weights of spin $j$. Define now $K_{1,2j+1}(p,L;x)$ as the
generating function of the highest weights of spin $j$, for given values of
$p$, $L$ and $x$. The partition function of the untwisted six-vertex model
with the spin $S$ (not $S_z$) fixed to $j$ is therefore $(2j+1) \,
K_{1,2j+1}(p,L;x)$. Imposing the twist, the corresponding contribution to the
partition function of the Potts model becomes $S_{j}(p) \, K_{1,2j+1}(p,L;x)$,
where the $q$-deformed number $S_{j}(p) \equiv (2j+1)_q$ is defined as follows
\be
  S_{j}(p) \;=\; \frac{ \sin (\pi (2j+1) /p) }{\sin (\pi/p) } \,.
\label{def_Sh}
\ee
$S$ has a simple interpretation in the FK representation as the number of
bridges, whereas it is $S_z$ which has a simple interpretation in the
six-vertex model representation as the conserved current.

Different representations correspond to choosing different basis states:  a
given cluster state is an eigenvector of $S$, but not
$S_z$, and a given vertex state is an eigenvector of
$S_z$, but not $S$.
The eigenvectors of the Hamiltonian are eigenvectors of both $S$
and $S_z$, and are thus combinations of vertex states (or of cluster states if
one works in the FK representation). But note that the dimensions of the
transfer matrix are not exactly the same in the vertex and the FK
representations, as the $2j+1$ possible values of $S_z$ for a given $S=j$ are
not taken into account in the same way:  in the vertex representation, it
corresponds to a degeneracy of the eigenvalues, whereas in the FK
representation it appears because of the initial and final vectors which
sandwich the transfer matrix in Eq.~(\ref{sandwich}).

The total partition function of the $Q$-state Potts model on a strip of size
$L_\text{F}\times N_\text{P}$ can therefore be {\em exactly} written as
\cite{Pasquier_87,Saleur_89,Pasquier_90}
\be
 Z_{L_\text{F}\times N_\text{P}}(Q;v) \;=\;
 Q^{L N/2} \sum_{j=0}^{L} S_{j}(p) \, K_{1,2j+1}(p,L;x)
\label{def_Z_uq}
\ee
Note that the summation is for $0 \le j \le L$, as the maximum number of
bridges is equal to the strip width $L$.

For $p$ rational, Eq.~(\ref{def_Z_uq}) is still correct, but can be
considerably simplified. In the context of this paper we only consider the
simplest case of $p$ integer. Indeed, note that using Eq.~(\ref{def_Sh}), we
obtain that, for any integer $n$,
\begin{subeqnarray}
 S_{(n+1)p-1-j}(p) &=& -S_{j}(p) \\
 S_{np+j}(p)       &=& S_{j}(p) \,.
 \label{twosyms}
\end{subeqnarray}
Therefore, after factorization, Eq.~(\ref{def_Z_uq}) can be rewritten as
\be
 Z_{L_\text{F}\times N_\text{P}}(Q;v) \;=\;
 Q^{L N/2} \sum_{j=0}^{\lfloor (p-2)/2 \rfloor} S_{j}(p) \,
 \chi_{1,2j+1}(p,L;x) \,,
\label{def_Z_RSOS}
\ee
where 
\be
 \chi_{1,2j+1}(p,L;x)\;=\;
 \sum_{n \ge 0} \left[ K_{1,2(np+j)+1}(p,L;x) -
                       K_{1,2((n+1)p-1-j)+1}(p,L;x) \right] \,.
\label{comb_K}
\ee
For convenience in writing Eq.~(\ref{comb_K}) we have defined
$K_{1,2j+1}(p,L;x) \equiv 0$ for $j > L$. Note that the summation in
Eq.~(\ref{def_Z_RSOS}) is now for $0 \le j \le \lfloor (p-2)/2 \rfloor$.
Furthermore, $\chi_{1,2j+1}(p,L;x)$ is a lot simpler that it seems. Indeed,
when $p$ is integer, the representations of $U_q(SU(2))$ mix different values
of $j$ related precisely by the transformations $j \to j+np$ and $j \to
(n+1)p-1-j$ [cf.~Eq.~\reff{twosyms}]. Therefore, a lot of eigenvalues cancel
each other in Eq.~(\ref{comb_K}). This is exactly why the transfer matrix in
the FK representation contains spurious eigenvalues, and is not adapted to the
case of $p$ integer.

The representation adapted to the case of $p$ integer is the so-called RSOS
representation. It can be proved that $\chi_{1,2j+1}$ is the partition function
of an RSOS model of the $A_{p-1}$ type \cite{Pasquier_87} with given boundary
conditions \cite{Saleur_89} (see below). In this model, heights
$h_i=1,2,\ldots,p-1$ are defined on the union of vertices and dual vertices of
the original Potts spin lattice. Neighboring heights are restricted to differ
by $\pm 1$ (whence the name RSOS = restricted solid-on-solid). The boundary
conditions on the heights are still periodic in the longitudinal direction,
but {\em fixed} in the transverse direction. More precisely, the cyclic strip
$L_{\rm F} \times N_{\rm P}$ has precisely two exterior dual vertices, whose
heights are fixed to $1$ and $2j+1$ respectively. It is convenient to draw the
lattice of heights as in Figures~\ref{fig_sq_L=2}--\ref{fig_tri_L=2} (showing
respectively a square and a triangular-lattice strip of width $L=2$), i.e.,
with $N$ exterior vertices above the upper rim, and $N$ exterior vertices
below the lower rim of the strip: all these exterior vertices close to a given
rim are then meant to be identified.

For a given lattice of spins, the weights of the RSOS model are most easily
defined by building up the height lattice face by face, using a transfer
matrix. The transfer matrix adding one face at position $i$ is denoted $H_i =
x I_i + e_i$ (resp.\ $V_i = I_i + x e_i$) if it propagates a height $h_i \to
h'_i$ standing on a direct (resp.\ a dual) vertex, where $I_i =
\delta(h_i,h'_i)$ is the identity operator, and $e_i$ is the Temperley-Lieb
generator in the RSOS representation \cite{Pasquier_87}:
\be
 e_i = \delta(h_{i-1},h_{i+1})
 \frac{\left[\sin(\pi h_j/p) \sin(\pi h'_j/p)\right]^{1/2}}
 {\sin(\pi h_{j-1}/p)} \,.
 \label{TLweights}
\ee

Note that all the amplitudes $S_j(p)$ entering in Eq.~(\ref{def_Z_RSOS}) are
strictly positive. Therefore, for a generic value of the temperature $x$, all
the eigenvalues associated with $\chi_{1,2j+1}(p,L;x)$ for $0<2j+1<p$
contribute to the partition function.%
\footnote{For exceptional values of $x$ there may still be cancellations
between eigenvalues with opposite sign. However, the pair of eigenvalues
that cancel must now necessarily belong to the {\em same} sector
$\chi_{1,2j+1}$.}
This is the very reason why we use the RSOS representation. Recall that there
are analogous results in conformal field theory \cite{Yellow_book}. In fact,
for $x$ equal to $x_{\rm FM}(Q)$ and in the continuum limit, $K_{1,2j+1}$
corresponds to the generating function of a generic representation of the
conformal symmetry with Kac-table indices $r=1$ and $s=2j+1$, whereas
$\chi_{1,2j+1}$ corresponds to the generating function (character) of a
minimal model. Thus, Eq.~(\ref{comb_K}) corresponds to the Rocha-Caridi
equation \cite{Rocha-Caridi}, which consists of taking into account the null
states. One could say that the FK representation does not identify all the
states differing by null states, whereas the RSOS representation does.
Therefore, the dimension of the transfer matrix is smaller in the RSOS
representation than in the FK representation.

The computation of the partition functions $\chi_{1,2j+1}(p,L;x)$ can be done 
in terms of transfer matrices $\T_{1,2j+1}$, denoted in the following simply
by $\T_{2j+1}$. In particular, for a strip of size $L\times N$, we have that
\be
 \chi_{1,2j+1}(p,L;x) \;=\; \tr \T_{2j+1}(p,L;x)^N 
 \label{def_chi}
\ee
Note that this is a completely standard untwisted trace. The transfer matrix
$\T_{2j+1}(L;x)$ acts on the space spanned by the vectors
$|h_0,h_1,\ldots,h_{2L}\rangle$, where the boundary heights $h_0=1$ and
$h_{2L}=2j+1$ are fixed. The dimensionality of this space is discussed in
Appendix~\ref{sec_dim}. For any fixed $h_0$ and $h_{2L}$, this dimensionality
grows asymptotically like $\sim Q^L$.

\medskip

\noindent
{\bf Remarks}. 1) Our numerical work is based on an automatized construction of
$\T_{2j+1}$. To validate our computer algorithm, we have verified that
Eq.~\reff{def_Z_RSOS} is indeed satisfied. More precisely, given $Q=B_p$, and
for fixed $L$ and $N$, we have verified that
\be
Z_{L_{\text F}\times N_\text{P}}(Q; v) \;=\; 
Q^{LN/2} \sum\limits_{0<2j+1<p} S_{j}(p)\,  \chi_{1,2j+1}(p,L;x) \;=\; 
Z_{N_\text{P}\times L_\text{F}}(Q;v)
\label{check_Z}
\ee
where $Z_{N_\text{P}\times L_\text{F}}(Q;v)$ is the partition function of the
$Q$-state Potts model on a strip of size $N_\text{P}\times L_\text{F}$ with
cylindrical boundary conditions, as computed in
Refs.~\cite{Tutte_sq,Tutte_tri}. We have made this check for $p=4,5,6$ and
for several values of $L$ and $N$.

2) For $p=3$ the RSOS model trivializes. Only the $\chi_{1,1}$ sector exists,
and $\T_1$ is one-dimensional for all $L$. Eq.~(\ref{def_Z_RSOS}) gives
simply
\be
 Z_{L_{\text F}\times N_\text{P}}(Q=1;x) = (1+x)^E \;,
\ee
where $E$ is the number of lattice edges (faces on the height lattice).
It is not possible to treat the bond percolation problem in the RSOS context,
since this necessitates taking $Q\to 1$ as a {\em limit}, and not to
sit directly at $Q=1$. Hence, the right representation for studying 
bond percolation is the FK representation.

\section{Square-lattice Potts model with free cyclic boundary conditions}
%
%

\subsection{Ising model ($\bm{p=4}$)}
\label{sec_sq_p=4}

The partition function for a strip of size $L_\text{F}\times N_\text{P}$ is
given in the RSOS representation as
\be
Z_{L_\text{F}\times N_\text{P}}(Q=2;x) \;=\; 2^{N L/2} \left[ 
              \chi_{1,1}(x)+ \chi_{1,3}(x) \right]
\ee
where $\chi_{1,2j+1}(x) = \tr \T_{2j+1}(p=4,L;x)^N$. 
The dimensionality of the transfer matrices can be obtained from the
general formulae derived in Appendix~\ref{sec_dim}: 
\be
\dim \T_k(p=4,L) \;=\; 2^{L-1} \,, \qquad k=1,3  
\ee 

We have computed the limiting curves $\mathcal{B}_L$ for $L=2,3,4$. These
curves are displayed in Figure~\ref{Curves_sq_p=4}(a)--(c). In
Figure~\ref{Curves_sq_p=4}(d), we show simultaneously all three curves for
comparison. In addition, we have computed the partition-function zeros
for finite strips of dimensions $L_\text{F}\times (\rho L)_\text{P}$ for 
aspect rations $\rho=10,20,30$. These zeros are also displayed in 
Figure~\ref{Curves_sq_p=4}(a)--(c).\footnote{
  After the completion of this work, we learned that Chang and Shrock
  had obtained the limiting curves for 
  $L=2$ \protect\cite[Figure~20]{Shrock_00b} and 
  $L=3$ \protect\cite[Figure~7]{Shrock_01a}. The eigenvalues and
  amplitudes for $L=2$ had been previoulsy published by Shrock 
  \protect\cite[Section~6.13]{Shrock_00a}. 
}
For $5\leq L\leq 8$, we have only computed selected features of
the corresponding limiting curves (e.g., the phase diagram for real $x$).

%
%
\subsubsection{$\bm{L=2}$} \label{sec.sq.ising.L=2}

This strip is displayed in Figure~\ref{fig_sq_L=2}. Let us denote the basis 
in the height space as $| h_1,h_2,h_3,h_4,h_5\rangle$, 
where the order is given as in Figure~\ref{fig_sq_L=2}.

The transfer matrix $\T_1$ is two-dimensional: 
in the basis $\{|1,2,1,2,1\rangle, |1,2,3,2,1\rangle\}$, it takes the form 
\be
\T_1(p=4,L=2) \;=\; \frac{1}{\sqrt{2}}\, 
   \left( \begin{array}{cc}
           Y_{2,0} & Y_{2,1} \\
           Y_{2,3} & Y_{2,2}  
           \end{array}
   \right)
\label{def_T_sq_p=4_L=2_k=1}
\ee
where we have used the shorthand notation
\be
Y_{L,k} \;=\;  x^k \, \left(x+\sqrt{2}\right)^{2L-1-k}\,, \qquad
               k=0,\ldots,2L-1
\label{def_Y}
\ee

The transfer matrix $\T_3$ is also two-dimensional: in the basis 
$\{ |1,2,1,2,3\rangle,|1,2,3,2,3\rangle\}$, it takes the form
\be
\T_3(p=4,L=2) \;=\; \frac{1}{\sqrt{2}}\, 
\left( \begin{array}{cc}
       Y_{2,1} & Y_{2,2} \\
       Y_{2,2} & Y_{2,1}
       \end{array}
\right)
\label{def_T_sq_p=4_L=2_k=3}
\ee

For real $x$, there is a single phase-transition point at 
\be
x_{\rm c} \;=\; -\frac{1}{\sqrt{2}} \approx -0.7071067812
\ee 
This point is actually a multiple point.%
\footnote{
    Throughout this paper a point on $\scrb_L$ of order $\ge 4$ is
    referred to as a multiple point.
}
There is an additional pair of complex conjugate multiple points at
$x=-e^{\pm i\pi/4} = -1/\sqrt{2} \pm i/\sqrt{2}$. We also find an isolated
limiting point at $x=-\sqrt{2}$ due to the vanishing of all the eigenvalues
(see Ref.~\cite{Sokal_04} for an explanation of this issue in terms of the
Beraha--Kahane--Weiss theorem).

The dominant sector on the real $x$-axis is always $\chi_{1,1}$, except at
$x=-\sqrt{2}$ and $x=-1/\sqrt{2}$; at these points the dominant eigenvalues
coming from each sector $\chi_{1,k}$ become equimodular. 
On the regions with null intersection with the real $x$-axis, the dominant
eigenvalue comes from the sector $\chi_{1,3}$.

%
%
\subsubsection{$\bm{L\geq 3}$}

For $3\leq L\leq 8$, we find two phase-transition points on the real $x$-axis: 
\begin{subeqnarray}
x_{{\rm c},1} &=& -\frac{1}{\sqrt{2}} \approx -0.7071067812\\
x_{{\rm c},2} &=& -\sqrt{2}           \approx -1.4142135624
\end{subeqnarray}  
Both points are actually multiple points (except $x_{{\rm c},2}$ for $L=3$).
There is an additional pair of complex conjugate multiple points at
$x=-e^{\pm i\pi/4}$.

For $x>x_{{\rm c},1}$, the dominant eigenvalue always belongs to the sector
$\chi_{1,1}$. For $x<x_{{\rm c},1}$, this property is true only for even
$L=4,6,8$; for odd $L=3,5,7$, the dominant eigenvalue for $x<x_{{\rm c},1}$
belongs to the $\chi_{1,3}$ sector.

%
%
\subsection{$\bm{Q=B_5}$ model ($\bm{p=5}$)}
\label{sec_sq_p=5}

The partition function for a strip of size $L_\text{F}\times N_\text{P}$ is
given in the RSOS representation as
\be
Z_{L_\text{F}\times N_\text{P}}(Q=B_5;x) \;=\; B_5^{N L/2} \, \left[ 
              \chi_{1,1}(x)+ \sqrt{B_5}\, \chi_{1,3}(x) \right] 
\ee
where $\chi_{1,2j+1}(x) = \tr \T_{2j+1}(p=5,L;x)^N$.

We have computed the limiting curves $\mathcal{B}_L$ for $L=2,3,4$. 
These curves are displayed in Figure~\ref{Curves_sq_p=5}(a)--(c). 
In Figure~\ref{Curves_sq_p=5}(d), we show all three curves for comparison. 
For $L=5,6$, we have only computed selected features of the 
corresponding limiting curves. 

%
%
\subsubsection{$\bm{L=2}$}

The transfer matrix $\T_1$ is two-dimensional: 
in the basis $\{ |1,2,1,2,1\rangle , |1,2,3,2,1\rangle \}$,  
it takes the form
\be
\T_1(p=5,L=2) \;=\; \left( \begin{array}{cc}
   \sqrt{B_5^*} X_3 & B_5^{1/4} \sqrt{B_5^*} \, x X_3^2 \\
   B_5^{*\, 1/4} \, x^3  & x^2(1+x)  
     \end{array}
     \right)
\label{def_T_sq_p=5_L=2_k=1}
\ee
where we have used the shorthand notation
\be
X_3 \;=\; x+\sqrt{B_5}  \,, \quad  X_3^* \;=\; x+\sqrt{B_5^*}
\ee 
in terms of $B_5$ and $B_5^*$ defined as 
\be
B_5   \;=\; \frac{3+\sqrt{5}}{2} \,, \qquad  B_5^* \;=\; \frac{3-\sqrt{5}}{2}  
\ee 

The transfer matrix $\T_3$ is three-dimensional. In the basis
$\{ |1,2,1,2,3\rangle,  |1,2,3,4,3\rangle$,  $|1,2,3,2,3\rangle\}$,  
it takes the form
\be
\T_3(p=5,L=2) \;=\; \left( \begin{array}{ccc}
 \sqrt{B_5^*} \, x X_3^2 & 0        & B_5^{*\, 1/4} \, x^2 X_3 \\
 \sqrt{B_5^*} \, x^2     & x X_3^*  & B_5^{*\, 1/4} \, x(1+x)  \\
 B_5^{*\, 1/4}\, x^2(1+x)&  B_5^{*\, 1/4}\, x & x(1+x)^2
     \end{array}
     \right)
\label{def_T_sq_p=5_L=2_k=3}
\ee

For real $x$, there is a single phase-transition point at
\be
x_{\rm c} \;=\; -\frac{\sqrt{B_5}}{2} \;=\; -\frac{1+\sqrt{5}}{4} 
                                \;\approx\; -0.8090169944
\ee
We have also found that the limiting curve contains a horizontal line
between $x=x_\text{BK}=-1$ and $x\approx -1.3843760945$. 
The latter point is a T point, and the former one, a multiple point.
There is an additional pair of complex conjugate multiple points at
\be
x\;=\;-e^{\pm i\pi/5} \;=\; 
-\frac{1+\sqrt{5}}{4} \pm \frac{i}{2} (5B_5^*)^{1/4}
\;\approx\; -0.8090169944 \pm 0.5877852523\,i 
\ee
We have found two additional pairs of complex conjugate T points at
$x\approx -1.5613823329 \pm 0.3695426938\,i$, and   
$x\approx -0.9270509831 \pm 0.3749352940\,i$.       
The dominant sectors on the real $x$-axis are 
\begin{itemize}
\item $\chi_{1,1}$ for $x\in(-\infty,-1.3843760945)\cup(-0.8090169944,\infty)$ 
\item $\chi_{1,3}$ for $x\in(-1.3843760945,-0.8090169944)$
\end{itemize}

%
%
\subsubsection{$\bm{L\geq 3}$}

For $L=3$ there are two real phase-transition points at
\begin{subeqnarray}
x_{{\rm c},1} &\approx& -2.1862990086 \\
x_{{\rm c},2} &\approx& -0.9176152641 
\end{subeqnarray}
The limiting curve contains a horizontal line
between two real T points $x\approx -1.2066212246$ and 
$x\approx -0.9713270390$. There are nine additional pairs of complex conjugate 
T points. 
The dominant sectors on the real $x$-axis are 
\begin{itemize}
\item $\chi_{1,1}$ for $x\in(-2.1862990086,-1.2066212246)\cup
                            (-0.9176152641,\infty)$
\item $\chi_{1,3}$ for $x\in(-\infty,-2.1862990086)\cup
                            (-1.2066212246,-0.9176152641)$
\end{itemize}

For $L=4$, the real transition points are located at
\begin{subeqnarray}
x_{{\rm c},1} &\approx& -1.3829734471 \\ 
x_{{\rm c},2} &\approx& -0.9475070976
\end{subeqnarray}
We have found that the curve $\mathcal{B}_4$ contains a horizontal line
between two real T points:
$x\approx -1.1982787848$ and $x\approx -0.9776507663$. 
Two points belonging to such line are actually 
multiple points: $x\approx -0.9923357481$ and $x\approx -0.9972135728$.
We have found 34 pairs of complex conjugate T points.
The phase diagram is rather involved, and we find several tiny closed
regions.
The dominant sectors on the real $x$-axis are 
\begin{itemize}
\item $\chi_{1,1}$ for $x\in(-\infty,x_{{\rm c},1})\cup
                            (-1.1982787848,-0.9972135728)\cup 
                            (x_{{\rm c},2},\infty)$
\item $\chi_{1,3}$ for $x\in(x_{{\rm c},1},-1.1982787848)\cup
                            (-0.9972135728,x_{{\rm c},2})$
\end{itemize}

For $L=5$, there are four real phase-transition points at
\begin{subeqnarray}
x_{{\rm c},1} &\approx& -2.4492425881\\
x_{{\rm c},2} &\approx& -1.2097913730\\
x_{{\rm c},3} &\approx& -1.1717714277\\ 
x_{{\rm c},4} &\approx& -0.9616402644 
\end{subeqnarray}
Again, $\mathcal{B}_5$ contains a horizontal line
between $x\approx -1.1323655119$ and $x\approx -0.9770339631$. 
The dominant sectors on the real $x$-axis are 
\begin{itemize}
\item $\chi_{1,1}$ for $x\in(x_{{\rm c},1},-0.9770339631)\cup(x_{{\rm c},4},\infty)$
\item $\chi_{1,3}$ for $x\in(-\infty,x_{{\rm c},1})\cup (-0.9770339631,x_{{\rm c},4})$
\end{itemize}

Finally, for $L=6$, there are five real phase-transition points at
\begin{subeqnarray}
x_{{\rm c},1} &\approx& -1.2750054535\\
x_{{\rm c},2} &\approx& -1.2712112920\\
x_{{\rm c},3} &\approx& -1.1323753929\\ 
x_{{\rm c},4} &\approx& -1.1052066740\\  
x_{{\rm c},5} &\approx& -0.9700021428 
\end{subeqnarray}
The limiting curve contains a horizontal line
between two real T points: $x\approx -1.0877465961$ and 
$x\approx -0.9792223546$. This line contains the multiple point 
$x\approx -1.0781213888$. 
The dominant sectors on the real $x$-axis are 
\begin{itemize}
\item $\chi_{1,1}$ for $x\in(-\infty,x_{{\rm c},1})\cup 
                            (x_{{\rm c},2},-1.0877465961)\cup
                            (-1.0781213888,\infty)$
\item $\chi_{1,3}$ for $x\in(x_{{\rm c},1},x_{{\rm c},2})\cup
                            (-1.0877465961,-1.0781213888)$
\end{itemize}

In all cases $2\leq L\leq 6$, there is a pair of complex conjugate 
multiple points at 
$ x\;=\;-e^{\pm i\pi/5} \approx -0.8090169944 \pm 0.5877852523\,i$. 

%
%
\subsection{Three-state Potts model ($\bm{p=6}$)}
\label{sec_sq_p=6}

The partition function for a strip of size $L_\text{F}\times N_\text{P}$ is
given in the RSOS representation as
\be
Z_{L_\text{F}\times N_\text{P}}(Q=3;x) \;=\; 3^{N L/2} \left[
              \chi_{1,1}(x)+ 2\chi_{1,3}(x) + \chi_{1,5}(x)\right]
\ee
where $\chi_{1,2j+1}(x)=\tr \T_{2j+1}(p=6,L;x)^N$.

We have computed the limiting curves $\mathcal{B}_L$ for $L=2,3,4$.
These curves are displayed in Figure~\ref{Curves_sq_p=6}(a)--(c).\footnote{
  After the completion of this work, we learned that the limiting curves 
  for the smallest widths had been already obtained 
  by Chang and Shrock: namely, $L=2$ \protect\cite[Figure~22]{Shrock_01a}, 
  and $L=3$ \protect\cite[Figure~8]{Shrock_01a}. 
  Please note that in the latter case, they used the variable
  $u=1/(v+1)=1/(x\sqrt{Q}+1)$, instead of our variable $x$.
}

In Figure~\ref{Curves_sq_p=6}(d), we show all three curves for comparison.
For $L=5,6,7$ we have only computed selected features of the
corresponding limiting curves.

%
%
\subsubsection{$\bm{L=2}$}

The transfer matrix $\T_5$ is one-dimensional, as there is a single 
basis vector $\{ |1,2,3,4,5\rangle\}$. The matrix is given by  
\be
\T_5(p=6,L=2) = x^2
\label{def_T_sq_p=6_L=2_k=5}
\ee

The transfer matrix $\T_1$ is two-dimensional: in the basis  
$\{ |1,2,1,2,1\rangle, |1,2,3,2,1\rangle\}$, it takes the form
\be
\T_1(p=6,L=2) \;=\; \frac{1}{\sqrt{3}} \, \left( \begin{array}{cc}
      X_1^3          & \sqrt{2}\,  x \, X_1^2 \\
     \sqrt{2}\, x^3  & x^2 X_2 
     \end{array}
     \right)
\label{def_T_sq_p=6_L=2_k=1}
\ee
where we have used the shorthand notation
\be
X_1 \;=\;  x+\sqrt{3} \,, \quad  X_2 \;=\; 2x+\sqrt{3} 
\ee

The transfer matrix $\T_3$ is three-dimensional. In the basis  
$\{ |1,2,1,2,3\rangle, |1,2,3,4,3\rangle$, $|1,2,3,2,3\rangle\}$,  
it takes the form
\be
\T_3(p=6,L=2) \;=\; \frac{1}{2\sqrt{3}}\, \left( \begin{array}{ccc}
 2\, x X_1^2       & 0                & 2 \sqrt{2}\, x^2 X_1 \\
 \sqrt{6}\, x^2    & \sqrt{3}\, x X_2 & \sqrt{3}\, x X_2     \\
 \sqrt{2}\, x^2 X_2& 3\, x            & x X_2^2
     \end{array}
     \right)
\label{def_T_sq_p=6_L=2_k=3}
\ee

For real $x$, there are two phase-transition points 
\begin{subeqnarray}
x_{{\rm c},1} &=& -\sqrt{3} \;=\; x_{-} \;\approx\; -1.7320508076\\ 
x_{{\rm c},2} &=& -\frac{\sqrt{3}}{2}         \;\approx\; -0.8660254038 
\end{subeqnarray}
There is one pair of complex conjugate T points at
$x\approx -1.6522167507 \pm 0.5104474197\,i$. There are three multiple 
points at $x=-\sqrt{3}/2$, and $x=-\sqrt{3}/2 \pm i/2 = -e^{\pm i\pi/6}$.
The dominant sectors on the real $x$-axis are 
\begin{itemize}
\item $\chi_{1,1}$ for $x\in(-\infty,-\sqrt{3})\cup(-\sqrt{3}/2,\infty)$ 
\item $\chi_{1,5}$ for $x\in (-\sqrt{3},-\sqrt{3}/2)$ 
\end{itemize}
On the regions with null intersection with the real $x$-axis, the dominant
eigenvalue comes from the sector $\chi_{1,3}$.

%
%
\subsubsection{$\bm{L\geq 3}$}

For $L=3$, there are three real phase-transition points 
\begin{subeqnarray}
x_{{\rm c},1} &\approx& -1.9904900679 \\ 
x_{{\rm c},2} &=& -\sqrt{3} \;=\; x_{-} \;\approx\; -1.7320508076\\ 
x_{{\rm c},3} &=& -\frac{\sqrt{3}}{2}         \;\approx\; -0.8660254038 
\end{subeqnarray}
The limiting curve contains a small horizontal segment running 
from $x\approx -1.0539518478$ to $x= x_\text{BK}=-1$. On this line, the 
two dominant equimodular eigenvalues come from the sector $\chi_{1,5}$.

We have found 15 T points (one real point and seven pairs of complex conjugate
T points). The real point is $x=-1$. 
The phase structure is vastly more complicated than that for $L=2$. 
In particular, it contains three non-connected pieces, and four bulb-like
regions.  
On the real $x$-axis, the dominant eigenvalue comes from 
\begin{itemize}
\item $\chi_{1,1}$ for $x\in(x_{{\rm c},1},-\sqrt{3})\cup(-\sqrt{3}/2,\infty)$
\item $\chi_{1,3}$ for $x\in(-\infty,x_{{\rm c},1})\cup (-\sqrt{3},-1.0539518478)$
\item $\chi_{1,5}$ for $x\in(-1.0539518478,-\sqrt{3}/2)$
\end{itemize} 

For $L=4$, there are four phase-transition points 
\begin{subeqnarray}
x_{{\rm c},1} &=& -\sqrt{3} \;=\; x_{-} \;\approx\; -1.7320508076\\ 
x_{{\rm c},2} &\approx& -1.3678583305 \\ 
x_{{\rm c},3} &\approx& -1.2237725061 \\ 
x_{{\rm c},4} &=& -\frac{\sqrt{3}}{2}         \;\approx\; -0.8660254038 
\end{subeqnarray}
This is the strip with smallest width for which a (complex conjugate)
pair of endpoints appears: 
$x\approx -0.9951436066 \pm 0.00444309186\, i$. These points are very close
to the transition point $x_\text{BK}=-1$.
We have found 36 pairs of conjugate T points. 
We have also found three multiple points at $x=-\sqrt{3}$, and 
$x=-\sqrt{3}/2 \pm i/2$.
The dominant sectors on the real $x$-axis are 
\begin{itemize}
\item $\chi_{1,1}$ for $x\in(-\infty,x_{{\rm c},3})\cup(-\sqrt{3}/2,\infty)$
\item $\chi_{1,5}$ for $x\in(x_{{\rm c},3},-\sqrt{3}/2)$
\end{itemize}

For $L=5$, there are six real phase-transition points 
\begin{subeqnarray}
x_{{\rm c},1} &\approx& -2.3018586529 \\ 
x_{{\rm c},2} &=& -\sqrt{3} \;=\; x_{-} \;\approx\; -1.7320508076\\ 
x_{{\rm c},3} &\approx& -1.4373407728 \\
x_{{\rm c},4} &\approx& -1.3412360954 \\ 
x_{{\rm c},5} &\approx& -1.2613579653 \\
x_{{\rm c},6} &=& -\frac{\sqrt{3}}{2}         \;\approx\; -0.8660254038 
\end{subeqnarray}
We have also found a horizontal line running between the T points  
$x\approx -1.0226306002$ and $x\approx -0.9984031794$. 
The dominant sectors on the real $x$-axis are 
\begin{itemize}
\item $\chi_{1,1}$ for $x\in(x_{{\rm c},1},x_{{\rm c},3})\cup(-\sqrt{3}/2,\infty)$
\item $\chi_{1,3}$ for $x\in(-\infty,x_{{\rm c},1})\cup(x_{{\rm c},3},-1.0226306002)$
\item $\chi_{1,5}$ for $x\in(-1.0226306002,-\sqrt{3}/2)$
\end{itemize}

For $L=6$, there are also six phase-transition points on the real axis 
\begin{subeqnarray}
x_{{\rm c},1} &=& -\sqrt{3} \;=\; x_{-} \;\approx\; -1.7320508076\\ 
x_{{\rm c},2} &\approx&  -1.2852299467   \\
x_{{\rm c},3} &\approx&  -1.2238569234   \\
x_{{\rm c},4} &\approx&  -1.1271443188   \\ 
x_{{\rm c},5} &\approx&  -1.0085262838   \\ 
x_{{\rm c},6} &=& -\frac{\sqrt{3}}{2}         \;\approx\; -0.8660254038 
\end{subeqnarray}
The dominant sectors on the real $x$-axis are 
\begin{itemize}
\item $\chi_{1,1}$ for $x\in(-\infty,x_{{\rm c},2})\cup(x_{{\rm c},3},x_{{\rm c},5})\cup
                            (-\sqrt{3}/2,\infty)$ 
\item $\chi_{1,5}$ for $x\in(x_{{\rm c},2},x_{{\rm c},3})\cup(x_{{\rm c},5},-\sqrt{3}/2)$ 
\end{itemize}

In all cases $2\leq L\leq 6$, we have found three multiple points at
$x= -\sqrt{3}$, and                         
$x= -\sqrt{3}/2 \pm i/2 = -e^{\pm i\pi/6}$. 

\subsection{Four-state Potts model ($p=\infty$)}

It follows from the RSOS constraint and the fact that $h_0=1$ is fixed, that
the maximal height participating in a state is $h_{\rm
max}=\text{max}(2L,p-1)$. In particular, for any fixed $L$ the number of
states stays finite when one takes the limit $p\to\infty$. Meanwhile,
the Boltzmann weight entering in Eq.~(\ref{TLweights}) has the well-defined
limit $(h_j h'_j)^{1/2}/h_{j-1}$, and the amplitudes (\ref{def_Sh}) tend
to $S_j(\infty)=2j+1$. We shall refer to this limit as the $p=\infty$
(or $Q=4$) model.

We have computed the limiting curves $\mathcal{B}_L$ for $L=2,3,4$.
These curves are displayed in Figure~\ref{Curves_sq_p=Infty}(a)--(c).
In Figure~\ref{Curves_sq_p=Infty}(d), we show all three curves for comparison.

%
%
\subsubsection{$\bm{L=2}$}

The transfer matrices are
\begin{subeqnarray}
\T_1 &=& \frac{1}{2}\,
\left( \begin{array}{cc}
  (x+2)^3        & \sqrt{3}\, x(x+2)^2 \\
  \sqrt{3}\, x^3 & x^2 (2+3x)   
       \end{array}
\right) \\
T_3 &=&  \frac{1}{6}\,
\left( \begin{array}{ccc}
3x(x+2)^2             &   0           & 3\sqrt{3}\, x^2 (x+2)\\
2\sqrt{6}\, x^2       & 2x(3x+4)      & 2\sqrt{2}\, x(3x+2) \\
\sqrt{3}\, x^2 (3x+2) & 4\sqrt{2}\, x & x(3x+2)^2  
\end{array}
\right) \\
T_5 &=& x^2
\end{subeqnarray}

For real $x$, we find a multiple point at $x=-1$, where all eigenvalues 
become equimodular with $|\lambda_i|=1$.
The dominant sector on the real $x$-axis is always $\chi_{1,1}$.

%
%
\subsubsection{$\bm{L\ge 3}$}

For $L=3$ there are two real phase-transition points: $x=-1$ (which is
a multiple point), and $x_c\approx -1.6424647621$. We have found ten pairs
of complex conjugate T points and a pair of complex conjugate endpoints.
The dominant sectors on the real $x$-axis are
$\chi_{1,3}$ for $x<-1$, and $\chi_{1,1}$ for $x> -1$. The sector $\chi_{1,5}$
is only dominant in two complex conjugate regions off the real $x$-axis,
and the sector $\chi_{1,7}$ is never dominant. 

For $L=4$ we only find a single real phase-transition point at $x=-1$.
We have also found 32 pairs of complex conjugate T points and two pairs
of complex conjugate endpoints. 
The dominant sector on the real $x$-axis is always $\chi_{1,1}$. There is
also two complex conjugate regions where the dominant eigenvalue comes from
the sector $\chi_{1,5}$, and the sectors $\chi_{1,7}$ and $\chi_{1,9}$ are
never dominant in the complex $x$-plane.

For $L=5$ we find four real phase-transition points at 
\begin{subeqnarray}
x_{c,1} &=& -1.9465787472 \\
x_{c,2} &=& -1.5202407889 \\
x_{c,3} &=& -1.3257163278 \\
x_{c,4} &=& -1 
\end{subeqnarray}
The dominant sectors are $\chi_{1,3}$ for 
$x\in(-\infty,x_{c,1}) \cup (x_{c,2},-1)$; and $\chi_{1,1}$ in the region 
$x\in(x_{c,1},x_{c,2})\cup (-1,\infty)$. 

For $L=4$ we only find a single real phase-transition point at $x=-1$.
The dominant sector on the real $x$-axis is always $\chi_{1,1}$.

In all cases $3\leq L \leq 5$, the point $x=-1$ is a multiple point where all
the eigenvalues are equimodular with $|\lambda_i|=1$.

%
%
\section{Common features of the square-lattice limiting curves with
free cyclic boundary conditions}
\label{sec_sq_results}

{}From the numerical data discussed in 
Sections~\ref{sec_sq_p=4}--\ref{sec_sq_p=6}, we can make the following
conjecture that states that certain points in the complex $x$-plane belong
to the limiting curve $\mathcal{B}_L$:

\begin{conjecture} \label{conj_sq_1}
For the square-lattice $Q$-state Potts model with $Q=B_p$ and widths $L \ge 2$:
\begin{enumerate}
\item The points $x=-e^{\pm i\pi/p}$ belong to the limiting curve.  
      At these points, all the eigenvalues are equimodular with 
      $|\lambda_i|=1$.\footnote{
         This property has been explicitly checked for all the widths reported
         in this paper.
      } 
      Thus, they are in general multiple points.
\item For even $p$, the point $x=-\sqrt{Q}/2$ always belongs to the limiting 
      curve $\mathcal{B}_L$.\footnote{
         This property has been verified for $p=8,10$ and $2\leq L \leq 6$.
      } 
      Furthermore, if $p=4,6$, then the point $x=-\sqrt{Q}$ also belongs 
      to $\mathcal{B}_L$.
\end{enumerate}
\end{conjecture}

The phase structure for the models considered above show certain regularities
on the real $x$-axis (which contains the physical regime of the model). 
In particular, we conclude 

\begin{conjecture} \label{conj_sq_2}
For the square-lattice $Q$-state Potts model with $Q=B_p$ and widths $L \ge 2$:
\begin{enumerate}
\item The relevant eigenvalue on the physical line $v\in[-1,\infty)$ comes 
      from the sector $\chi_{1,1}$. 
\item For even $L$, the leading eigenvalue for real $x$ comes always from 
      the sector $\chi_{1,1}$, except perhaps in an interval contained in 
      $[-\sqrt{Q},-\sqrt{Q}/2]$. 
\item For odd $L$, the leading eigenvalue for real $x$ comes from the sector
      $\chi_{1,3}$ for all $x< x_0\leq -\sqrt{Q}$, and from the sector 
      $\chi_{1,1}$ for all $x\geq -\sqrt{Q}/2$. 
\end{enumerate}
\end{conjecture}

In the limiting case $p=\infty$ the RSOS construction
simplifies. Namely, the quantum group $U_q(SU(2))$ reduces to the
classical $U(SU(2))$ (i.e., $q\to 1$), and its representations no
longer couple different $K_{1,2j+1}$,
cf.~Eq.~(\ref{twosyms}). Accordingly we have simply
$K_{1,2j+1}=\chi_{1,2j+1}$.  When increasing $p$ along the line
$x_{\rm BK}(Q)$, the sector $K_{1,2j+1}$ which dominates for {\em
irrational} $p$ will have higher and higher spin $j$ \cite{Saleur_91};
this is even true throughout the Berker-Kadanoff phase.%
\footnote{See Ref.~\cite{transfer4} for numerical evidence along the
chromatic line $x=-1/\sqrt{Q}$ which intersects the BK phase up to
$p=12$ \cite{transfer3}.}  One would therefore expect that the
$p=\infty$ RSOS model will have a dominant sector $\chi_{1,2j+1}$ with
$j$ becoming larger and larger as one approaches $x_{BK}(Q=4)=-1$.

This argument should however be handled with care. Indeed, for
$p\to\infty$ the BK phase contracts to a point, $(Q,v)=(4,-2)$, and
this point turns out to be a very singular limit of the Potts model.
In particular, one has $x_{\rm BK}=x_\pm$ for $Q=4$, and very different
results indeed are obtained depending on whether one approaches
$(Q,v)=(4,-2)$ along the AF or the BK curves (\ref{values_sq_xc}).
This is visible, for instance, on the level of the central charge,
with $c\to 2$ in the former and $c\to -\infty$ in the latter case.
To wit, taking $x\to -1$ after having fixed $p=\infty$ in the RSOS model
is yet another limiting prescription, which may lead to different results.

The phase diagrams for $Q=4$ ($p\to\infty$) do agree with the above 
general conjectures \ref{conj_sq_1}-\ref{conj_sq_2}. In particular, when
$p\to\infty$, the multiple points $-e^{\pm i\pi/p}\to -1=x_\text{BK}$ 
(Conjecture~\ref{conj_sq_1}.1) and this coincides with the point 
$-\sqrt{Q}/2$ (Conjecture~\ref{conj_sq_1}.2). On the other hand, the 
sector $\chi_{1,1}$ is the dominant one on the physical line
$v\in[-1,\infty)$ (Conjecture~\ref{conj_sq_2}.1), and we observe a parity
effect on the unphysical regime $v\in(-\infty,-1)$. For even $L$, the 
only dominant sector is $\chi_{1,1}$ in agreement with 
Conjecture~\ref{conj_sq_2}.2 (although there is no interval inside 
$[-2,-1]$ where $\chi_{1,3}$ becomes dominant). For odd $L$, 
Conjecture~\ref{conj_sq_2}.3 also holds with $x_0=-\sqrt{Q}=-2$ (at least for
$L=3,5$). For $L=2,3,4$, we find that in addition to the sectors $\chi_{1,1}$
and $\chi_{1,3}$, only the sector $\chi_{1,5}$ becomes relevant in some
regions in the complex $x$-plane.

\subsection{Asymptotic behavior for $\bm{|x|\to\infty}$}

Figures~\ref{Curves_sq_p=4}--\ref{Curves_sq_p=Infty} show a rather uncommon 
scenario: the limiting curves contain outward branches. As a matter of fact,
these branches extend to infinity (i.e., they are unbounded%
\footnote{An unbounded branch is one which does not have a finite endpoint.}),
in sharp contrast with the {\em bounded}\/ limiting curves obtained using 
{\em free} longitudinal boundary conditions \cite{Tutte_sq,Tutte_tri}.
It is important to remark that this phenomenon also holds in the 
limit $p\to\infty$, as shown in Figure~\ref{Curves_sq_p=Infty}. 
 
As $|x|\to\infty$ these branches converge to rays with definite slopes.
More precisely, our numerical data suggest the following conjecture:\footnote{
   Chang and Shrock \protect\cite{Shrock_01a} observed for $L=3$ that 
   if we plot the limiting curve in the variabe $u=1/(x\sqrt{Q} +1)$, then
   the point $u=0$ is approached at specific 
   angles $\arg u$ consistent with our Conjecture~\protect\ref{conj_sq_3}. 
}

\begin{conjecture} \label{conj_sq_3}
For any value of $p$, the limiting curve $\mathcal{B}_L$ for a square-lattice
strip has exactly $2L$ outward branches. As $|x|\to\infty$, these branches
are asymptotically rays with
\be
 \arg x \; \equiv \; \theta_n(L) \;=\;
 \pi\left( \frac{n}{L} -\frac{1}{2L}\right) \,,  
 \qquad n=1,2,\ldots,2L
\label{def_asymptotics_sq}
\ee
\end{conjecture}

By inspection of Figures~\ref{Curves_sq_p=4}--\ref{Curves_sq_p=Infty},
it is also clear that the only two sectors that are relevant
in this regime are $\chi_{1,1}$ and $\chi_{1,3}$. In particular, the
dominant eigenvalue belongs to the $\chi_{1,1}$ sector for large positive
real $x$, and each time we cross one of these outward branches, the dominant
eigenvalue switches the sector it comes from. In particular, we conjecture that

\begin{conjecture} \label{conj_sq_4}
The dominant eigenvalue for a square-lattice strip of width $L$ in the
large $|x|$ regime comes from the sector $\chi_{1,1}$ in the 
asymptotic regions
\be
\arg x \;\in\; \left(\theta_{2n-1}(L),\theta_{2n}(L)\right)\,, 
   \qquad n=1,2,\ldots,L 
\ee
In the other asymptotic regions the dominant eigenvalue
comes from the sector $\chi_{1,3}$.
\end{conjecture}

In particular, this means that for large positive $x$ the dominant sector
is always $\chi_{1,1}$. However, for large negative $x$ the dominant eigenvalue
comes from $\chi_{1,1}$ is $L$ is even, and from $\chi_{1,3}$ if $L$ is
odd. Thus, this conjecture is compatible with Conjecture~\ref{conj_sq_2}. 

An empirical explanation of this fact comes from the computation of the
asymptotic expansion for large $|x|$ of the leading eigenvalues in each
sector. It turns out that there is a unique leading eigenvalue in each sector 
$\chi_{1,1}$ and $\chi_{1,3}$ when $|x|\to\infty$. As there is a unique
eigenvalue in this regime, we can obtain it by the power method 
\cite{Golub_book}. Our numerical 
results suggest the following conjecture

\begin{conjecture} \label{conj_sq_5}
Let $\lambda_{\star,1}(L)$ (resp.\ $\lambda_{\star,3}$(L)) be the leading 
eigenvalue of the sector $\chi_{1,1}$ (resp.\  $\chi_{1,3}$) in the 
regime $|x|\to \infty$. Then
\begin{subeqnarray}
\lambda_{\star,1}(L) &=& Q^{(L-1)/2} \, x^{2L-1} \, \left[
    1 + \sum\limits_{k=1}^\infty \frac{ a_k(L) }{Q^{k/2}} \, x^{-k} \right]\\
\lambda_{\star,1}(L) - \lambda_{\star,3}(L) &=& \sqrt{Q} \, x^{L-1} + 
                       3(L-1) x^{L-2} + O(x^{L-3}) 
\end{subeqnarray} 
Furthermore, we have that 
\begin{subeqnarray}
a_1(L) &=& 2L-1 \,,      \qquad L\geq 2 \\
a_2(L) &=& 2L^2 -3L + 1 \,, \qquad L\geq 3 
\label{conj.sq.ak}
\end{subeqnarray}
\end{conjecture}

The first coefficients $a_k(L)$ are displayed in Table~\ref{table_sq_coef};
the patterns displayed in \reff{conj.sq.ak} are easily verified. The 
coefficients $a_k(L)$ also depend on $p$ for $k\geq 3$. 

Indeed, the above conjecture explains easily the observed pattern for the
leading sector when $x$ is real. But it also explains the observed pattern
for all the outward branches. These branches are defined by the equimodularity
of the two leading eigenvalues
\be
|\lambda_{\star,1}| \;=\; |\lambda_{\star,3}| \;=\; 
\left|\lambda_{\star,1} - \sqrt{Q} x^{L-1} + O(x^{L-2})\right| 
\ee
This implies that
\be
\Re\left[ \lambda_{\star,1} \, \widebar{x}^{L-1} \right] \;=\; 0
\ee
where $\widebar{x}$ is the complex conjugate of $x$. Then, if 
$x= |x| e^{i\theta}$, then the above equation reduces to
\be
\cos\left( \theta L \right) \;=\; 0 \;\Rightarrow\; 
\theta_n = \frac{\pi}{2L}(2n-1) \,, \quad n=1,2,\dots,2L
\label{theta_n1}
\ee
in agreement with Eq.~\reff{def_asymptotics_sq}. 

\bigskip

\noindent
{\bf Remark.} The existence of unbounded outward branches for the 
limiting curve of the Potts model with cyclic boundary conditions is 
already present for the simplest case $L=1$. Here, the strip is just
the cyclic graph of $n$ vertices $C_n$. Its partition function is 
given exactly by
\be
Z_{C_n}(Q,v) \;=\; (Q+v)^n + (Q-1) \, v^n  
\ee
Then, we have two eigenvalues $\lambda_1 = Q+v=Q+x\sqrt{Q}$ and 
$\lambda_2 = v=x\sqrt{Q}$, which grow like $\sim x^{2L-1}=x$ and whose
difference is $Q=O(x^{L-1})$, in agreement with 
Conjecture~\ref{conj_sq_5}. Furthermore, the limiting curve is the line
$\Re x = -\sqrt{Q}/2$, which, as $|x|\to\infty$, has slopes given
by $\pm \pi/2$, in agreement with Conjecture~\ref{conj_sq_3}.

\subsection{Other asymptotic behaviors} 

For the Ising case ($p=4$) the points $x=-\sqrt{2}$ and $x=-1/\sqrt{2}$
are in general multiple points and we observe a pattern similar to the one
observed for $|x|\to\infty$.

For $x=-1/\sqrt{2}$, we find that, if we write $x=-1/\sqrt{2} + \epsilon$
with $|\epsilon|\ll 1$, within each sector there is only one leading
eigenvalue $\lambda_{\star,j}(L)\sim O(1)$. More precisely, for $L\geq 3$, 
\begin{subeqnarray}
\lambda_{\star,1}(L) &=& 2^{-L/2} + O(\epsilon^3)\\
\lambda_{\star,1}(L) - \lambda_{\star,3}(L) &=& 2 \epsilon^L + 
                    O(\epsilon^{L+1})  
\end{subeqnarray}
Again, the equimodularity condition when $|\epsilon|\to 0$ implies that
$\Re(\epsilon^L) = 0$, whence $\arg \epsilon = \theta_n$ with
$\theta_n$ given by Eq.~(\ref{theta_n1}).

The case $x=-\sqrt{2}$ is more involved. If we write
$x=-\sqrt{2}+\epsilon$ with $|\epsilon|\ll 1$, we find that in the sector
$\chi_{1,1}$ there are two eigenvalues of order $O(\epsilon)$, and the rest are
of order at least $O(\epsilon^2)$. The same conclusion is obtained from 
the sector $\chi_{1,3}$. If we call $\lambda_{\star,j}^{(i)}$ ($i=1,2$)
the dominant eigenvalues coming from sector $\chi_{1,j}$, then we find
for $L\geq 3$ that
\begin{subeqnarray}
\lambda_{\star,1}^{(1)}(L) &=& 2^{(L-1)/2} \epsilon + O(\epsilon^2) \;\approx\;
  -\lambda_{\star,1}^{(2)}(L) \\
\lambda_{\star,1}^{(1)}(L) + \lambda_{\star,1}^{(2)}(L) &=& 
   \begin{cases}
      \sqrt{2} \epsilon^{L-1} & \quad \text{$L$ even}\\
      2(L-1)   \epsilon^L     & \quad \text{$L$ odd}
   \end{cases} \\
\lambda_{\star,3}^{(1)}(L) &=& -2^{(L-1)/2} \epsilon + O(\epsilon^2) \;\approx\;
  -\lambda_{\star,3}^{(2)}(L) \\
\lambda_{\star,3}^{(1)}(L) + \lambda_{\star,3}^{(2)}(L) &=& 
   \begin{cases}
      -2(L-1)   \epsilon^L      & \quad \text{$L$ even}\\
      -\sqrt{2} \epsilon^{L-1}  & \quad \text{$L$ odd}
   \end{cases} \\
\lambda_{\star,1}^{(1)}(L) + \lambda_{\star,3}^{(1)}(L) &=& 
\frac{(-1)^L}{\sqrt{2}} \epsilon^{L-1} + O(\epsilon^L) 
\end{subeqnarray}
The equimodularity condition implies that
\be
\Re \left[ \epsilon \, \widebar{\epsilon}^{L-1} \right] \;=\;0 
\;\Rightarrow\; \cos(\theta(L-2)) \;=\; 0
\ee
Thus, the same asymptotic behavior is obtained as for $x=-1/\sqrt{2}$, 
except that $L\to L-2$:
\be
\theta_n \;=\; \frac{\pi}{2(L-2)}(2n-1) \,, \qquad n=1,\ldots,2(L-2)
\ee

\section{Triangular-lattice Potts model with free cyclic boundary conditions}
%
%
\subsection{Ising model ($\bm{p=4}$)}
\label{sec_tri_p=4}

For this model we know \cite{Stephenson_64,Blote_82b,Nienhuis_84b} the exact
transition temperature for the antiferromagnetic model $v_{{\rm
c},\text{AF}}=-1=v_{{\rm c},\text{BK}}$. The partition function is given by a
formula similar to that of the square lattice, and the dimensionality of
$\T_j(2,L)$ is the same as for the square lattice. In what follows we give the
different matrices in the same bases as for the square lattice.

We have computed the limiting curves $\mathcal{B}_L$ for $L=2,3,4$.
These curves are displayed in Figure~\ref{Curves_tri_p=4}(a)--(c).\footnote{
  After the completion of this work, we learned that Chang and Shrock
  had obtained the limiting curve for 
  $L=2$ \protect\cite[Figure~18]{Shrock_00b}. 
}
In Figure~\ref{Curves_tri_p=4}(d), we show all three curves for comparison.

%
%
\subsubsection{$\bm{L=2}$}

This strip is drawn in Figure~\ref{fig_tri_L=2}. 
The transfer matrices are
\begin{subeqnarray}
\T_1 &=& \frac{1}{2}\, 
   \left( \begin{array}{cc}
 2x^4 + 5\sqrt{2}x^3+12x^2 +8\sqrt{2}x+4 & x(2x^3+5\sqrt{2}x^2 +8x+2\sqrt{2})\\ 
  x^2 (2x^2+3\sqrt{2}x+2) & x^2 (2x^2+3\sqrt{2}x+2)  
           \end{array}
   \right)
\slabel{def_T_tri_p=4_L=2_k=1} \\
T_3 &=& \frac{x}{2}\, 
\left( \begin{array}{cc}
2x^3+5\sqrt{2}x^2 +8x+2\sqrt{2} & 2x^3+5\sqrt{2}x^2 +8x+2\sqrt{2} \\
x(2x^2+3\sqrt{2}x+2)            & 8x^3+3\sqrt{2}x^2 +6x+2\sqrt{2}
       \end{array}
\right)
\slabel{def_T_tri_p=4_L=2_k=3}
\end{subeqnarray} 

For real $x$, there is a single phase-transition point at 
\be
x_{\rm c} \;=\; -1/\sqrt{2} \;\approx\; -0.7071067812 
\ee 
We have found that the entire line 
\be
\Re x \;=\; -1/\sqrt{2} 
\ee
belongs to the limiting curve. Furthermore, $\mathcal{B}_2$ is symmetric with
respect to this line. Finally, there are two complex conjugate multiple 
points at $x=-1/\sqrt{2} \pm i/\sqrt{2}=-e^{\pm i\pi/4}$.

The dominant sector on the real $x$-axis is $\chi_{1,1}$ for
$x>-1/\sqrt{2}$, and $\chi_{1,3}$ for $x<-1/\sqrt{2}$.
Note that $x_{\rm c} = -1/\sqrt{2}$ gives the right bulk critical
temperature for this model in the antiferromagnetic regime.

%
%
\subsubsection{$\bm{L\geq 3}$}

For $L=3,4$ we have found that a) The line $\Re x = -1/\sqrt{2}$
belongs to the limiting curve; b) $\mathcal{B}_L$ is symmetric under
reflection with respect to that line; c) $\mathcal{B}_L$ contains a
pair of  multiple points at $x=-e^{\pm i\pi/4}$; and d) The dominant
sector on the real $x$-axis is $\chi_{1,1}$ for $x>-1/\sqrt{2}$, and
$\chi_{1,3}$ for $x<-1/\sqrt{2}$.

For $L=3$, there is another pair of multiple points at
$x\approx -1/\sqrt{2} \pm 0.7257238112\,i$; for $L=4$ this pair is located at 
$x\approx -1/\sqrt{2} \pm 0.7647261156\,i$. 

For $L=5,6,7$, we have found that there is a single real phase-transition
point at $x=-1/\sqrt{2}$, and that the dominant sector for $x>-1/\sqrt{2}$ 
(resp.\  $x<-1/\sqrt{2}$) is $\chi_{1,1}$ (resp.\   $\chi_{1,3}$).

%
%
\subsection{$\bm{Q=B_5}$ model ($\bm{p=5}$)}
\label{sec_tri_p=5}

We have computed the limiting curves $\mathcal{B}_L$ for $L=2,3,4$.
These curves are displayed in Figure~\ref{Curves_tri_p=5}(a)--(c).
In Figure~\ref{Curves_tri_p=5}(d), we show all three curves for comparison.

%
%
\subsubsection{$\bm{L=2}$}

The transfer matrices are  
\begin{subeqnarray}
\T_1 &=& 
   \left( \begin{array}{cc}
B_5 + x(2x + 4X_3 + x^2 X_4^\star) & 
x B_5^{1/4}(\sqrt{B_5}+4x+x^2 X_5^\star) \\
x^2 B_5^{1/4}(1 + 3 \sqrt{B_4^\star} x + x^2) &
x^2 (1 + 3x + 3\sqrt{B_5} x^2) 
           \end{array}
   \right)
\slabel{def_T_tri_p=5_L=2_k=1} \\
\T_3 &=& x\,  
   \left( \begin{array}{ccc}
\sqrt{B_5} + 4x +x^2 X_4^\star & 
\sqrt{B_5^\star} X_3 & 
B_5^{1/4}(1+x\sqrt{5 B_5} + x^2 X_4^\star)\\
\sqrt{B_5^\star}x & 
X_3^\star & 
(B_5^\star)^{1/4}(1+x) \\
(B_5^\star)^{1/4} x (1+3x+ \sqrt{B_5} x^2) &
(B_5^\star)^{1/4}(1+x) &
1+3x+3x^2 + \sqrt{B_5} x^3
           \end{array}
   \right) \nonumber \\
 & & 
\slabel{def_T_tri_p=5_L=2_k=3} 
\end{subeqnarray} 
where we have defined the shorthand notations
\begin{subeqnarray}
X_4^\star &=& 1 + 3\sqrt{B_5^\star} + X_3^\star \\
X_5^\star &=& 1 + 4\sqrt{B_5^\star} + X_3^\star  
\label{def_X45star}
\end{subeqnarray}

For real $x$, there are two phase-transition points at
\begin{subeqnarray}
x_{{\rm c},1} &\approx& -0.9630466372 \\
x_{{\rm c},2} &\approx& -0.5908569607 
\end{subeqnarray} 
In fact both points are T points and the whole interval 
$[x_{{\rm c},1},x_{{\rm c},2}]$ belongs to the limiting curve $\mathcal{B}_2$.
Finally, there are two complex conjugate multiple 
points at $x=-e^{\pm i\pi/5}$, as for the square-lattice case.
The dominant sector on the real $x$-axis is $\chi_{1,1}$ for
$x>x_{{\rm c},1}$, and $\chi_{1,3}$ for $x<x_{{\rm c},1}$.

%
%
\subsubsection{$\bm{L\geq 3}$}

For $L=3$, there are two real phase-transition points at 
\begin{subeqnarray}
x_{{\rm c},1} &\approx&  -1.0976251052\\ 
x_{{\rm c},2} &\approx&  -0.6376476917
\end{subeqnarray}  
We have found two pairs of complex conjugate endpoints at 
$x\approx -0.4297467004 \pm 0.6445268125\,i$, and 
$x\approx -0.3955590901 \pm 0.8536454650\,i$. 
There are nine pairs of complex conjugate T points, and two complex conjugate
multiple points at $x=-e^{\pm i\pi/5}$.  
The dominant sectors on the real $x$-axis are
$\chi_{1,1}$ for $x>x_{{\rm c},1}$, and $\chi_{1,3}$ for $x<x_{{\rm c},1}$

For $L=4$, there are three real phase-transition points at 
\begin{subeqnarray}
x_{{\rm c},1} &\approx& -1.0953543257\\   
x_{{\rm c},2} &\approx& -0.9708876996\\
x_{{\rm c},3} &\approx& -0.6102005246
\end{subeqnarray}  
The points $x_{{\rm c},2}$ and $x_{{\rm c},3}$ are T points, and they
define a line belonging to the limiting curve. This line contains two
multiple points at $x\approx -0.6319374252$, and 
$x\approx -0.7685805289$.  
We have found two additional pairs of complex conjugate endpoints at 
$x\approx -0.9270404586 \pm 0.3749352143\,i$, and 
$x=-e^{\pm i\pi/5}$. 
In addition, there are 22 pairs of complex conjugate T points. 
The dominant sectors on the real $x$-axis are
\begin{itemize}
\item $\chi_{1,1}$ for $x\in(x_{{\rm c},1},x_{{\rm c},2})\cup
                            (x_{{\rm c},3},\infty)$ 
\item $\chi_{1,3}$ for $x\in(-\infty,x_{{\rm c},1})\cup
                                    (x_{{\rm c},2},x_{{\rm c},3})$ 
\end{itemize}

For $L=5$, we have found five real phase-transition points at
\begin{subeqnarray}
x_{{\rm c},1} &\approx&  -1.0945337809\\ 
x_{{\rm c},2} &\approx&  -1.0615208835\\
x_{{\rm c},3} &\approx&  -0.8629689747\\ 
x_{{\rm c},4} &\approx&  -0.6393693994\\ 
x_{{\rm c},5} &\approx&  -0.6362471039 
\end{subeqnarray}
The dominant sectors on the real $x$-axis are
\begin{itemize}
\item $\chi_{1,1}$ for $x\in(x_{{\rm c},1},x_{{\rm c},2})\cup
                            (x_{{\rm c},3},\infty)$ 
\item $\chi_{1,3}$ for $x\in(-\infty,x_{{\rm c},1})\cup
                                    (x_{{\rm c},2},x_{{\rm c},3})$ 
\end{itemize}

For $L=6$ the amount of memory needed for the computation of the phase
diagram on the real $x$-axis is very large,
so we have focused on trying to obtain the largest 
real phase-transition point. The result is 
$x_{{\rm c},1} \approx -0.6221939194<-1/\sqrt{B_5}$. 
The sector $\chi_{1,1}$ dominates for all $x> x_{{\rm c},1}$; and for 
$x\ltapprox x_{{\rm c},1}$, the sector $\chi_{1,3}$ dominates. 

%
%
\subsection{Three-state Potts model ($\bm{p=6}$)}
\label{sec_tri_p=6}

For this model we also know that there is a first-order phase 
transition in the antiferromagnetic regime at \cite{Adler_95,Tutte_tri}
\be
x_{{\rm c},\text{AF}}(q=3) \;=\; -0.563512(14) 
\ee

We have computed the limiting curves $\mathcal{B}_L$ for $L=2,3,4$.
These curves are displayed in Figure~\ref{Curves_tri_p=6}(a)--(c).\footnote{
  After the completion of this work, we learned that Chang and Shrock
  had obtained the limiting curve for 
  $L=2$ \protect\cite[Figure~19]{Shrock_00b}.
}
In Figure~\ref{Curves_tri_p=6}(d), we show all three curves for comparison.

%
%
\subsubsection{$\bm{L=2}$}

The transfer matrices are  
\begin{subeqnarray}
\T_1 &=& \frac{1}{2}\, 
   \left( \begin{array}{cc}
x^4 + 2\sqrt{3}x^3+6x^2+4\sqrt{3}x + 3 & 
x\sqrt{2}(x^3+2\sqrt{3}x^2 +4x+\sqrt{3})\\
x^2\sqrt{2}(x^2+ \sqrt{3}x +1) & 
x^2(2x^2+ 2\sqrt{3}x +1)
           \end{array}
   \right)
\slabel{def_T_tri_p=6_L=2_k=1} \\
\T_3 &=& \frac{x}{2}\, 
   \left( \begin{array}{ccc}
2(x^3+2\sqrt{3}x^2+4x+\sqrt{3}) & \sqrt{2} X_1 & 
\sqrt{2}(2x^3+4\sqrt{3}x^2 + 7x +\sqrt{3}) \\
\sqrt{2}x & X_2 & X_2 \\
x\sqrt{2}(2x^2 + 2\sqrt{3} x +1) & X_2 & 4x^3+4\sqrt{3}x^2 +6x+\sqrt{3} 
           \end{array}
   \right) \nonumber \\
 & & 
\slabel{def_T_tri_p=6_L=2_k=3} \\ 
\T_5 &=& x^2  
\slabel{def_T_tri_p=6_L=2_k=5} 
\end{subeqnarray} 

For real $x$, there are two phase-transition points at 
\begin{subeqnarray}
x_{{\rm c},1} &=& -2/\sqrt{3} \;\approx\; -1.1547005384\\
x_{{\rm c},2} &=& -1/\sqrt{3} \;\approx\; -0.5773502692
\end{subeqnarray}  
The latter one is actually a multiple point. There are also a pair
of complex conjugate multiple points at 
$x=-e^{\pm i\pi/6}=-\sqrt{3}/2 \pm i/2$.
The dominant sectors on the real $x$-axis are: 
$\chi_{1,1}$ for $x>-1/\sqrt{3}$,  
$\chi_{1,3}$ for $x<-2/\sqrt{3}$, and  
$\chi_{1,5}$ for $x\in(-2/\sqrt{3},-1/\sqrt{3})$.

%
%
\subsubsection{$\bm{L\geq 3}$}

For $L=3$, there are three real phase-transition points at 
\begin{subeqnarray}
x_{{\rm c},1} &=& -2/\sqrt{3} \;\approx\; -1.1547005384\\ 
x_{{\rm c},2} &\approx&                   -0.9712924104\\ 
x_{{\rm c},3} &=& -1/\sqrt{3} \;\approx\; -0.5773502692
\end{subeqnarray}  
The latter one is actually a multiple point.
We have found two pairs of complex conjugate endpoints at 
$x\approx -0.3495004588 \pm 0.6911735024\,i$, and 
$x\approx -0.2862942369 \pm 0.8514701201\,i$. 
There are 16 pairs of complex conjugate T points. 
The dominant sectors on the real $x$-axis are
$\chi_{1,1}$ for $x>-1/\sqrt{3}$,  
$\chi_{1,3}$ for $x<-2/\sqrt{3}$, and  
$\chi_{1,5}$ for $x\in(-2/\sqrt{3},-1/\sqrt{3})$.

For $L=4$, there are five real phase-transition points at 
\begin{subeqnarray}
x_{{\rm c},1} &=& -2/\sqrt{3} \;\approx\; -1.1547005384\\ 
x_{{\rm c},2} &\approx&                   -1.0219801955\\ 
x_{{\rm c},3} &\approx&                   -1.0041094453\\
x_{{\rm c},4} &\approx&                   -0.7664034488\\ 
x_{{\rm c},5} &=& -1/\sqrt{3} \;\approx\; -0.5773502692
\end{subeqnarray}  
The points $x_{{\rm c},3}$ and $x_{{\rm c},4}$ are T points, 
while $x_{{\rm c},5}$ is a 
multiple point. We have found a pair of complex conjugate endpoints at 
$x\approx -0.3857232364 \pm 0.6652216322\,i$. 
In addition, there are 14 pairs of complex conjugate T points. 
The dominant sectors on the real $x$-axis are
\begin{itemize}
\item $\chi_{1,1}$ for $x>x_{{\rm c},4}$ 
\item $\chi_{1,3}$ for $x<-2/\sqrt{3}$ and $x\in(x_{{\rm c},2},x_{{\rm c},3})$ 
\item $\chi_{1,5}$ for $x\in(-2/\sqrt{3},x_{{\rm c},2}) \cup 
                            (x_{{\rm c},3},x_{{\rm c},4})$
\end{itemize}

For $L=5$, there are five real phase-transition points at
\begin{subeqnarray}
x_{{\rm c},1} &=& -2/\sqrt{3} \;\approx\; -1.1547005384\\
x_{{\rm c},2} &\approx&                   -0.9326923327\\ 
x_{{\rm c},3} &\approx&                   -0.7350208125\\
x_{{\rm c},4} &\approx&                   -0.6186679617\\ 
x_{{\rm c},5} &=& -1/\sqrt{3} \;\approx\; -0.5773502692
\end{subeqnarray}
The dominant sectors on the real $x$-axis are
\begin{itemize}
\item $\chi_{1,1}$ for $x\in(x_{{\rm c},2},x_{{\rm c},3})\cup
                            (x_{{\rm c},4},\infty)$ 
\item $\chi_{1,3}$ for $x<x_{{\rm c},2}$ 
\item $\chi_{1,5}$ for $x\in (x_{{\rm c},3},x_{{\rm c},4})$
\end{itemize}

For $L=6$, there are three real phase-transition points at
\begin{subeqnarray}
x_{{\rm c},1} &=& -2/\sqrt{3} \;\approx\; -1.1547005384\\
x_{{\rm c},2} &\approx&                   -1.0504774228\\ 
x_{{\rm c},3} &=& -1/\sqrt{3} \;\approx\; -0.5773502692
\end{subeqnarray}
We have also found a small horizontal line belonging to the limiting curve
$\mathcal{B}_6$ and bounded by the T points
\begin{subeqnarray}
x_{{\rm c},4} &\approx& -0.7688389273\\
x_{{\rm c},5} &\approx& -0.7646464215  
\end{subeqnarray}
The dominant sectors on the real $x$-axis are
\begin{itemize}
\item $\chi_{1,1}$ for $x\in(-1/\sqrt{3},\infty) \cup 
                            (x_{{\rm c},4},x_{{\rm c},5})$ 
\item $\chi_{1,3}$ for $x\in(-\infty,-2/\sqrt{3})\cup
                            (x_{{\rm c},2},x_{{\rm c},4})$
\item $\chi_{1,5}$ for $x\in (-2/\sqrt{3},x_{{\rm c},2}) \cup 
                             (x_{{\rm c},5},-1/\sqrt{3})$
\end{itemize}

In all cases $3\le L\le 6$, there is a pair of complex conjugate multiple 
points at $x=-e^{\pm i\pi/6}$.

%
%
\subsection{Four-state Potts model ($p=\infty$)}
\label{sec_tri_p=Infty}

We have computed the limiting curves $\mathcal{B}_L$ for $L=2,3,4$.
These curves are displayed in Figure~\ref{Curves_tri_p=Infty}(a)--(c).
In Figure~\ref{Curves_tri_p=Infty}(d), we show all three curves for comparison.

%
%
\subsubsection{$\bm{L=2}$}

The transfer matrices are
\begin{subeqnarray}
\T_1 &=& \frac{1}{2}\,
\left( \begin{array}{cc}
   X_8\, (2x^3+3x^2+6x+4) & \sqrt{3}\, x\, X_8\, X_7 \\
   \sqrt{3}\, x^2 \, X_7  &            x^2\, X_6  
       \end{array}
\right) \\
T_3 &=&  \frac{1}{6}\,
\left( \begin{array}{ccc}     
3x X_8 (2x^2+3x+2) & 2 \sqrt{6}\, x X_8 &\sqrt{3}\, x X_8 X_6 \\
2\sqrt{6}\, x^2    & 2x(4+3x)           & 2\sqrt{2}\, x (2+3x) \\
\sqrt{3}\, x^2 X_6 & 2\sqrt{2}\, x(3x+2)& x X_9  
\end{array}
\right) \\
T_5 &=& x^2
\end{subeqnarray}
where we have defined the short-hand notations
\begin{subeqnarray}
X_6 &=& 6x^2+9x+2 \\
X_7 &=& 2x^2+3x+2\\
X_8 &=& x + 2 \\
X_9 &=& 18x^3+27x^2+18x+4 
\end{subeqnarray}

For real $x$, we find a multiple point at $x=-1$, and a T point at 
$x_c \approx -0.5808613334$. The limiting curve $\mathcal{B}_2$ contains the
real interval $[-1,x_c]$. At $x=-1$, all eigenvalues become equimodular
with $|\lambda_i|=1$. 

We have found two additional pairs of complex conjugate T points at
$x\approx -0.9882427690 \pm  0.0896233991\,i$, and
$x\approx -3/4 \pm  0.6614378278\,i$. The dominant sectors on the real
$x$-axis are $\chi_{1,1}$ for $x>x_c$, and $\chi_{1,3}$ for $x<x_c$. 
We have found no region in the complex $x$-plane where the sector $\chi_{1,5}$
is dominant. 
 
%
%
\subsubsection{$\bm{L\ge 3}$}

For $L=3$ there are two real phase-transition points: $x=-1$ (which is 
a multiple point), and $x_c\approx -0.8953488450$. The limiting curve
contains two connected pieces, two pairs of complex conjugate endpoints,
12 complex conjugate T points, and one additional pair of complex conjugate
multiple points at $x\approx -3/4 \pm 0.6614378278\,i$. 
The dominant sectors on the real $x$-axis are
$\chi_{1,3}$ for $x<-1$; $\chi_{1,5}$ for $x\in(-1,x_c)$; and   
$\chi_{1,1}$ for $x> x_c$.
We have found no region where the sector $\chi_{1,7}$ is dominant.

For $L=4$ there are two real phase-transition points at $x=-1$ and 
$x = x_c \approx -0.7107999762$, which is a T point. The real line
$[-1,x_c]$ belongs to the limiting curve. 
The dominant sectors on the real $x$-axis are:
$\chi_{1,3}$ for $x<-1$; $\chi_{1,7}$ for $x\in(-1,x_c)$; and 
$\chi_{1,1}$ for $x> x_c$. We have found a few small regions with dominant
eigenvalue coming from the sector $\chi_{1,5}$; but we have found 
no region where the sector $\chi_{1,9}$ is dominant. 

For $L=5$ there are again two real phase-transition points at $x=-1$ and 
$x = x_c \approx -0.8004698444$, which is a T point. The real line
$[-1,x_c]$ belongs to the limiting curve. 
The dominant sectors on the real $x$-axis are:
$\chi_{1,3}$ for $x<-1$; $\chi_{1,9}$ for $x\in(-1,x_c)$; and 
$\chi_{1,1}$ for $x> x_c$.

For $L=6$ there are two real phase-transition points at $x=-1$ and 
$x = x_c \approx -0.7033434642$, which is a T point. The real line
$[-1,x_c]$ belongs to the limiting curve. 
The dominant sectors on the real $x$-axis are:
$\chi_{1,3}$ for $x<-1$; $\chi_{1,11}$ for $x\in(-1,x_c)$; and 
$\chi_{1,1}$ for $x> x_c$. 

In all cases, the point $x=-1$ is a multiple point where all the eigenvalues
are equimodular with $|\lambda_i|=1$.

%
%
\section{Common features of the triangular-lattice limiting curves with
free cyclic boundary conditions}
\label{sec_tri_results}

The results discussed in  
Sections~\ref{sec_tri_p=4}--\ref{sec_tri_p=6} allow us to make the following
conjecture (in the same spirit as Conjecture~\ref{conj_sq_1} for the 
square-lattice case) that states that certain points in the complex 
$x$-plane belong to the limiting curve $\mathcal{B}_L$:

\begin{conjecture} \label{conj_tri_1}
For the triangular-lattice $Q$-state Potts model with $Q=B_p$ 
and width $L \ge 2$:
\begin{enumerate}
\item The points $x=-e^{\pm i\pi/p}$ belong to the limiting curve.  
      At these points, all the eigenvalues are equimodular with
      $|\lambda_i|=1$. Thus, they are in general multiple points. 
      
\item For even $p\geq 6$, the point $x=-2/\sqrt{Q}$ always belongs to the 
      limiting curve $\mathcal{B}_L$.\footnote{
         This property has been verified for $p=6$ and $2\leq L\leq 7$, and
         for $p=8,10$ and $2\leq L \leq 5$.
      } 
      Furthermore, if $p=4,6$, then the point $x=-1/\sqrt{Q}$ also belongs 
      to $\mathcal{B}_L$.
\end{enumerate}
\end{conjecture}

The phase diagram on the real $x$-axis (which contains the physical regime
of the model) shows certain regularities that allow us to make the following 
conjecture: 

\begin{conjecture} \label{conj_tri_2}
For the triangular-lattice $Q$-state Potts model with $Q=B_p$
and width $L \ge 2$:
\begin{enumerate}
\item For even $p$, the relevant eigenvalue on the physical line 
      $v\in[-1,\infty)$ comes from the sector $\chi_{1,1}$.
      For odd $p$, the same conclusion holds for all $L\geq L_0$.\footnote{
      For $p=5$, we find that $L_0=5$. For $L=2,4$, the relevant eigenvalue
      belongs to the sector $\chi_{1,3}$ on a small portion of the 
      antiferromagnetic physical line $v\in[-1,v_0]$. 
      }
\item The relevant eigenvalue belongs to the sector $\chi_{1,3}$ 
      for all real $x<-2/\sqrt{Q}$. 
\end{enumerate}
\end{conjecture}

The above conjectures also apply to the limiting case $p\to\infty$ (i.e.,
$Q=4$). As for the square-lattice case, the multiple points 
$-e^{\pm i\pi/p}\to -1$ as $p\to\infty$ (Conjecture~\ref{conj_tri_1}.1) in
agreement with the fact that $x=-1$ is a multiple point for $Q=4$. 
Furthermore, this is also in agreement with Conjecture~\ref{conj_tri_1}.2,
as in this limit, $-2/\sqrt{Q}=-1$. 
The dominant sectors for $p\to\infty$ also agree with 
Conjecture~\ref{conj_tri_2}: on the physical line $v\in[-1,\infty)$ the
dominant sector is $\chi_{1,1}$, and for $x<-1$, the dominant sector is
$\chi_{1,3}$. More precisely, we can state the following conjecture
based on the empirical observations reported above:  

\begin{conjecture} \label{conj_tri_2bis}
For the triangular-lattice $4$-state Potts model 
defined on a semi-infinite strip of width $L \ge 2$, there exists 
some $x_c(L)>-1$ such that 
$\chi_{1,1}$ is dominant for $x > x_c(L)$,
$\chi_{1,2L-1}$ is dominant for $-1 < x < x_c(L)$,
$\chi_{1,3}$ is dominant for $x < -1$.
\end{conjecture}

\subsection{Asymptotic behavior for $\bm{|x|\to\infty}$}

Figures~\ref{Curves_tri_p=4}--\ref{Curves_tri_p=Infty} show a similar scenario
to the one discussed in Section~\ref{sec_sq_results}:
There are several unbounded outward branches with a clear asymptotic behavior 
for large $|x|$. Again, this scenario also holds in the limit $p\to\infty$
(See Figure~\ref{Curves_tri_p=Infty}). 
However there are quantitative differences with the scenario 
found for the square lattice. We should modify Conjecture~\ref{conj_sq_5} 
as follows: 

\begin{conjecture}  
\label{conj_tri_3}
Let $\lambda_{\star,1}(L)$ (resp.\ $\lambda_{\star,3}$(L)) be the leading 
eigenvalue of the sector $\chi_{1,1}$ (resp.\  $\chi_{1,3}$) in the 
regime $|x|\to \infty$. Then
\begin{subeqnarray}
\lambda_{\star,1}(L) &=& Q^{L-1} \, x^{3L-2} \, \left[
    1 + \sum\limits_{k=1}^\infty \frac{ b_k(L) }{Q^{k/2}} \, x^{-k} \right]\\
\lambda_{\star,1}(L) - \lambda_{\star,3}(L) &=& 2^{L-1}\sqrt{Q}\, x^{L-1} + 
                       (L-1)\, 2^{L-1} \, x^{L-2} + 
                       O(x^{L-3}) 
\end{subeqnarray} 
Furthermore, we have that
\begin{subeqnarray}
b_1(L) &=& 3L-2 \,,      \qquad L\geq 2 \\
b_2(L) &=& \frac{9}{2} L^2 -\frac{15}{2} L + 3 \,, \qquad L\geq 2\\
b_3(L) &=& \frac{9}{2} L^3 -\frac{27}{2} L^2 + 13 L -4 \,, \qquad L\geq 3
\label{conj.sq.bk}
\end{subeqnarray}
\end{conjecture}

The first coefficients $b_k(L)$ are displayed in Table~\ref{table_tri_coef};
the patterns displayed in \reff{conj.sq.bk} are easily verified. The
coefficients $b_k(L)$ also depend on $p$ for $k\geq 4$.

Conjecture~\ref{conj_tri_3} explains the number of outward branches in the
triangular-lattice case, as well as the observed pattern 
for the outward branches. Again, these branches are defined by the 
equimodularity of the two leading eigenvalues
\be
|\lambda_{\star,1}| \;=\; |\lambda_{\star,3}| \;=\; 
\left|\lambda_{\star,1} - \text{const.\ } x^{L-1} + O(x^{L-2})\right| 
\ee
This implies that
\be
\Re\left[ \lambda_{\star,1} \widebar{x}^{L-1} \right] \;=\; 0
\ee
Then, if $x= |x| e^{i\theta}$, the above equation reduces to
\be
\cos\left( \theta (2L-1) \right) \;=\; 0 \;\Rightarrow\; 
\theta_n = \frac{\pi}{2(2L-1)}(2n-1) \,, \quad n=1,2,\dots,2(2L-1)
\label{theta_n1_tri}
\ee
Thus, we get the same asymptotic behavior as for the square lattice with the 
replacement $L\to 2L-1$. 

\section{Discussion of the results with free cyclic boundary conditions}
\label{sec_discussion}

The results obtained give indications on the phase diagram of the
Potts model, as the accumulating points of the zeros of the partition 
function correspond to singularities of the free energy. 

Extrapolating the curves obtained to $L\to\infty$ in not an easy matter,
given that we have only access to relatively small $L$. However, in
Sections~\ref{sec_sq_results} and \ref{sec_tri_results} we have noted
a number of features which hold for all $L$ considered, and hence presumably
for all finite $L$ and also in the thermodynamic limit.

\subsection{Ising model}
\label{sec:Ising}

The most transparent case is that of the Ising model ($p=4$) on the
square lattice. Let $D(x,r)$ denote the disk centered in $x$ and
of radius $r$. There are then four different domains of interest:
\begin{subeqnarray}
 \scrd_1 &\;=\;& D(0,1) \setminus D(-\sqrt{2},1) \\
 \scrd_2 &\;=\;& D(0,1) \cap D(-\sqrt{2},1) \\
 \scrd_3 &\;=\;& D(-\sqrt{2},1) \setminus D(0,1) \\
 \scrd_4 &\;=\;& \C \setminus \left( D(0,1) \cup D(-\sqrt{2},1) \right)
 \label{domainsd}
\end{subeqnarray}
The $L \times N$ strips with even $N$ are bipartite, whence the Ising model
possesses the exact gauge symmetry $J \to -J$ (change the sign of the spins on
the even sublattice). Since the limit $N\to\infty$ can be taken through even
$N$ only, the limiting curves $\scrb_L$ should be gauge invariant. In terms of
$x$ the gauge transformation reads
\be
 x \to - \frac{x}{1+x \sqrt{2}} \,.
\ee
Note that it exchanges $\scrd_2 \leftrightarrow \scrd_4$, while leaving
$\scrd_1$ and $\scrd_3$ invariant. In particular, the structures of
$\scrb_L$ around $x=-1/\sqrt{2}$ and $|x|=\infty$ discussed in
Section~\ref{sec_sq_results} are equivalent.

On the other hand, the duality transformation $x \to 1/x$ is {\em not}
a symmetry of $\scrb_L$: this is due to the fact that the boundary
conditions prevent the lattice from being selfdual. Note that the
duality exchanges $\scrd_1 \leftrightarrow \scrd_4$ and $\scrd_2
\leftrightarrow \scrd_3$.  But whilst there are many branches of
$\scrb_L$ in $\scrd_4$, there are none in $\scrd_1$.

The Ising model being very simple, we do however expect the fixed
point structure on the real $x$-axis to satisfy duality. Combining
the gauge and duality transformations one can connect all critical
fixed points:
\be
  x_\text{FM} \; \stackrel{\rm gauge}{\longrightarrow}   \; x_+ 
              \; \stackrel{\rm duality}{\longrightarrow} \; x_-
              \; \stackrel{\rm gauge}{\longrightarrow}   \; x_\text{BK},
 \label{4repulsive}
\ee
and the first and the last points in the series are selfdual.
In the same way, all the non-critical (trivial) fixed points are
connected:
\be
  x=0 \; \stackrel{\rm duality}{\longrightarrow}\; |x|=\infty 
      \; \stackrel{\rm gauge}{\longrightarrow}  \; x=-1/\sqrt{2}
      \; \stackrel{\rm duality}{\longrightarrow}\; x=-\sqrt{2},
 \label{4attractive}
\ee
and the first and the last points in the series are gauge invariant.

The reason that we discuss these well-known facts in detail is that
the square-lattice Ising model is really the simplest example of how
taking $p$ rational (here, in fact, integer) profoundly modifies and
enriches the fixed/critical point structure of the Potts model, as compared
to the generic case of $p$ irrational. Taking the limit $p\to 4$ through
irrational values we would have had three equivalent $c=1/2$ critical
points, RG repulsive in $x$, situated at $x_\text{FM}$ and $x_\pm$;
one $c=-25/2$ critical point, RG attractive in $x$, situated at $x_\text{BK}$;
and two non-critical (trivial) fixed points, RG attractive in $x$,
situated at $x=0$ and $|x|=\infty$. This makes up for a phase diagram
on the real $x$-axis which is consistent in terms of renormalization
group flows (see the top part of Fig.~\ref{fig:Ising}).

Conversely, sitting directly at $p=4$ replaces this structure by the
four repulsive $c=1/2$ critical points (\ref{4repulsive}) and the four
attractive non-critical fixed points (\ref{4attractive}). This again
gives a consistent scenario, in which notably the BK phase has
disappeared (see the bottom part of Fig.~\ref{fig:Ising}).
In other cases than the Ising model ($p>4$ integer) we
could expect the emergence of even more new (as compared to the case
of irrational $p$) fixed points (critical or non-critical), which will
in general be inequivalent (due in particular to the absence of the
Ising gauge symmetry).

Going back to the case of complex $x$ we can now conjecture:

\begin{conjecture} \label{conj_ising_sq}
 Let $\scrd_1$ be the domain defined in Eq.~(\ref{domainsd}d). Then
 \begin{itemize}
 \item The points $x$ such that
  \be
    Z_{L_{\rm F} \times N_{\rm P}}(Q=2;x) = 0 \quad (\text{\rm square lattice})
  \ee
  for some $L$ and $N$ are dense in $\C \setminus \scrd_1$.
 \item There are no such points in $\scrd_1$.
 \end{itemize}
\end{conjecture}

We now turn to the Ising model on the triangular lattice. We first
note that all the limiting curves $\scrb_L$ are symmetric under the
combined transformation $x \leftrightarrow -x-\sqrt{2}$ and
$\chi_{1,1} \leftrightarrow \chi_{1,3}$. On the level of the coupling
constant this can also be written $\exp(J) \to -\exp(J)$.

We also conjecture that

\begin{conjecture} \label{conj_ising_tri}
  Let $\scrd_{\rm tri}$ be the interior of the ellipse
  \be
    \left( \Re x + 1/\sqrt{2} \right)^2 + 3 \left( \Im x \right)^2 = 3/2 \,.
  \ee
  Then
 \begin{itemize}
 \item The points $x$ such that
  \be
    Z_{L_{\rm F} \times N_{\rm P}}(Q=2;x) = 0 \quad (\text{\rm triangular lattice})
  \ee
  for some $L$ and $N$ are dense in $\C \setminus \scrd_{\rm tri}$.
 \item There are no such points in $\scrd_{\rm tri}$.
 \end{itemize}
\end{conjecture}

\subsection{Models with $p>4$}

For square-lattice models with $p>4$ the phase diagram in the thermodynamic
limit is expected to be more complicated. We can nevertheless conjecture that
the four values $x_{\rm c}$ given by Eq.~(\ref{values_sq_xc}), and denoted by
solid squares in the figures, correspond to phase transition points even for
$Q=B_p$ a Beraha number. Accordingly, these points are expected to be
accumulation points for the limiting curves $\scrb_L$, when $L\to\infty$.

But these four values of $x$ are not the only fixed points. There is a
complex fixed point structure between $x_{-}(Q)$ and $x_{\rm BK}(Q)$, and
between $x_{\rm BK}(Q)$ and $x_{+}(Q)$. This is because for $Q$ equal to a
Beraha number, the thermal operator is repulsive at $x_{\rm BK}(Q)$ (and
not attractive as it would have been in the BK phase for irrational
$p$), whereas it remains repulsive at $x_{-}(Q)$ and $x_{+}(Q)$.
Therefore, there must at the very least be one attractive fixed point
in each of the two intervals mentioned, in order for a consistent
phase diagram to emerge. Indeed, for $p$ even, there are two new fixed
points, one of them being conjectured as $-\sqrt{Q}/2$ for all even
$p$, and the other being equal to $-\sqrt{Q}$ only for $p=4$ and
$p=6$. But our results for finite $L$ are in favor of an even more
complicated structure, involving more new fixed points.  The structure
of the phase diagram for $p$ odd is further complicated by the
emergence of segments of the real $x$-axis belonging to $\scrb_L$. It
is however uncertain, whether these segments will stay of finite
length in the $L\to\infty$ limit.

In the models with $p=5,6,\infty$ and on both the square and triangular 
lattices, we have found strong numerical evidence to conjecture that 
the partition-function zeros are {\em dense} in the whole complex $x$-plane 
with the exception of the interior of some domain. The shape of this
domain depends on both $p$ and the lattice structure; and unlike in the 
Ising case ($p=4$), we do not have enough evidence to conjecture its 
algebraic expression [c.f., Conjectures~\ref{conj_ising_sq} 
and~\ref{conj_ising_tri}]. For the square lattice and fixed $p$, the
limiting curves $\mathcal{B}_L$ seem to approach (from the outside) 
the circles  
\reff{circles_sq_x}, especially in the ferromagnetic regime $\Re x \ge 0$.
For the triangular lattice and $p=\infty$, the limiting curves 
in Figure~\ref{Curves_tri_p=Infty}  seem to approach the circle 
\be 
\left( \Re x + \frac{1}{4} \right)^2 + (\Im x)^2 \;=\; 
\left(\frac{3}{4}\right)^2 
\label{circles_tri_x_p=infty}
\ee
which goes through the bulk critical points $x=-1$ and $x=1/2$.

\subsection{The region $|x| \gg 1$}

The emergence of unbounded branches of $\scrb_L$ in the region of $|x| \gg 1$
is at first sight rather puzzling. Because when $|x|$ is large enough, we
should expect the system to be non-critical, and thus be described by a {\em
unique} leading eigenvalue of the transfer matrix. This is at least what
happens for the $q$-state Potts model on a strip with cylindrical or free
boundary conditions using the Fortuin--Kasteleyn representation
\cite{Tutte_sq,Tutte_tri}.

One of the main reasons for studying the limiting curves in the first
place is that we wish to use them to detect the critical points of the
models at hand.  At a conformally invariant critical point there
should be an infinite spectrum of transfer-matrix eigenvalues
$|\Lambda_0| \ge |\Lambda_1| \ge \ldots$ that become degenerate according
to \cite{CFT_Cardy} $|\Lambda_i/\Lambda_0| \sim \exp(-2\pi x_i/L)$ when
$L\to\infty$, where $x_i$ are critical exponents. The limiting curves
just tell us that the {\em two} dominant eigenvalues become
degenerate, and not even with what finite-size corrections. Therefore
the fact that a point $x$ (even on the real axis) is an accumulation
point of $\scrb_L$ is not sufficient for $x$ to be a critical point in
the sense of the above scaling behavior.

The observed behavior for $|x| \gg 1$ just shows that the leading eigenvalues
in sectors with different boundary conditions ($\chi_{1,1}$ and $\chi_{1,3}$)
come close. This is most transparent in the Ising case, where there is a
bijection between RSOS heights and {\em dual} spins. It is easily seen that
$\chi_{1,1}$ (resp.\ $\chi_{1,3}$) corresponds to {\em fixed} boundary
conditions in the spin representation, with all the dual spins on the
upper/lower rim being fixed as $+/+$ (resp.\ $+/-$). On the other hand, within
a given sector there should be a finite gap between the leading and
next-leading eigenvalues, in the region $|x| \gg 1$, signaling non-critical
behavior.

\subsection{Fixed cyclic boundary conditions}
\label{sec:fixed_cyclic}

To avoid the (from the point of view of detecting critical behavior)
spurious coexistence between two different boundary conditions, we
should rather pick boundary conditions that break the $Z_Q$ symmetry
of the $Q$-state Potts model explicitly. We now illustrate this possibility
by making a particular choice of fixed boundary conditions, which has the
double advantage of generalizing those for the Ising case (as discussed above)
{\em and} enabling the corresponding Potts model partition function
$Z_{L_{\rm X} \times N_{\rm P}}(Q;v)$ to be written as a sum of RSOS
model partition functions.

Consider first the Potts model partition function $\tilde{Z}$ on the dual
lattice, with spins $S_+$ and $S_-$ on the upper and lower
exterior dual sites, and at the dual coupling $\tilde{J}$. 
Recall that the duality relation reads simply $v \tilde{v} = Q$.
If we impose {\em free} boundary conditions on $S_\pm$, we have by the
fundamental duality relation \cite{Wu_82}
\be
 Q^{V-E/2-1} x^E \tilde{Z}(Q;Q/v) = Z(Q;v),
\ee
where $E$ (resp.\ $V$) is the total number of lattice edges (resp.\ direct
sites). Note that $V=LN$, and that $E=2V-N$ (resp.\ $E=3V-2N$) for the square
(resp.\ triangular) lattice. We now claim that this object with {\em fixed and
equal values} for $S_\pm$ can again be expressed in terms of $K_{1,2j+1}$, for
a generic $p$. The precise relation reads
\be
 Z_{L_{\rm X} \times N_{\rm P}}(Q;v) \equiv
 Q^{V-E/2} x^E \left. \tilde{Z}(Q;Q/v) \right|_{S_+=S_-} =
 Q^{LN/2} \sum_{j=0}^L \beta_{j}(p) \, K_{1,2j+1}(p,L;x),
 \label{Q0=1}
\ee
which should be compared with Eq.~(\ref{def_Z_uq}). We henceforth refer to
$Z_{L_{\rm X} \times N_{\rm P}}(Q;v)$ as the partition function of the Potts
model with fixed cyclic boundary conditions (even though it would be more
precise to say that it is actually the two exterior dual spins that get
fixed). The amplitudes read
\be
 \beta_{j}(p)= \frac{S_{j}(p)}{Q} +(-1)^{j} \left(1-\frac{1}{Q} \right) \;.
\ee
Note that for arbitrary values of $Q$, the partition function
$\left. \tilde{Z}(Q;Q/v) \right|_{S_+=S_-}$ can be defined by its
FK cluster expansion on the dual lattice, by giving a weight $Q$
to clusters that do not contain any of the two exterior sites, and a weight
$1$ to clusters containing at least one of two exterior sites.
Eq.~(\ref{Q0=1}) is a special case of a more general relation which
will be proved and discussed elsewhere.

Now, for $p$ integer, we would like to express $Z_{L_{\rm X} \times N_{\rm
P}}(Q;v)$ in terms of the $\chi_{1,2j+1}(p,L;x)$ as we did in the case of free
cyclic boundary conditions. But because of the $(-1)^{j}$ in the expression of
$\beta_{j}(p)$, we have $\beta_{np+j}(p)=\beta_{j}(p)$ and
$\beta_{(n+1)p-1-j}=-\beta_{j}$, cf.~Eq.~(\ref{twosyms}) for the case of
$S_j(p)$, only if $p$ is even. For $p$ even, we can express
\be
 Z_{L_{\rm X} \times N_{\rm P}}(Q;v) =
 Q^{LN/2} \sum_{j=0}^{\lfloor (p-2)/2 \rfloor}
 \beta_{j}(p) \, \chi_{1,2j+1}(p,L;x) \qquad (p \ \mbox{even})
 \label{QQ0=1}
\ee
which should be compared with Eq.~(\ref{def_Z_RSOS}).
For $p$ odd, there does not appear to exist an expansion of
$Z_{L_{\rm X} \times N_{\rm P}}$ in terms of $\chi_{1,2j+1}$.

Note in particular that $\beta_1(p)=0$ for any $p$. This has the consequence
of eliminating the $\chi_{1,3}$ sector from the partition function,
and, as we now shall see, modify the $|x| \gg 1$ behavior of the phase
diagram.

\section{Square-lattice Potts model with fixed cyclic boundary conditions}
\label{sec:sq_fixed}

The limiting curves $\scrb_L$ with fixed cyclic boundary conditions
(see Figs.~\ref{Curves_sq_p=4_1}--\ref{Curves_tri_p=6_1}) are very
similar to those obtained in Ref.~\cite{Tutte_sq} for the Potts model
with fully free boundary conditions. On the other hand, we have
already seen that the $\scrb_L$ with free cyclic boundary conditions
are very different.

Before presenting the results for fixed cyclic boundary conditions in
detail we wish to explain this similarity. We proceed in two
stages. First we present an argument why the limiting curves corresponding
to just the sector $\chi_{1,1}$ almost coincide with those for fully
free boundary conditions. Second, we take into account the effect of
adding other sectors $\chi_{1,2j+1}$.

Let $\T_{\rm FK}$ be the transfer matrix in the FK representation
with zero bridges (cf.~footnote \ref{fn_FK}), and let $\lambda_i$
be its eigenvalues.%
\footnote{We label the $\lambda_i$ by letting $\lambda_0$ be
the eigenvalue which dominates for $x$ real and positive, and
using lexicographic ordering \cite{transfer1} for the remaining
eigenvalues.}
Then one has, with cyclic boundary conditions
\be
 K_{1,1} = \tr \T_\text{FK}^N = \sum_i \lambda_i^N.
\ee
Due to the coupling of $K_{1,2j+1}$, given by Eq.~(\ref{comb_K}), the
eigenvalues of $\T_1$ (i.e., the transfer matrix that generates
$\chi_{1,1}$, cf.~Eq.~(\ref{def_chi})) form only a subset of the
eigenvalues of $\T_{\rm FK}$.  More precisely,
\be
 \chi_{1,1} = \sum_i \tilde{\alpha}_i \lambda_i^N,
\ee
where $\tilde{\alpha}_i = 0$ or $1$ are {\em independent} of $x$.
Note that when $L<p-1$, Eq.~(\ref{comb_K}) gives simply $\chi_{1,1}=K_{1,1}$,
and so in that case all $\tilde{\alpha}_i =1$.

Meanwhile, the partition function of the Potts model with fully free
boundary conditions is given by \cite{transfer1}
\be
 Z_\text{free} = \langle f | \T_{\rm FK}^N | i \rangle =
 \sum\limits_{i\geq 1} \alpha_i \, \lambda_i^N,
\ee
where the amplitudes $\alpha_i$ are due to the free {\em longitudinal}
boundary conditions. Note that some of the $\alpha_i$ could vanish
identically, and indeed many of them {\em do} vanish. For example, in
the case of the square lattice, the vectors $|i\rangle$ and $\langle
f|$ are symmetric under a reflection with respect to the axis of the
strip, whence only the $\lambda_i$ corresponding to eigenvectors which
are symmetric under this reflection will contribute to
$Z_\text{free}$.

For $x>0$ real and positive, it follows from simple probabilistic
arguments that the dominant eigenvalue $\lambda_0$ will reside in the
zero-bridge sector $K_{1,1}$ and is not canceled by eigenvalues
coming from other sectors. Therefore $\tilde{\alpha}_0 = 1$. On the
other hand, the Perron-Frobenius theorem and the structure of the
vectors $|i\rangle$ and $\langle f|$ implies that $\alpha_0 > 0$. We
conclude that the dominant term in the expansions of $\chi_{1,1}$ and
$Z_{\rm free}$ are proportional. By analytic continuation the same
conclusion holds true in some domain in the complex $x$-plane
containing the positive real half-axis. Moving away from that
half-axis, a first level crossing will take place when $\lambda_0$
crosses another eigenvalue $\lambda_i$. If none of the functions
$\alpha_i$ and $\tilde{\alpha}_i$ are {\em identically} zero, the
corresponding branch of the limiting curve $\scrb_L$ coincides in the
two cases. Further away from the positive real half-axis other level
crossings may take place, and the limiting curves remain identical
until a level crossing between $\lambda_j$ and $\lambda_k$ takes place
in which either $\alpha_j=0$ and $\tilde{\alpha}_j \neq 0$, or
conversely $\alpha_j \neq 0$ and $\tilde{\alpha}_j = 0$.  When $L<p-1$
the only possibility is the former one, since all $\tilde{\alpha}_i =
1$.

If we now compare the limiting curves of $Z_\text{free}$ and
$Z_\text{RSOS}$, the latter being defined as some linear combination
of $\chi_{1,2j+1}$ (containing $\chi_{1,1}$), the above argument
will be invalidated if the first level crossing in $Z_\text{RSOS}$
when moving away from the positive half-axis involves an eigenvalue
from $\chi_{1,2j+1}$ with $j>0$.

With free cyclic boundary conditions, $Z_\text{RSOS}$ contains
$\chi_{1,3}$. The first level crossing involves eigenvalues from
$\chi_{1,1}$ and $\chi_{1,3}$ (cf.~the observed unbounded branches)
and is situated very ``close'' [cf.~Eqs.~(\ref{theta_n1}) and
(\ref{theta_n1_tri}) with $n=1$] to the positive real
half-axis. Accordingly, the limiting curves $\scrb_L$ do not at all
resemble those with fully free boundary conditions.  On the other
hand, when $\chi_{1,3}$ is excluded (i.e., in the case of fixed cyclic
boundary conditions) the first level crossing is between two different
eigenvalues from the $\chi_{1,1}$ sector (see
Figs.~\ref{Curves_sq_p=4_1}--\ref{Curves_tri_p=6_1}).

%
%
\subsection{Ising model ($p=4$)}
\label{sec_sq_p=4_bis}

We have studied the limiting curves given by the sector $\chi_{1,1}$ in 
the square-lattice Ising case. The results are displayed in 
Figure~\ref{Curves_sq_p=4_1}. It is clear that there are no outward
branches, as there is a unique dominant eigenvalue in the region 
$|x|\gg 1$. Indeed, this agrees with the expected non-critical phase. 
These curves are very similar to those obtained using the Fortuin-Kasteley
representation for a square-lattice strip with {\em free}
boundary conditions \cite{Tutte_sq}. In particular, for even $L=2,4$ we find
that these curves do in fact {\em coincide}. However, for $L=3$ we find
disagreements; but only in the region $\Re v <-1$. Namely, the
complex conjugate closed regions defined by the multiple points 
$x=-e^{-i\pi/4}$ and $x=-\sqrt{2}$ (see Figure~\ref{Curves_sq_p=4_1}b) 
are replaced by two complex conjugate arcs emerging from $x=-e^{-i\pi/4}$. 
These arcs bifurcate at two complex conjugate T points. 

For $L=2$ we find two pairs of complex conjugate endpoints at
$x \approx -0.5558929703 \pm 0.1923469388\,i$, and 
$x \approx  0.5558929703 \pm 1.6065605012\,i$.
There is a double endpoint at $x=-\sqrt{2}$.

For $L=3$ we also find two pairs of complex conjugate endpoints at
$x \approx -0.5054436896 \pm 0.1404486742\,i$, and 
$x \approx  0.9624601506 \pm 1.1627733180\,i$.
There is a multiple point at $q=-\sqrt{2}$, and a pair of complex
conjugate multiple points at $q=-e^{-\pm \pi i/4}$. These multiple points
also appear in $L=4$. 

For $L=4$ we find two connected components in the limiting curve. There
are two pairs of complex conjugate T points at
$q \approx -1.1111427356 \pm 0.8231882219\,i$, and 
$q \approx -0.9473515724 \pm 0.4894779296\,i$.  We also find
four complex conjugate pairs of endpoints at
$q \approx -0.6052879436 \pm 0.3554255102\,i$, 
$q \approx -0.4820292937 \pm 0.1111133833\,i$, 
$q \approx -0.3346743307 \pm 1.3000737077\,i$, and 
$q \approx  1.0790506924 \pm 0.8817674400\,i$. 

%
%
\subsection{Three-state Potts model ($p=6$)}
\label{sec_sq_p=6_bis}

We have studied the limiting curves given by the sectors $\chi_{1,1}$
and $\chi_{1,5}$, cf.~Eq.~(\ref{QQ0=1}). The results are displayed in
Figure~\ref{Curves_sq_p=6_1}.
We have compared these curves with those obtained for a square-lattice
strip with free boundary conditions \cite{Tutte_sq}. We find that they
agree almost perfectly in the region $\Re x \geq -1$. The only
exceptions are the tiny complex conjugate branches emerging from the
multiple points $-e^{-i\pi/6}$ for $L=3,4$ and pointing to
$x_\text{BK}$. The differences are in both cases rather small and they
are away from the real $x$-axis.  In the region $\Re x < -1$, however,
the differences between the two boundary conditions are sizeable. For
free boundary conditions the closed regions tend to disappear, or, at
least, to diminish in number and size.

\section{Triangular-lattice Potts model with fixed cyclic boundary conditions}
\label{sec:tri_fixed}

%
%
\subsection{Ising model ($p=4$)}
\label{sec_tri_p=4_bis}

We have studied the limiting curves given by the sector $\chi_{1,1}$ in the
triangular-lattice Ising case. The results are displayed in
Figure~\ref{Curves_tri_p=4_1}, and they are the same than those obtained with
the Fortuin-Kasteleyn representation \cite{Tutte_tri}, with free boundary
conditions, for all $L$. Therefore, we see a non-trivial effect of the
lattice: for the triangular lattice, the dominant eigenvalues always comes from
$K_{1,1}$, contrary to the case of the square lattice.

For $L=2$ we find two real endpoints at $q=-\sqrt{2}$ and $q=-1/\sqrt{2}$, and
an additional pair of complex conjugate endpoints at $x \approx 0.3535533906
\pm 0.9354143467\,i$. At $x=-1$ there is a crossing between the two branches
of the limiting curve.

For $L=3$ we find two real endpoints at $q=-\sqrt{2}$ and $q=-1/\sqrt{2}$,
and four pairs of complex conjugate endpoints at
$x \approx -1.4346151869 \pm 0.9530458628\,i$, 
$x \approx  0.5477064083 \pm 0.6206108204\,i$,
$x \approx -1/\sqrt{2} \pm 0.4918781633\,i$, and
$x \approx -1/\sqrt{2} \pm 0.9374415716\,i$. 
The limiting curve contains two complex conjugate vertical lines determined 
by the latter two pairs of endpoints, and a horizontal line determined
by the two real endpoints. 
We have found three pairs of complex conjugate T points at
$x\approx -1\sqrt{2} \pm 0.5353475100\,i$,
$x\approx -1\sqrt{2} \pm 0.7246267519\,i$, and
$x\approx -1\sqrt{2} \pm 0.8539546894\,i$.
Finally, there is a multiple point at $x\approx -0.9681295813$.

For $L=4$, we again find a horizontal real line bounded by two
real endpoints at $x=-\sqrt{2}$, and $x=-1/\sqrt{2}$, and a pair
of complex conjugate vertical lines bounded by the endpoints
$x\approx -1\sqrt{2} \pm 1.0514178378\,i$, and 
$x\approx -1\sqrt{2} \pm 0.3816638845\,i$. 
We have found and additional pair of endpoints at
$x\approx 0.5890850526 \pm 0.4519358255\,i$.
There are five pairs of T points; two of them are located on the 
line $\Re x=-1\sqrt{2}$. These are
$x\approx -1\sqrt{2} \pm 0.4336035301\,i$, and
$x\approx -1\sqrt{2} \pm 0.7394246716\,i$.
The other three pairs are
$x\approx -1.0712333535 \pm 0.7555078808\,i$, 
$x\approx -1.6123945698 \pm 0.8042942359\,i$, and
$x\approx -0.3186094544 \pm 0.9388965869\,i$.
We find four bulb-like regions around the latter two pairs of T points.
Finally, there is a multiple point at $x\approx -0.9415556904$, and 
a complex conjugate pair of multiple points at 
$q = -e^{\pm i\pi/4}$. 

We have compared the above-described limiting curves with those of a 
triangular-lattice model with {\em free} boundary conditions \cite{Tutte_tri}.
The agreement is perfect on the whole complex $x$-plane for $L=2,3,4$.  

%
%
\subsection{Three-state Potts model ($p=6$)}
\label{sec_tri_p=6_bis}

We have studied the limiting curves given by the sectors $\chi_{1,1}$
and $\chi_{1,5}$, cf.~Eq.~(\ref{QQ0=1}). The results are displayed in
Figure~\ref{Curves_tri_p=6_1}.  As for the square-lattice case
discussed in Section~\ref{sec_sq_p=6_bis}, the limiting curves
coincide with those obtained with free boundary conditions in a domain
containing the real positive $v$-axis. In particular, the agreement is
perfect in the first regime $\Re v \ge 0$. In the second
regime $-1\le \Re v \leq 0$, the coincidence holds
except on a small region close to $\Re v = -1$, and $|\Im v|$ small
for $L=3,4$.  In both cases, the branches that emerge from
$x=-1/\sqrt{3}$ and penetrate inside the second regime (and
defining a closed region), change their shape for free boundary
conditions (and in particular, the aforementioned closed regions are
no longer closed).  Finally, in the third regime $\Re v < -1$,
the limiting curves for both types of boundary conditions clearly
differ. As for the square-lattice three-state model, free boundary
conditions usually imply less and smaller closed regions.

%
%
\section{Conclusion and outlook}\label{sec_final}

We have studied the complex-temperature phase diagram of the $Q$-state Potts
model on the square and triangular lattices with $Q=4\cos^2(\pi/p)$ and $p$
integer. The boundary conditions were taken to be cyclic so as to make contact
with the theory of quantum groups
\cite{Pasquier_87,Saleur_89,Pasquier_90,Saleur_90,Saleur_91}, which provides a
framework for explaining how a large amount of the eigenvalues of the cluster
model transfer matrix---defined for generic values of $p$---actually do not
contribute to the partition function $Z$ for $p$ integer. Moreover, for $p$
integer, the exact equivalence (\ref{def_Z_RSOS}) between the Potts and the
$A_{p-1}$ RSOS model provides an efficient way of computing exactly those
eigenvalues that do contribute to $Z$. Using the Beraha-Kahane-Weiss theorem
\cite{BKW_theorem}, this permitted us to compute the curves $\scrb_L$ along
which partition function zeros for cyclic strips of finite width $L$
accumulate when the length $N\to\infty$.

The RSOS model has the further advantage of associating a quantum number $j$
with each eigenvalue, which is related to the number of clusters of
non-trivial topology with respect to the periodic direction of the lattice and
to the spin $S_z$ of the associated six-vertex model. This number then
characterizes each of the phases (enclosed regions) defined by $\scrb_L$.

The curves $\scrb_L$ turn out to exhibit a remarkable regularity in
$L$---at least in some respects---thus enabling us to make a number of
conjectures about the thermodynamic limit $L\to\infty$. On the other
hand, even a casual glance at the many figures included in this paper
should convince the reader that the $L\to\infty$ limit of the models
at hand might well conceal many complicated features and exotic phase
transitions. Despite of these complications, we venture to summarize
our essential findings, by regrouping them in the same way as in the
list of open issues presented in the Introduction:

\begin{enumerate}
\item The points $x_{\rm FM}(Q)$ and $x_{-}(Q)$ (and for the square
  lattice also its dual $x_+(Q)$), that act as phase transition points
  in the generic phase diagram, should play a similar role for integer
  $p$. This can be verified from the figures in which it is
  more-or-less obvious that the corresponding red solid squares will
  be traversed, or pinched, by branches of $\scrb_L$ in the
  $L\to\infty$ limit. What is maybe more surprising is that also
  $x_{\rm BK}(Q)$ has a similar property, despite of the profoundly
  changed physics inside the BK phase. Indeed, in most cases, $x_{\rm
    BK}(Q)$ is either exactly on or very close to a traversing branch
  of $\scrb_L$. It remains an open question to characterize exactly
  the nature of the corresponding phase transition.
\item It follows from Conjecture~\ref{conj_sq_1}.2 that for the square
  lattice, $\scrb_\infty$ will contain $x=-\sqrt{Q}/2$ for $p$ integer
  and $x=-\sqrt{Q}$ for $Q$ integer. For the triangular lattice the
  corresponding Conjecture~\ref{conj_tri_1}.2 involves the points
  $x=-2/\sqrt{Q}$ for $p$ integer and $x=-1/\sqrt{Q}$ for $Q$ integer.
  Thus, both lattices exhibit a phase transition on the chromatic line
  $x=-1/\sqrt{Q}$ or its dual, but only for integer $Q$. It is
  tempting to speculate that the chromatic line and its dual might
  play symmetric roles upon imposing fully periodic boundary
  conditions, but that remains to be investigated.
\item We have found that with free cyclic boundary conditions,
  partition functions zeros are dense in a substantial region of the
  phase diagram, including the region $|x| \gg 1$. See in particular
  Conjectures \ref{conj_sq_3}--\ref{conj_sq_4} for the square lattice
  and Conjecture \ref{conj_tri_3} for the triangular lattice.  For the
  Ising model ($Q=2$), the finite-size data is conclusive enough to
  make a precise guess as to the extent of that region,
  cf.~Conjectures \ref{conj_ising_sq}--\ref{conj_ising_tri}. We have
  argued (in Section \ref{sec:fixed_cyclic}) and observed explicitly
  (in Sections \ref{sec:sq_fixed}--\ref{sec:tri_fixed}) that this
  feature is completely modified by changing to fixed cyclic boundary
  conditions.  Another example of the paramount role of the boundary
  conditions has been provided with the argument of Section
  \ref{sec:sq_fixed} that when restricting to the sector $\chi_{1,1}$
  one sees essentially the physics of free longitudinal boundary
  conditions.
\item It is an interesting exercise to compare the limiting curves
  found here with the numerically evaluated effective central charge
  shown in Figs.~23--25 of Ref.~\cite{JS_05}. In particular, for $p=5$
  it does not seem far-fetched that the two new phase transitions
  identified in Fig.~23 of that paper might be located exactly at
  $x=-1/\sqrt{Q}\simeq -0.618$ and $x=-\sqrt{Q}/2\simeq -0.809$.
  These points (for the former point, actually its dual, but we remind
  that the transverse boundary conditions in Ref.~\cite{JS_05} are
  periodic) are among the special points discussed in item 2 above.
\item We have provided some evidence that {\em on the triangular
    lattice} for $Q=4$ (i.e., $p=\infty$) phases with arbitrary high
  $j$ will exist close to the point $x=-1$. For the square lattice we
  have only found phases with $j\le 5$. This should be compared with
  the arbitrarily high values of $S_z$ taken when approaching
  $(Q,x)=(4,-1)$ from within the BK phase in the generic case
  \cite{Saleur_91,transfer4}.
\end{enumerate}

It would be interesting to extend the study to fully periodic (toroidal)
boundary conditions. This would presumably diminish the importance of
finite-size corrections, but note that the possibility of the non-trivial
clusters having a more complicated topology makes the link to the quantum
group more subtle.

Another line of investigation would be to study the Potts model for a generic
value of $Q$, i.e., to transpose what we did for the $\chi_{1,2j+1}$ to the
$K_{1,2j+1}$. Indeed, studies for $v$ given in the complex $Q$-plane have
already been made, for example in Ref.~\cite{transfer4} for $v=-1$, but to our
knowledge no study exists for $Q$ given in the complex $v$-plane. Note that
the results are very different in these two cases. For example, with $L$ fixed
and finite, the Beraha number $Q=B_p$ are limiting points in the complex
$Q$-plane for fixed $v=-1$ (and presumably everywhere in the Berker-Kadanoff
phase), but $v=-1$ is not a limiting point in the complex $v$-plane for fixed
$Q=B_p$ ($p>4$). This is just one example that different limits may not
commute and the very concept of ``a thermodynamic limit'' for
antiferromagnetic models has to be manipulated with great care.

\subsection*{Acknowledgments}

We thank to Hubert Saleur for useful comments on the first stage of this
work, Robert Shrock for correspondence, and Alan Sokal for discussions 
on closely related projects. 
J.S. thanks the warm hospitality of the members of the LPTMS, where 
part of this work was done. 
This research has been partially supported by U.S. National Science 
Foundation grants PHY-0116590 and PHY-0424082, and by MEC (Spain) 
grants MTM2004-01728 and FIS2004-03767. 

%
%
\appendix
\section{Dimension of the transfer matrix} \label{sec_dim}

The dimension of the transfer matrices $\T_k(p,L)$ can be obtained
in closed form. First note that for given $p$, $k=2j+1$, and $L$, the
dimension of the transfer matrix $\T_k(p,L)$   
\be
d_k^{(p)}(L) \;=\; \dim \T_k(p,L)
\label{def_dim_T}
\ee
equals the number of random walks (with up and down steps) of length $2L$ 
steps that start at height $1$ and end at height $k$. This random walks
have to evolve between a ``ceiling'' $1$ and a ``roof'' $m=p-1$.

Let us now proceed in steps. For $k=1$ and $m=\infty$ we have just the
Catalan numbers. Thus, if $z$ is the fugacity of a single step, then
the ordinary generating function (o.g.f.) is
\be
f(z) \;=\; \frac{1 - \sqrt{1-4z^2}}{2z^2} \;=\; 
             1 + \sum\limits_{L=1}^\infty C_L \, z^{2L}
\ee

We now keep $k=1$, and we introduce the ``roof'' $m$. A walk is either
empty or consists of two independent parts. The first part is between the
very first step (necessarily up) and the first down step that hits the
ceiling (i.e., $1$); the second part is the rest of the walk (which may be 
empty). For instance, if $p=4$ ($m=3$) and $L=3$, a possible walk can be
$1$--$2$--$3$--$2$--$1$--$2$--$1$. The first part of this walk is 
$1$--$2$--$3$--$2$--$1$; while the second part of the walk is 
$1$--$2$--$1$. 
If we take away the first and last steps of the first part 
(i.e., we are left with $2$--$3$--$2$), we have a walk with $m\to  m-1$
(as this is equivalent to $1$--$2$--$1$). Thus, the o.g.f.\  $f(m,z)$ 
satisfies the equation 
\be
f(m,z) \;=\; 1 + z^2 \, f(m-1,z)\, f(m,z) 
\ee
which is solved by the recurrence
\begin{subeqnarray}
f(m,z) &=& \frac{1}{1-z^2 \, f(m-1,z)} \\
f(1,z) &=& 1 
\label{def_fmz}
\end{subeqnarray}

Finally, let us consider the general case with $k>1$. In this case, the walk
cannot be empty, and the first step is necessarily up. There are two classes
of walks. In the first one, the walk never hits the ceiling $1$ again. 
For instance if $p=4$, $L=3$, and $k=3$, a walk belonging to this class
is given by $1$--$2$--$3$--$2$--$3$--$2$--$3$. So it consists in one step
and a walk with a raised ceiling (i.e., $2$--$3$--$2$--$3$--$2$--$3$ is 
equivalent to $1$--$2$--$1$--$2$--$1$--$2$ with roof $m=2$). In the second 
class, the walk does hit the ceiling somewhere for the first time, so we
can split the walk into two independent parts as in the preceding paragraph.
Thus, the o.g.f.~satisfies the equation
\be
f(m,k,z) \;=\; z \, f(m-1,k-1,z) + z^2 \, f(m-1,z) \, f(m,k,z) 
\ee
which can be solved by the recurrence
\begin{subeqnarray}
f(m,k,z) &=& \frac{z\, f(m-1,k-1,z)}{1-z^2 \, f(m-1,z)} \\
f(m,1,z) &=& f(m,z) 
\label{def_fmkz}
\end{subeqnarray}
where $f(m,z)$ is given by \reff{def_fmz}. The dimensions $d_k^{(p)}(L)$ can
be read off immediately 
\be
f(m,k,z) \;=\; \sum\limits_{L=0}^\infty d_k^{(p)}(L)\, z^{2L} 
\ee

In the particular case $p=4$, we easily find that
\begin{subeqnarray}
f(3,1,z) &=& \frac{1-z^2}{1-2z^2} \;=\; 
1 + \sum\limits_{L=1}^\infty 2^{L-1}\, z^{2L}\\
f(3,3,z) &=& \frac{z^2}{1-2z^2}\;=\; 
\sum\limits_{L=1}^\infty 2^{L-1}\, z^{2L}
\label{dim_p=4}
\end{subeqnarray}
For the other cases, we can get closed formulas for the generating functions,
and obtain the result
\be
 d_k^{(p)}(L)\;=\;
 \sum_{n \ge 0} \left( \gamma_{np+j}(L) - \gamma_{(n+1)p-1-j}(L) \right) \,.
 \label{dkpL}
\ee
where $k=2j+1$ and we have defined $\gamma_{j}(L)\equiv 0$ for $j>L$.
The $\gamma_j(L)$ are given by
\be
\gamma_{j}(L)= {2L \choose L-j} - {2L \choose L-j-1} =
\frac{2j+1}{L+j+1} {2L \choose L-j} \;.
\ee
This result can also be obtained by another method, which consists of
calculating the number $\gamma_{j}(L)$ of states of highest weight with spin
$S=S_z=j$ for the vertex model and taking into account the coupling of
$U_q(SU(2))$ between different $j$ for $p$ integer \cite{Pasquier_90}.
Yet another method consists in relating $d_k^{(p)}$ to the number of paths on
the Dynkin diagram $A_{p-1}$ going from $1$ to $2j+1$ and using the
eigenvectors of the adjacency matrix \cite{Saleur_89}.

The $\gamma_j(L)$ can also be interpreted as the dimension of the transfer
matrix in the FK representation with $j$ bridges, i.e., for a generic
(irrational) value of $p$. In that context, Eq.~(\ref{dkpL}) represents the
reduction of the dimension that takes case at $p$ integer when going from the
FK to the RSOS representation (with spin $j$), and thus, is completely
analogous to Eq.~(\ref{comb_K}) for the generation functions.

On the chromatic line $x=-1/\sqrt{Q}$, $\gamma _{j}(L)$ is replaced by
a smaller dimension $\Gamma_{j}(L)$, because the operator $V=\prod
V_i$ is a projector ($V^2=V$) that projects out nearest-neighbor
connectivities (i.e.  the action of $V$ on states with nearest
neighbours connected gives zero). We do not know of any explicit
expression for $\Gamma_{j}(L)$, but it verifies the following
recursion relation \cite{refhan}
\begin{subeqnarray}
\Gamma_{0}(L+1) &=& \Gamma_{1}(L)  \\
\Gamma_{j}(L+1) &=&  \Gamma_{j-1}(L) + \Gamma_{j}(L) + \Gamma_{j+1}(L) \mbox{ for $j>0$}
\end{subeqnarray}
with the convention that $\Gamma_{j}(L)=0$ for $j<0$ and the
conditions $\Gamma_{j}(L)=0$ for $j>L$, $\Gamma_{L}(L)=1$, and
$\Gamma_{0}(1)=0$. In particular, it can be shown that
$\Gamma_{0}(L)=M_{L-1}$, where $M_{L-1}$ is a Motzkin number and
corresponds to the number of non-crossing non-nearest neighbor
partitions of $\{1,\ldots,L\}$ (i.e., it is the dimension of the
cluster transfer matrix in the case of free longitudinal boundary
conditions and $x=-1/\sqrt{Q}$).  Note that in the RSOS representation
(in the case of $p$ integer), we {\em cannot} reduce the dimension of
the $\T_{2j+1}$, since although $V$ is a projector in the RSOS
representation too the states which are projected out are linear
combinations of the basis states (corresponding to a given
configuration of the heights), and not simply basis states as in the
case of the FK representation. But because of Eq.~(\ref{comb_K}), the
number $d_j(L)$ of non null eigenvalues of $T_{2j+1}$ is given by
Eq.~(\ref{dkpL}) with $\gamma_{j}(L)$ replaced by $\Gamma_{j}(L)$. In
particular, for $Q=3$ we find using the recursion relation that
$d_{1}(L)=d_{5}(L)=2^{L-2}$ and $d_{3}(L)=2^{L-1}$. Indeed, for
$x=-1/\sqrt{Q}$, the three-state Potts model is equivalent to a
homogeneous six-vertex model with all the weights equal to $1$
\cite{Lenard_67} (note that this six-vertex model is different from
the one we considered before).

\clearpage
%
%
\begin{table}
\centering
\begin{tabular}{|rr|r|r|r|r|r|r|r|}
\hline\hline
$p$ & $L$ & $a_1$ & $a_2$ & $a_3$ & $a_4$ & $a_5$ & $a_6$ & $a_7$ \\
\hline
4 & 2 &  3 &  4 &     &     &      &      &      \\ 
  & 3 &  5 & 10 &  13 &     &      &      &      \\ 
  & 4 &  7 & 21 &  37 &  48 &      &      &      \\ 
  & 5 &  9 & 36 &  86 & 143 &  186 &      &      \\ 
  & 6 & 11 & 55 & 167 & 352 &  564 &  739 &      \\
  & 7 & 13 & 78 & 288 & 742 & 1444 & 2256 & 2973 \\
\hline
5 & 2 &  3 &     $  3+ \sqrt{B_5}$ &     &     &       &      &      \\ 
  & 3 &  5 & 10 &$ 10+3\sqrt{B_5}$ &     &     &       &      \\
  & 4 &  7 & 21 &$ 35+2\sqrt{B_5}$ &$ 35+13\sqrt{B_5}$ &  & & \\ 
  & 5 &  9 & 36 &$ 84+2\sqrt{B_5}$ &$126+17\sqrt{B_5}$ &$128+ 60\sqrt{B_5}$& &\\
  & 6 & 11 & 55 &$165+2\sqrt{B_5}$ &$330+22\sqrt{B_5}$ &$464+102\sqrt{B_5}$& 
                                                        $479+277\sqrt{B_5}$& \\
\hline
6 & 2 &  3 &  5 &     &     &      &      &      \\ 
  & 3 &  5 & 10 &  16 &     &      &      &      \\ 
  & 4 &  7 & 21 &  39 &  61 &      &      &      \\ 
  & 5 &  9 & 36 &  88 & 160 &  250 &      &      \\ 
  & 6 & 11 & 55 & 169 & 374 &  670 & 1050 &      \\
  & 7 & 13 & 78 & 290 & 769 & 1605 & 2838 & 4470 \\
\hline
$\infty$ & 
    2 &  3 &  6 &     &     &      &      &      \\ 
  & 3 &  5 & 10 &  19 &     &      &      &      \\ 
  & 4 &  7 & 21 &  41 &  70 &      &      &      \\ 
  & 5 &  9 & 36 &  90 & 177 &  318 &      &      \\ 
  & 6 & 11 & 55 & 171 & 396 &  780 & 1395 &      \\
\hline\hline
\end{tabular}
\caption{\label{table_sq_coef}
First $L$ coefficients $a_k$ for the leading eigenvalue 
$\lambda_{\star,1}(L)$ coming from the sector $\chi_{1,1}$ for a
square-lattice strip of width $L$. 
}
\end{table}

\clearpage
\begin{table}
\centering
\begin{tabular}{|rr|r|r|r|r|r|r|r|}
\hline\hline
$p$ & $L$ & $b_1$ & $b_2$ & $b_3$ & $b_4$ & $b_5$ & 
            $b_6$ & $b_7$  \\
\hline
4 & 2 &  4 &   6 &   6 &      &       &       &       \\ 
  & 3 &  7 &  21 &  35 &   37 &    31 &       &       \\ 
  & 4 & 10 &  45 & 120 &  212 &   264 &   244 &   184 \\ 
  & 5 & 13 &  78 & 286 &  717 &  1305 &  1793 &  1919 \\
  & 6 & 16 & 120 & 560 & 1822 &  4392 &  8146 & 11940 \\
  & 7 & 19 & 171 & 969 & 3878 & 11658 & 27349 & 51389 \\
  & 8 & 22 & 231 &1540 & 7317 & 26370 & 74927 &172304 \\
\hline
5 & 2 &  4 &   6 &      $   4+  2\sqrt{B_5}$ & & &  &\\ 
  & 3 &  7 &  21 &  35 &$  35+  2\sqrt{B_5}$ &$   21+ 10\sqrt{B_5}$&  & \\ 
  & 4 & 10 &  45 & 120 &$ 210+  2\sqrt{B_5}$ &$  252+ 12\sqrt{B_5}$&
                        $ 210+ 34\sqrt{B_5}$ &$  122+ 64\sqrt{B_5}$  \\
  & 5 & 13 &  78 & 286 &$ 715+  2\sqrt{B_5}$ &$ 1287+ 18\sqrt{B_5}$&
                        $1716+ 77\sqrt{B_5}$ &$ 1718+203\sqrt{B_5}$ \\
  & 6 & 16 & 120 & 560 &$1820+  2\sqrt{B_5}$ &$ 4368+ 24\sqrt{B_5}$& 
                        $8008+138\sqrt{B_5}$ &$11442+500\sqrt{B_5}$\\
\hline
6 & 2 &  4 &   6 &   8 &      &       &       &       \\ 
  & 3 &  7 &  21 &  35 &   39 &    41 &       &       \\ 
  & 4 & 10 &  45 & 120 &  214 &   276 &   278 &   252 \\ 
  & 5 & 13 &  78 & 286 &  719 &  1323 &  1870 &  2126 \\
  & 6 & 16 & 120 & 560 & 1824 &  4416 &  8284 & 12444 \\
  & 7 & 19 & 171 & 969 & 3880 & 11688 & 27566 & 52394 \\
\hline
$\infty$
  & 2 &  4 &   6 &  10 &      &       &       &       \\ 
  & 3 &  7 &  21 &  35 &   41 &    51 &       &       \\ 
  & 4 & 10 &  45 & 120 &  216 &   288 &   312 &   324 \\ 
  & 5 & 13 &  78 & 286 &  721 &  1341 &  1947 &  2337 \\
  & 6 & 16 & 120 & 560 & 1826 &  4440 &  8422 & 12952 \\
\hline\hline
\end{tabular}
\caption{\label{table_tri_coef}
First $\min(2L-1,7)$ coefficients $b_k$ for the leading eigenvalue 
$\lambda_{\star,1}(L)$ coming from the sector $\chi_{1,1}$ for a
triangular-lattice strip of width $L$. 
}
\end{table}

\clearpage
%
%

%
%

%
%

%
%
\begin{figure}
\centering
 \psset{xunit=1pt}
 \psset{yunit=1pt}
 \pspicture(0,0)(400,250)
    \psline[linewidth=1.0pt,linecolor=blue,linestyle=dashed]( 50, 50)(350, 50)
    \psline[linewidth=1.0pt,linecolor=blue,linestyle=dashed]( 50,150)(350,150)
    \psline[linewidth=1.0pt,linecolor=blue,linestyle=dashed]( 50, 50)( 50,150)
    \psline[linewidth=1.0pt,linecolor=blue,linestyle=dashed](150, 50)(150,150)
    \psline[linewidth=1.0pt,linecolor=blue,linestyle=dashed](250, 50)(250,150)
    \psline[linewidth=1.0pt,linecolor=blue,linestyle=dashed](350, 50)(350,150)

    \psline[linewidth=1.5pt,linecolor=blue](  0,100)(100,200)
    \psline[linewidth=1.5pt,linecolor=blue]( 50, 50)(200,200)
    \psline[linewidth=1.5pt,linecolor=blue](100,  0)(300,200)
    \psline[linewidth=1.5pt,linecolor=blue](200,  0)(350,150)
    \psline[linewidth=1.5pt,linecolor=blue](300,  0)(350, 50)

    \psline[linewidth=1.5pt,linecolor=blue](100,  0)(  0,100)
    \psline[linewidth=1.5pt,linecolor=blue](200,  0)( 50,150)
    \psline[linewidth=1.5pt,linecolor=blue](300,  0)(100,200)
    \psline[linewidth=1.5pt,linecolor=blue](350, 50)(200,200)
    \psline[linewidth=1.5pt,linecolor=blue](350,150)(300,200)
   
    \multirput{0}(50,  50)(0,100){2}{\pscircle*(0,0){4pt}} 
    \multirput{0}(150, 50)(0,100){2}{\pscircle*(0,0){4pt}} 
    \multirput{0}(250, 50)(0,100){2}{\pscircle*(0,0){4pt}} 
    \multirput{0}(350, 50)(0,100){2}{\pscircle*(0,0){4pt}} 

    \psline[linewidth=4pt,linecolor=black]{->}(360,100)(400,100)

    \rput{0}(97,210){$h_1=1$}
    \rput{0}(15,100){$h_3$}
    \rput{0}(100,-10){$h_5$}
    \rput{0}(35,50){$h_4$}
    \rput{0}(35,152){$h_2$}
 \endpspicture
\vspace*{1cm}
\caption{\label{fig_sq_L=2}
RSOS lattice (solid thick lines) and label convention for the basis in the 
height space for a square-lattice of width $L=2$ (dashed thinner lines). 
The thick black arrow shows the transfer direction (to the right).
}
\end{figure}

\clearpage
%
%
\begin{figure}
\centering
 \psset{xunit=1pt}
 \psset{yunit=1pt}
 \pspicture(0,0)(400,250)
    \psline[linewidth=1.0pt,linecolor=blue,linestyle=dashed]( 50, 50)(350, 50)
    \psline[linewidth=1.0pt,linecolor=blue,linestyle=dashed]( 50,150)(350,150)
    \psline[linewidth=1.0pt,linecolor=blue,linestyle=dashed]( 50, 50)( 50,150)
    \psline[linewidth=1.0pt,linecolor=blue,linestyle=dashed](150, 50)(150,150)
    \psline[linewidth=1.0pt,linecolor=blue,linestyle=dashed](250, 50)(250,150)
    \psline[linewidth=1.0pt,linecolor=blue,linestyle=dashed](350, 50)(350,150)
    \psline[linewidth=1.0pt,linecolor=blue,linestyle=dashed]( 50,150)(150, 50)
    \psline[linewidth=1.0pt,linecolor=blue,linestyle=dashed](150,150)(250, 50)
    \psline[linewidth=1.0pt,linecolor=blue,linestyle=dashed](250,150)(350, 50)

    \psline[linewidth=1.5pt,linecolor=blue]( 50,150)(100,200)(150,150)
    \psline[linewidth=1.5pt,linecolor=blue](150,150)(200,200)(250,150)
    \psline[linewidth=1.5pt,linecolor=blue](250,150)(300,200)(350,150)

    \psline[linewidth=1.5pt,linecolor=blue]( 50, 50)(100,  0)(150, 50)
    \psline[linewidth=1.5pt,linecolor=blue](150, 50)(200,  0)(250, 50)
    \psline[linewidth=1.5pt,linecolor=blue](250, 50)(300,  0)(350, 50)
   
    \psline[linewidth=1.5pt,linecolor=blue]( 50, 50)( 80, 80)( 50,150)
    \psline[linewidth=1.5pt,linecolor=blue](150, 50)( 80, 80)
    \psline[linewidth=1.5pt,linecolor=blue](150, 50)(180, 80)(150,150)
    \psline[linewidth=1.5pt,linecolor=blue](250, 50)(180, 80)
    \psline[linewidth=1.5pt,linecolor=blue](250, 50)(280, 80)(250,150)
    \psline[linewidth=1.5pt,linecolor=blue](350, 50)(280, 80)

    \psline[linewidth=1.5pt,linecolor=blue]( 50, 50)( 20,120)( 50,150)
    \psline[linewidth=1.5pt,linecolor=blue](150, 50)(120,120)(150,150)
    \psline[linewidth=1.5pt,linecolor=blue]( 50,150)(120,120)
    \psline[linewidth=1.5pt,linecolor=blue](250, 50)(220,120)(250,150)
    \psline[linewidth=1.5pt,linecolor=blue](150,150)(220,120)
    \psline[linewidth=1.5pt,linecolor=blue](350, 50)(320,120)(350,150)
    \psline[linewidth=1.5pt,linecolor=blue](250,150)(320,120)

    \psline[linewidth=4pt,linecolor=black]{->}(360,100)(400,100)


    \multirput{0}(50,  50)(0,100){2}{\pscircle*(0,0){4pt}} 
    \multirput{0}(150, 50)(0,100){2}{\pscircle*(0,0){4pt}} 
    \multirput{0}(250, 50)(0,100){2}{\pscircle*(0,0){4pt}} 
    \multirput{0}(350, 50)(0,100){2}{\pscircle*(0,0){4pt}} 

    \rput{0}(97,210){$h_1=1$}
    \rput{0}(10,120){$h_3$}
    \rput{0}(100,-10){$h_5$}
    \rput{0}(35,50){$h_4$}
    \rput{0}(35,152){$h_2$}
 \endpspicture
\vspace*{1cm}
\caption{\label{fig_tri_L=2}
RSOS lattice (solid thick lines) and label convention for the basis in the 
height space for a triangular-lattice of width $L=2$ (dashed thinner lines). 
The thick black arrow shows the transfer direction (to the right).
}
\end{figure}

\clearpage
%
%
\begin{figure}
\centering
\begin{tabular}{cc}
  \includegraphics[width=200pt]{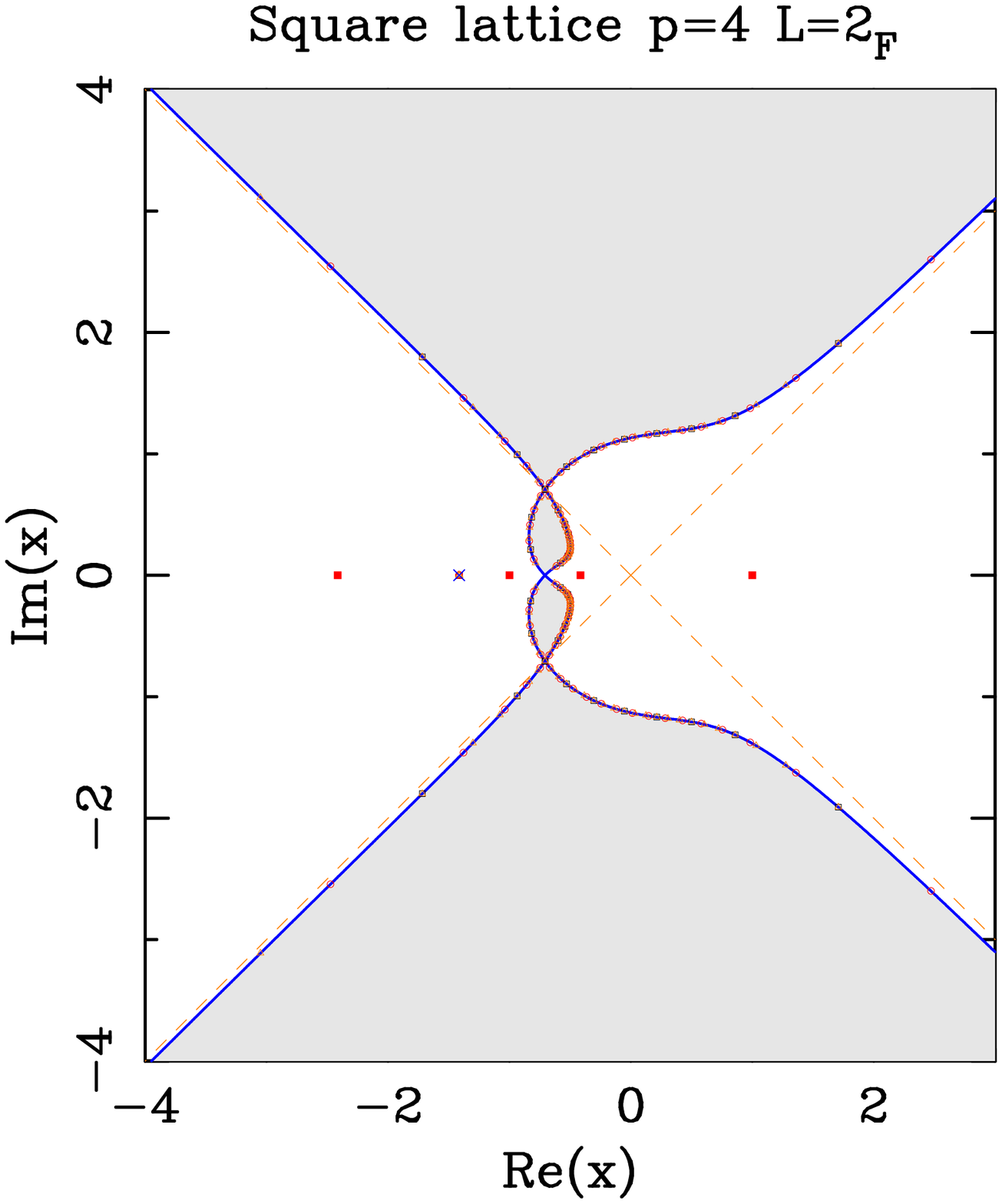} &
  \includegraphics[width=200pt]{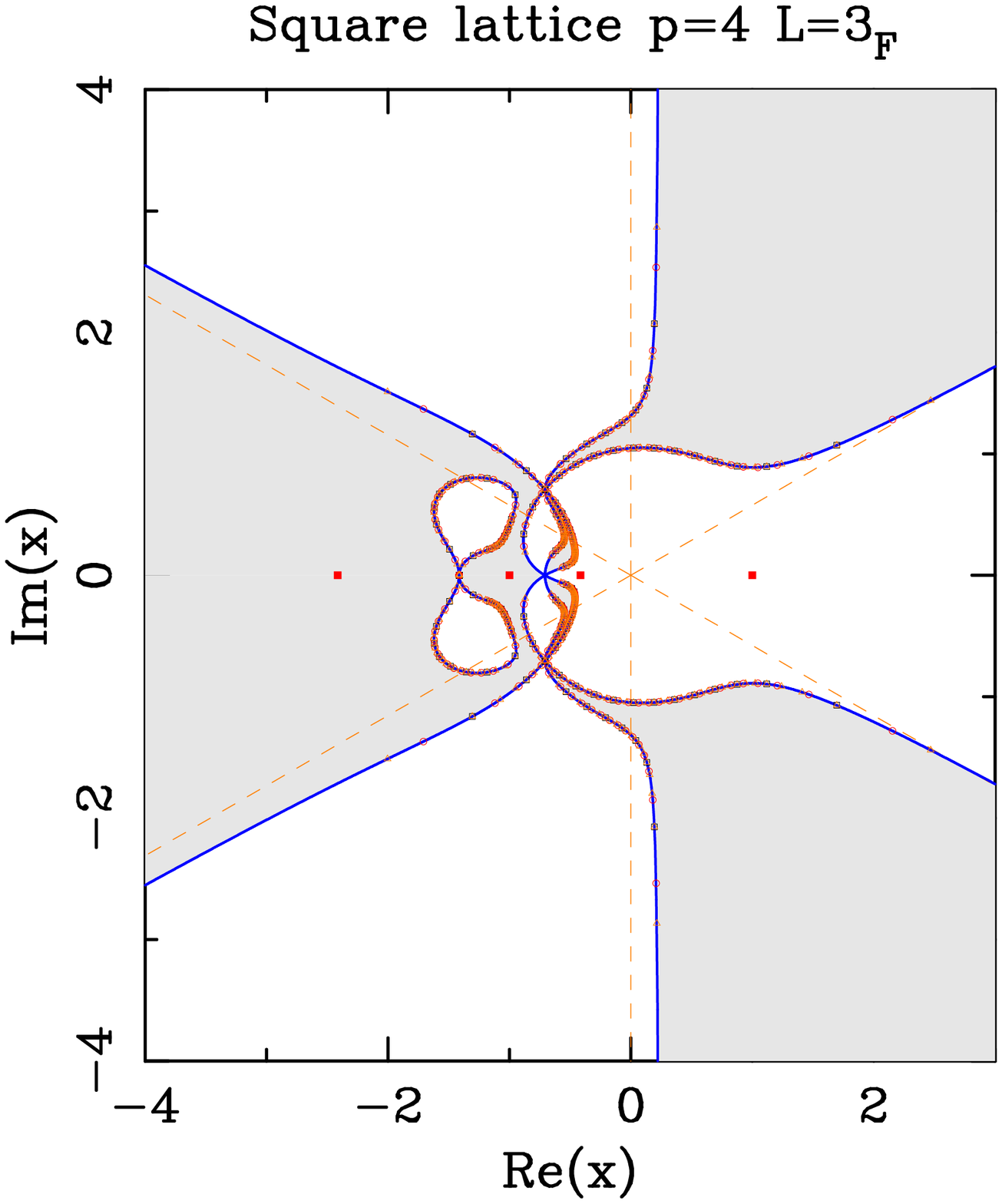} \\[2mm]
  \phantom{(((a)}(a) & \phantom{(((a)}(b)\\[5mm]
  \includegraphics[width=200pt]{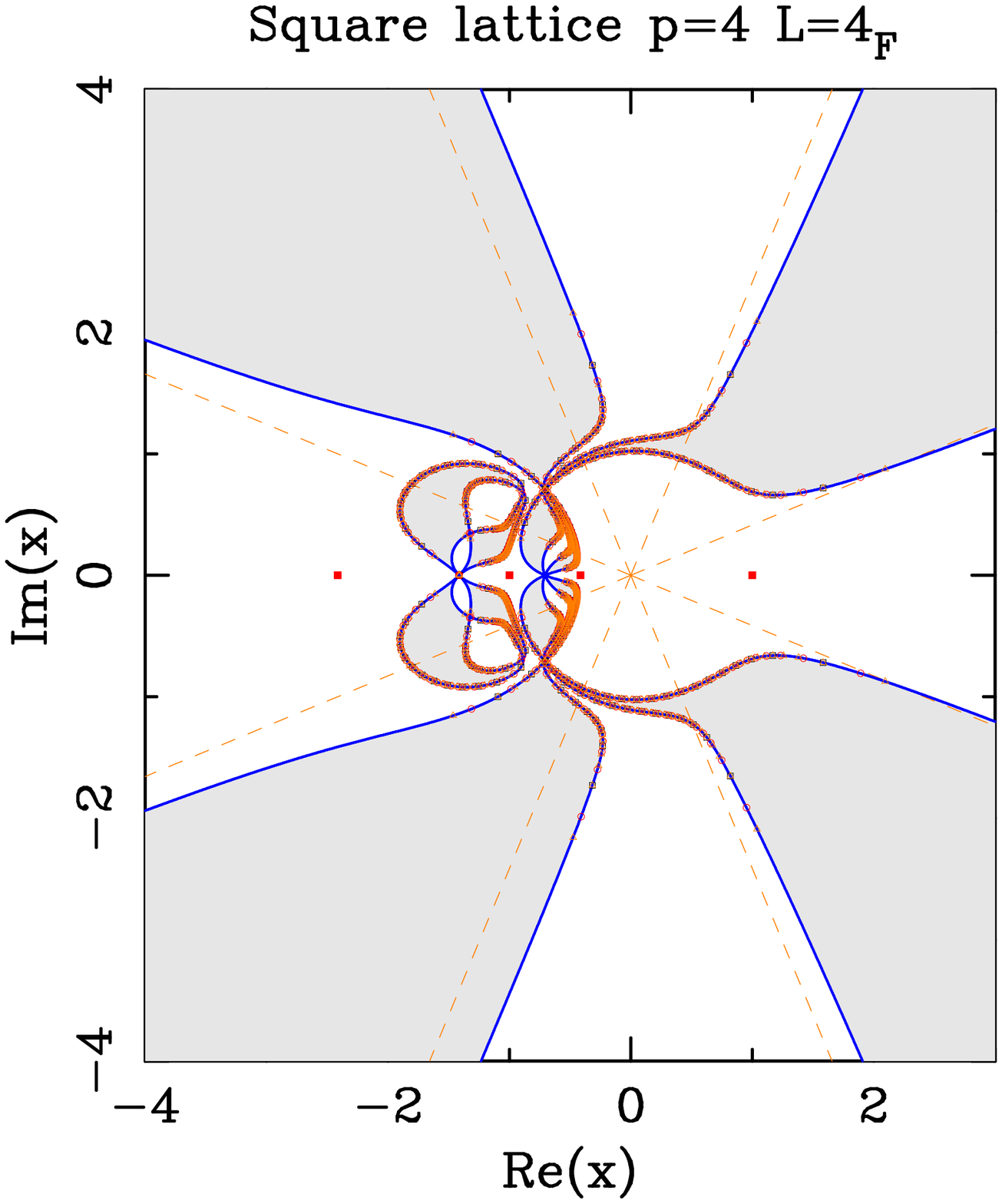} &
  \includegraphics[width=200pt]{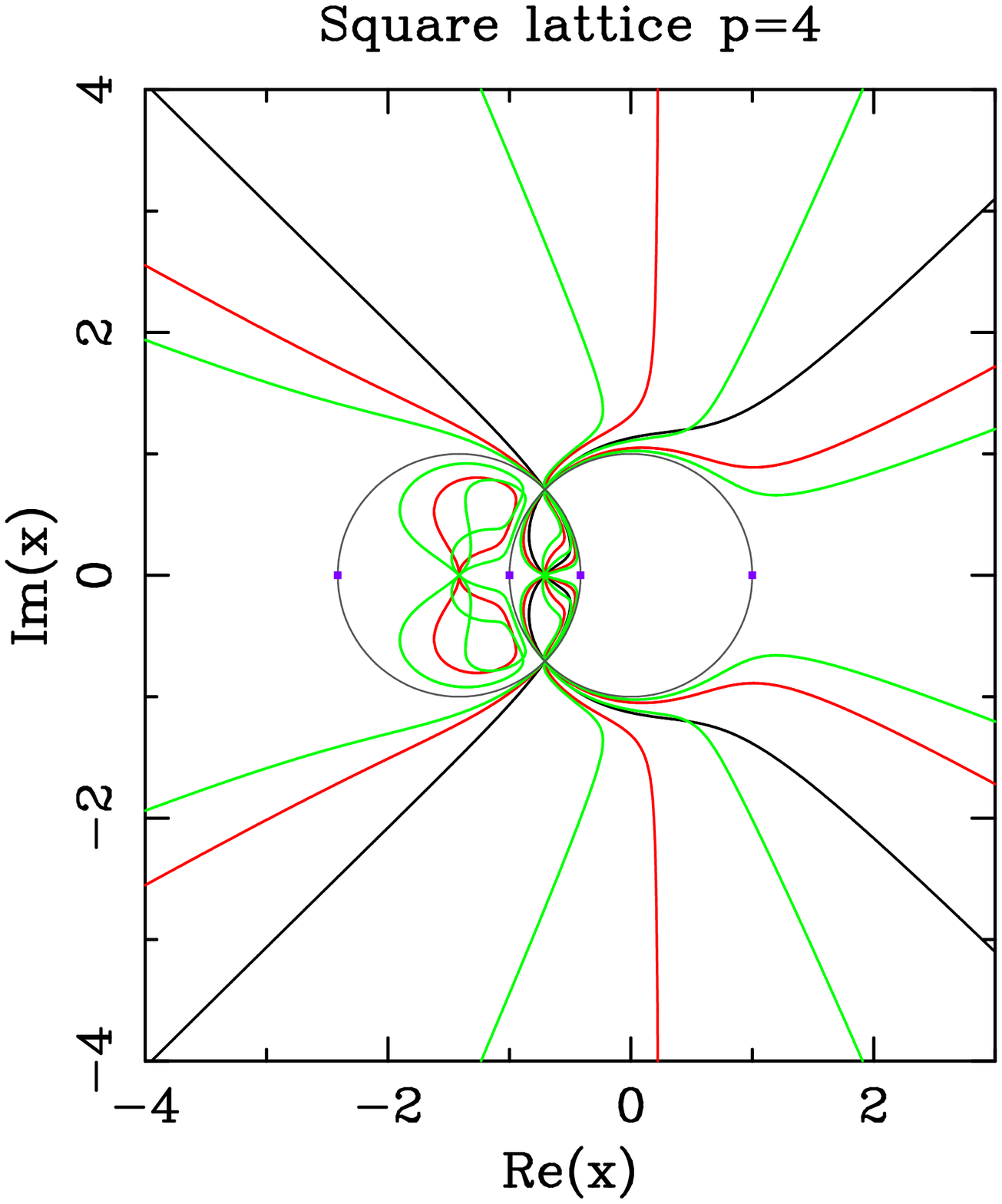} \\[2mm]
  \phantom{(((a)}(c) &  \phantom{(((a)}(d)
\end{tabular}
\caption{\label{Curves_sq_p=4}
Limiting curves for the square-lattice RSOS model with $p=4$ and 
several widths: $L=2$ (a), $L=3$ (b), and $L=4$ (c). For each width $L$, we
also show the partition-function zeros for finite strips of dimensions
$L_\text{F}\times (10L)_\text{P}$ (black $\square$), 
$L_\text{F}\times (20L)_\text{P}$ (red $\circ$), and
$L_\text{F}\times (30L)_\text{P}$ (brown $\triangle$). 
Figure~(d) shows all these limiting curves together: $L=2$ (black), $L=3$ (red),
$L=4$ (green). 
The solid squares $\blacksquare$ show the values where Baxter found 
the free energy.
The symbol $\times$ in (a) marks the position of the found isolated limiting
point.
In the regions displayed in gray (resp.\  white)
the dominant eigenvalue comes from the sector $\chi_{1,3}$ 
(resp.\ $\chi_{1,1}$). 
The dark gray circles correspond to \protect\reff{circles_sq_x}
}
\end{figure}

\clearpage
%
%
\begin{figure}
\centering
\begin{tabular}{cc}
  \includegraphics[width=200pt]{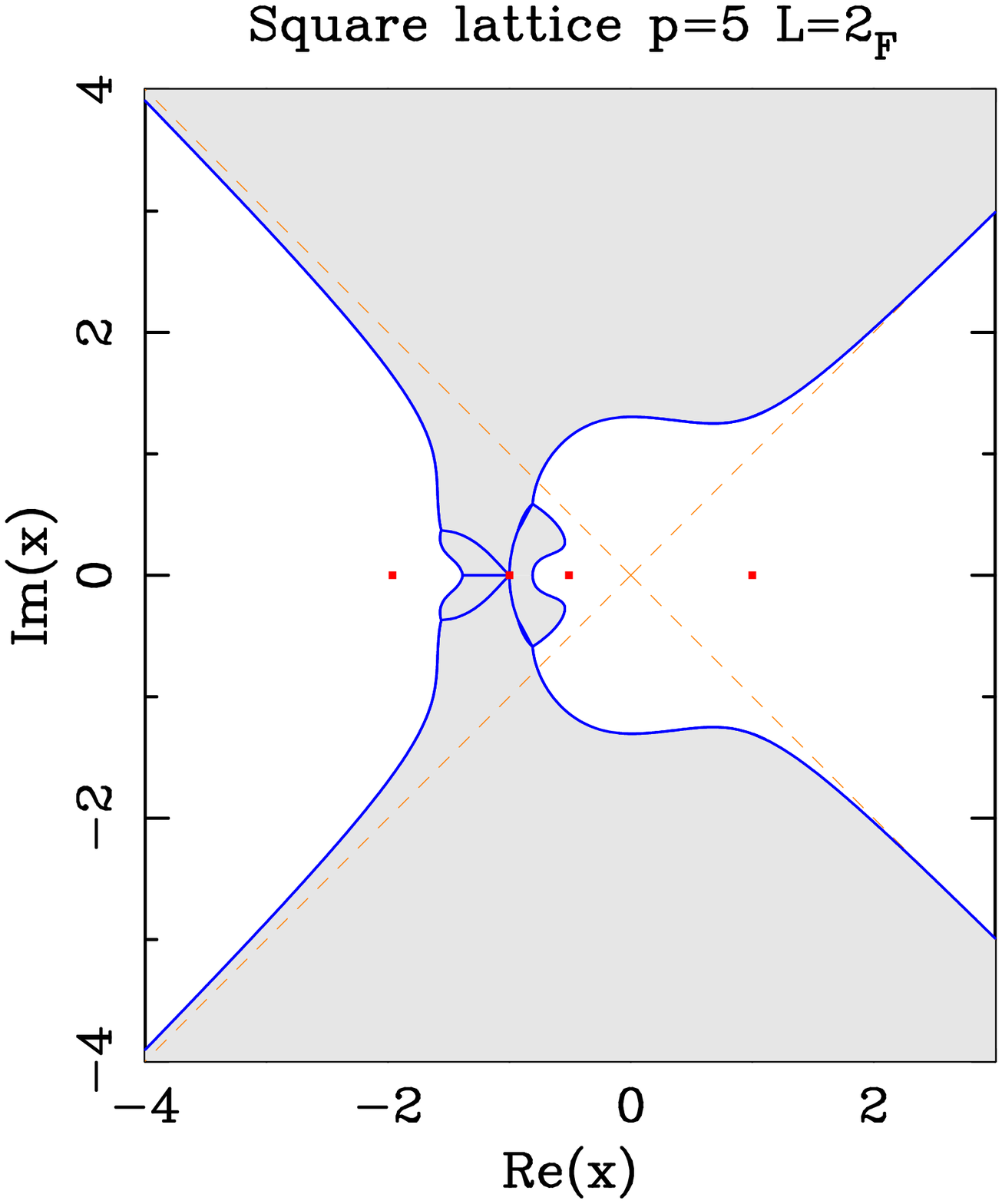} &
  \includegraphics[width=200pt]{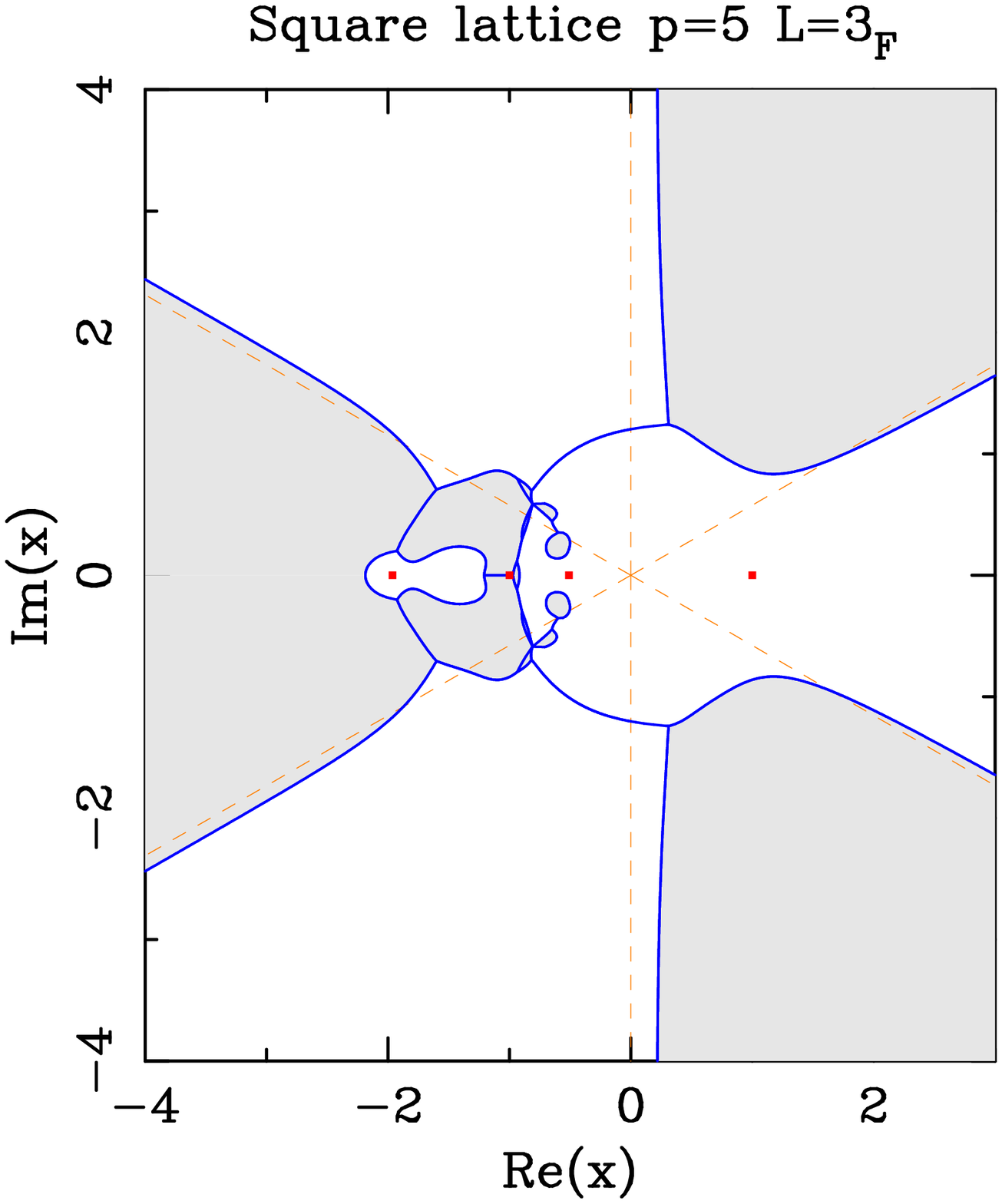} \\[2mm]
  \phantom{(((a)}(a) & \phantom{(((a)}(b)\\[5mm]
  \includegraphics[width=200pt]{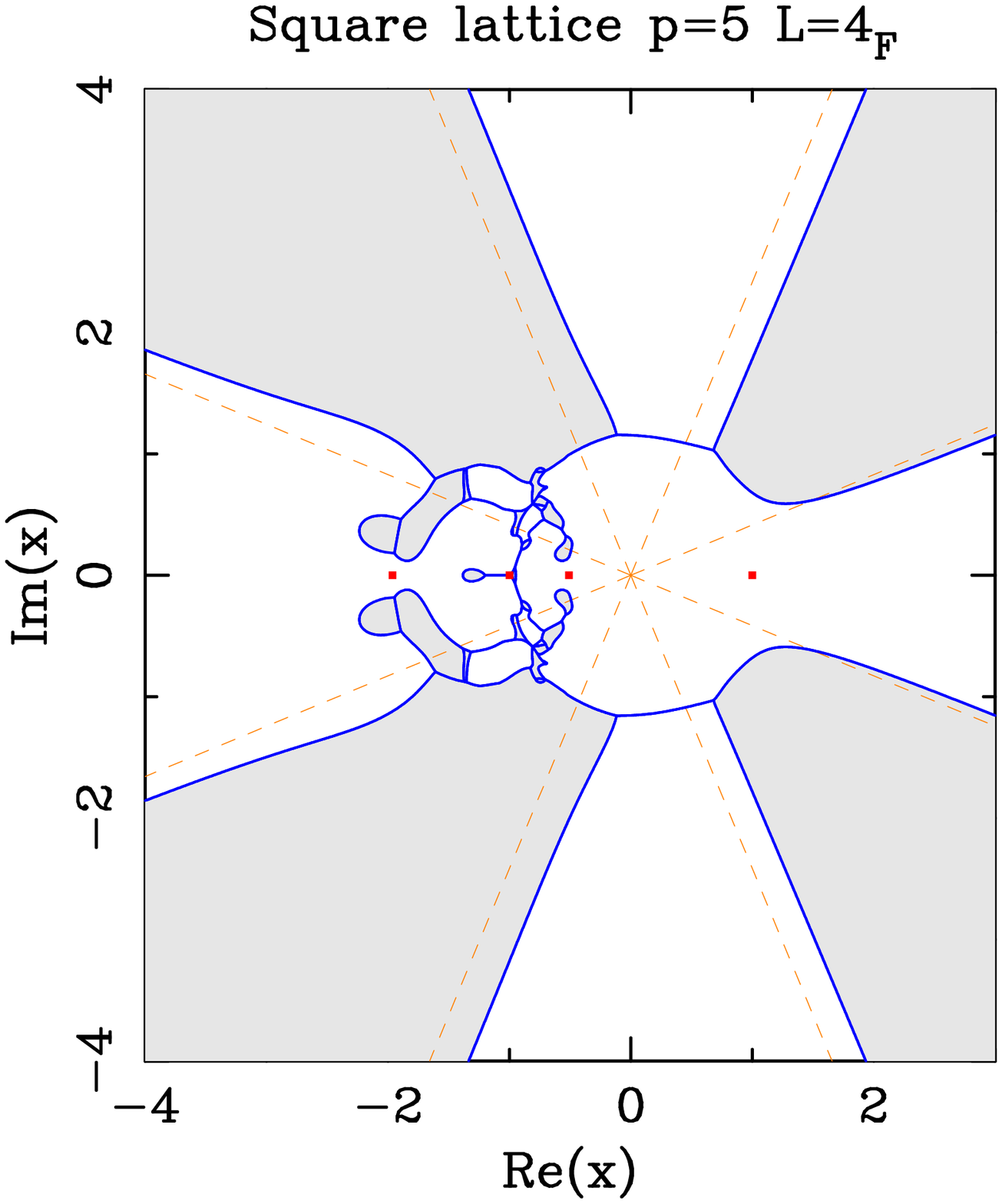} &
  \includegraphics[width=200pt]{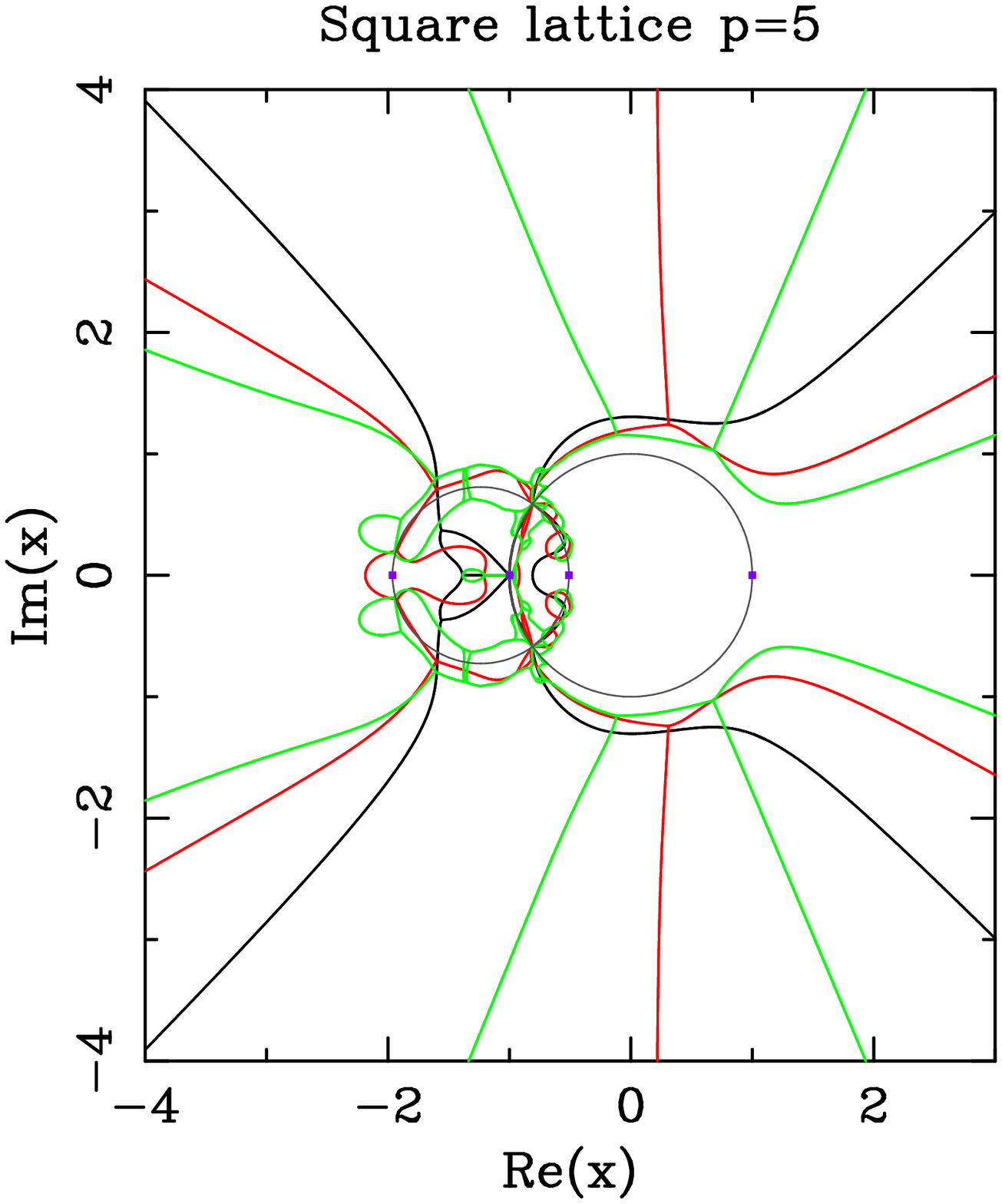} \\[2mm]
  \phantom{(((a)}(c) &  \phantom{(((a)}(d)
\end{tabular}
\caption{\label{Curves_sq_p=5}
Limiting curves for the square-lattice RSOS model with $p=5$ and several widths:
$L=2$ (a), $L=3$ (b), and $L=4$ (c). 
Figure~(d) shows all these curves together: $L=2$ (black), $L=3$ (red),
$L=4$ (green).
The solid squares
$\blacksquare$ show the values where Baxter found the free energy.
In the regions displayed in light gray (resp.\  white)
the dominant eigenvalue comes from the sector $\chi_{1,3}$ 
(resp.\ $\chi_{1,1}$). 
The dark gray circles correspond to \protect\reff{circles_sq_x}
}
\end{figure}

\clearpage
%
%
\begin{figure}
\centering
\begin{tabular}{cc}
  \includegraphics[width=200pt]{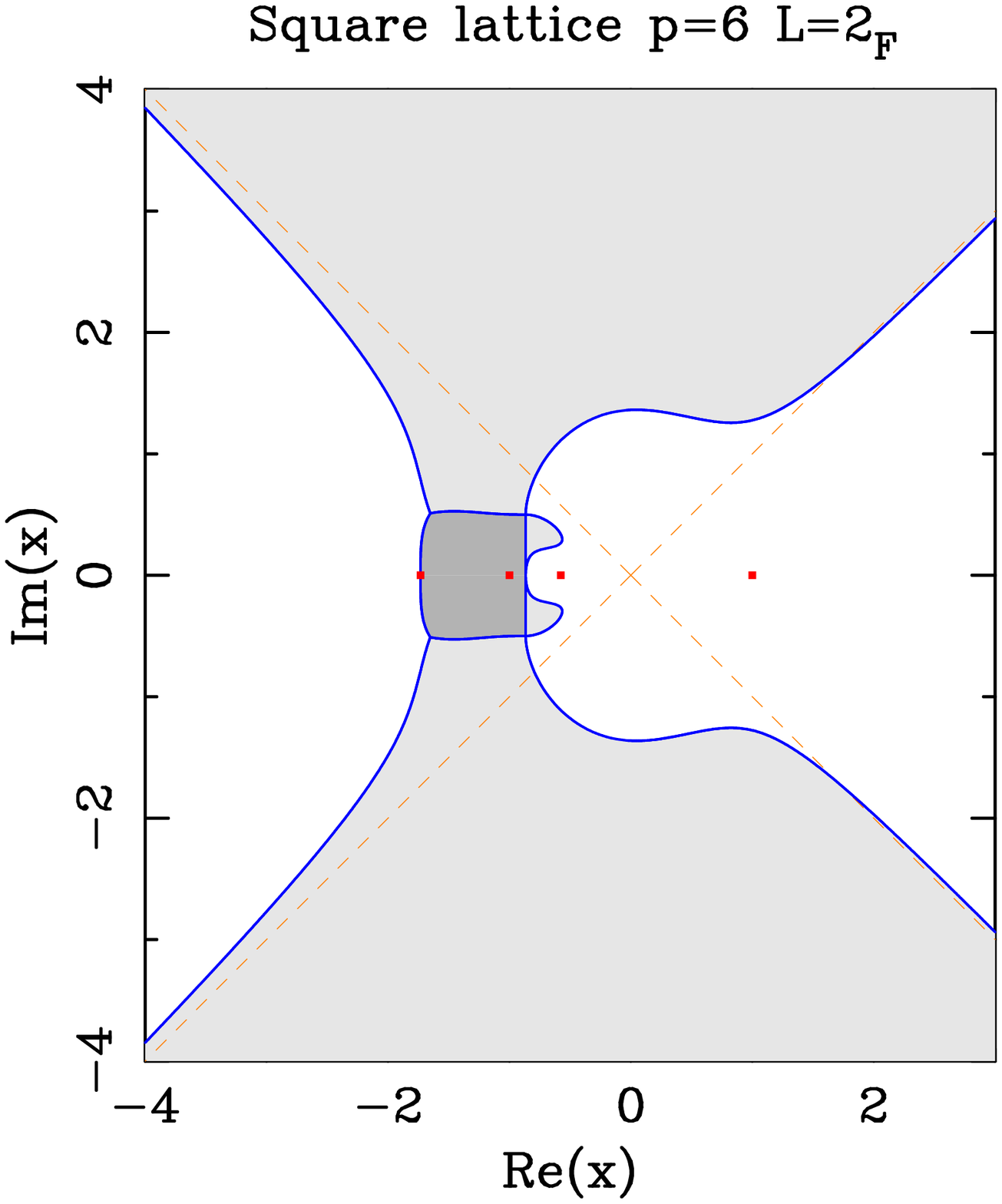} &
  \includegraphics[width=200pt]{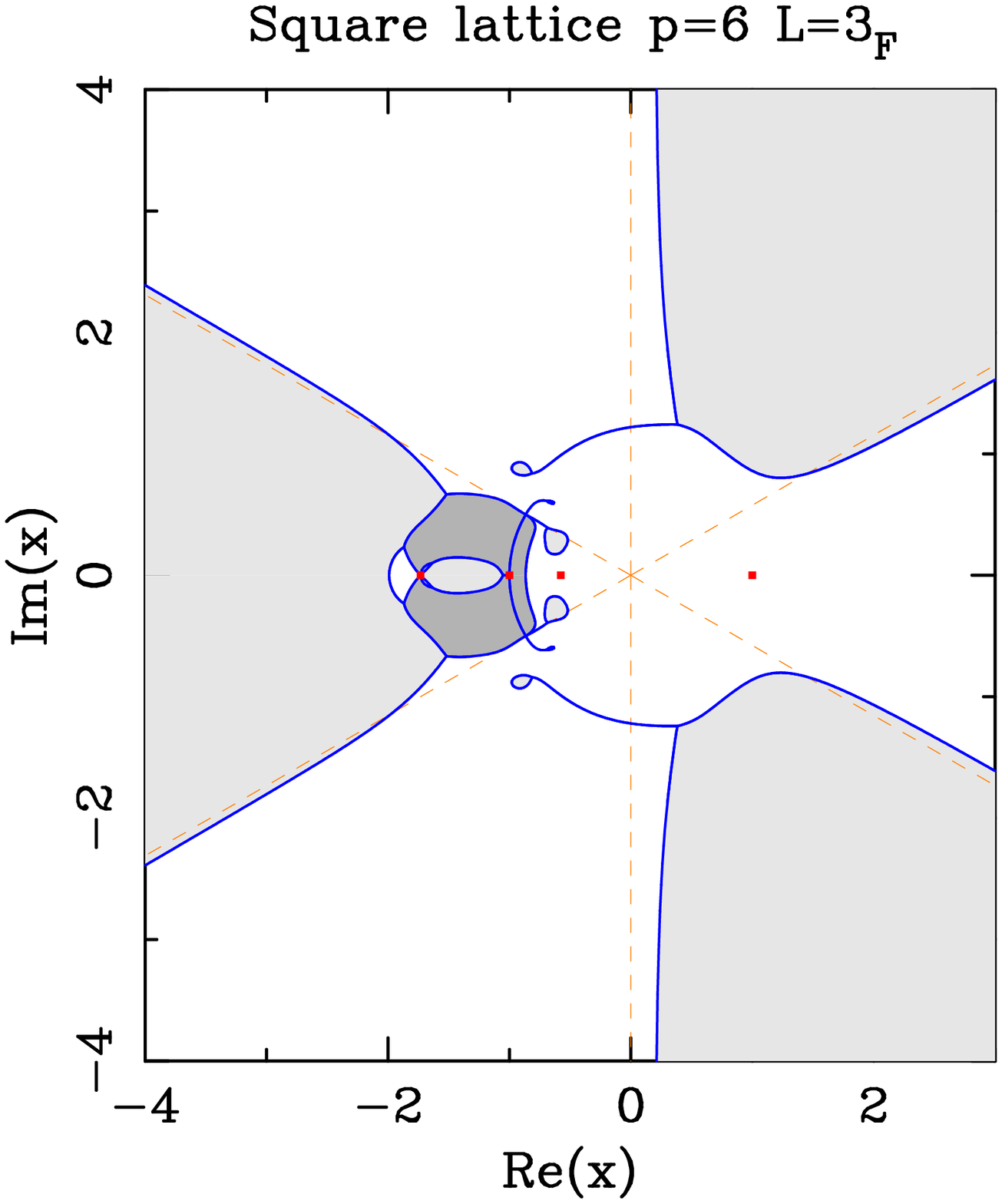} \\[2mm]
  \phantom{(((a)}(a) & \phantom{(((a)}(b)\\[5mm]
  \includegraphics[width=200pt]{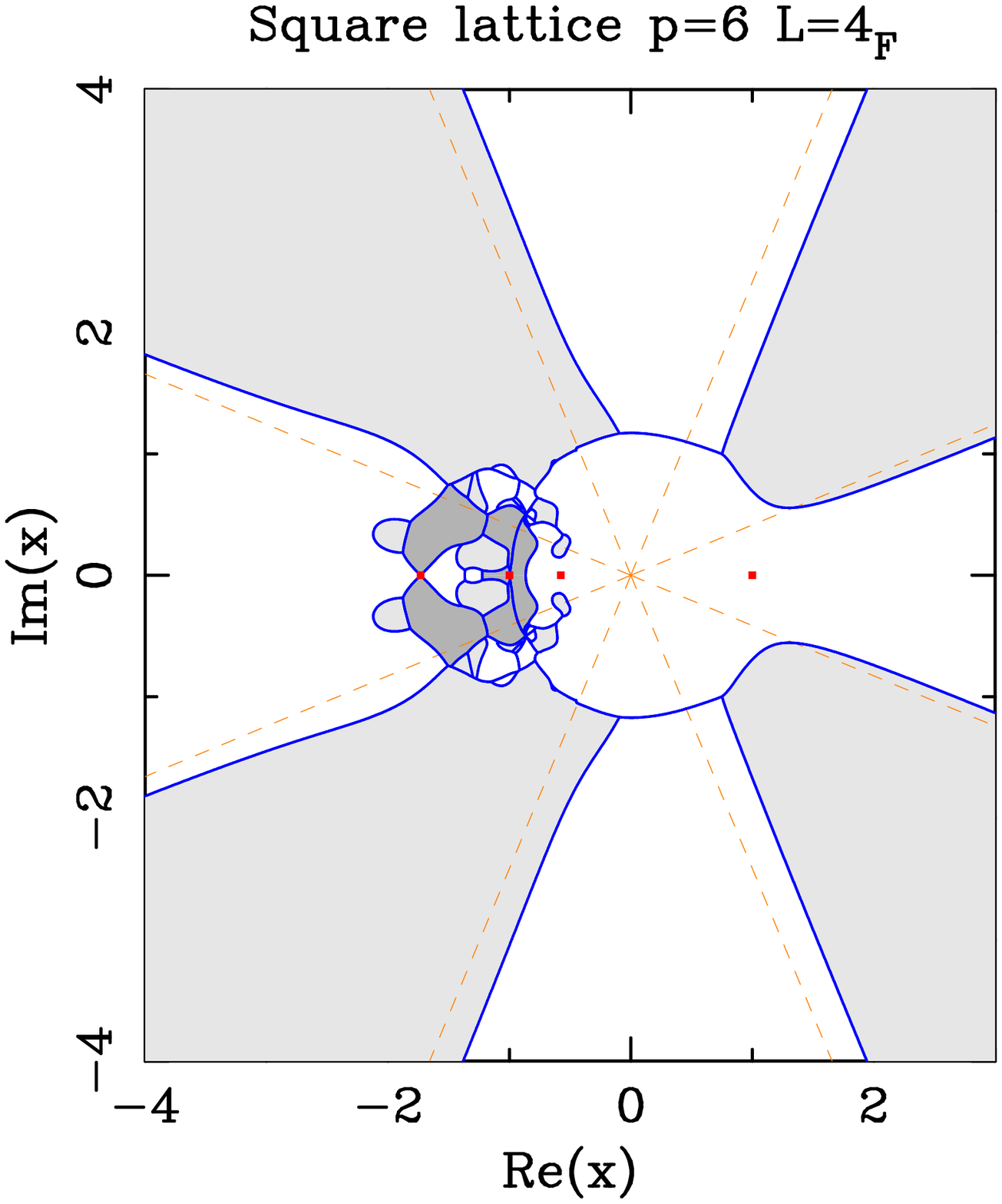} &
  \includegraphics[width=200pt]{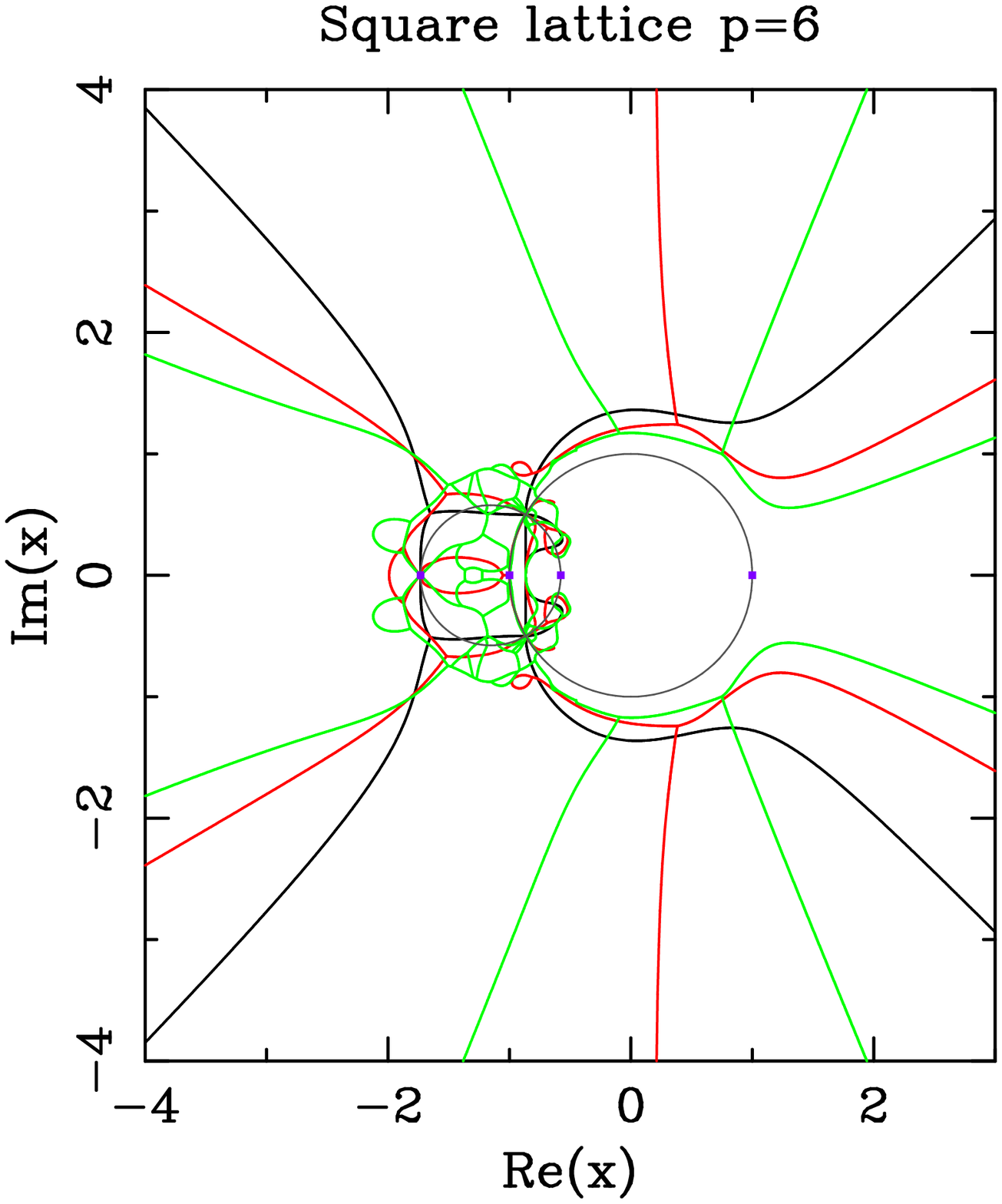} \\[2mm]
  \phantom{(((a)}(c) &  \phantom{(((a)}(d)
\end{tabular}
\caption{\label{Curves_sq_p=6}
Limiting curves for the square-lattice RSOS model with $p=6$ and several widths:
$L=2$ (a), $L=3$ (b), and $L=4$ (c). 
Figure~(d) shows all these curves together: $L=2$ (black), $L=3$ (red),
$L=4$ (green).
The solid squares
$\blacksquare$ show the values where Baxter found the free energy.
In the regions displayed in light gray (resp.\  white)
the dominant eigenvalue comes from the sector $\chi_{1,3}$ 
(resp.\ $\chi_{1,1}$). In the regions displayed in a darker gray
the dominant eigenvalue comes from the sector $\chi_{1,5}$. 
The dark gray circles correspond to \protect\reff{circles_sq_x}
}
\end{figure}

\clearpage
%
%
\begin{figure}
\centering
\begin{tabular}{cc}
  \includegraphics[width=200pt]{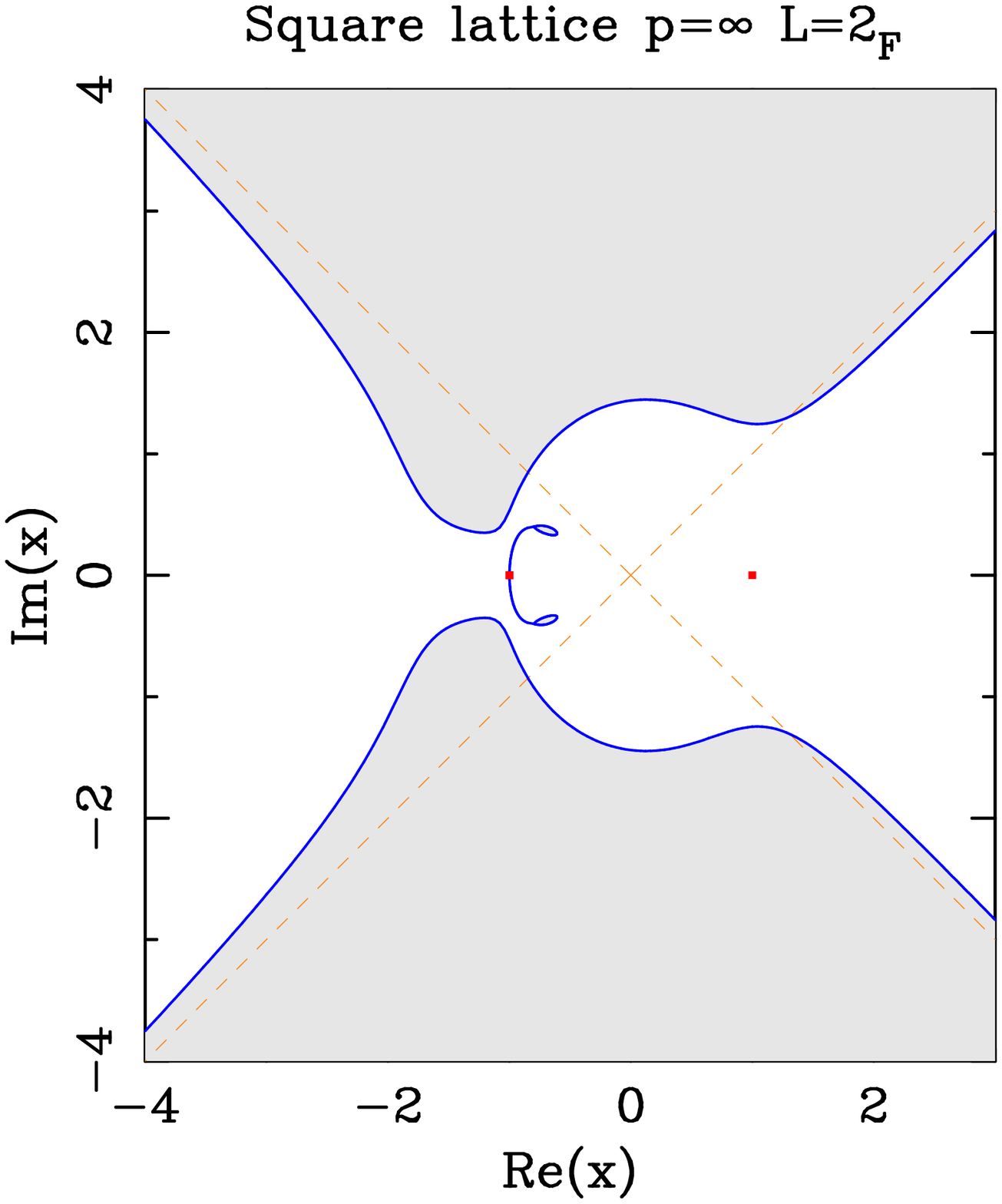} &
  \includegraphics[width=200pt]{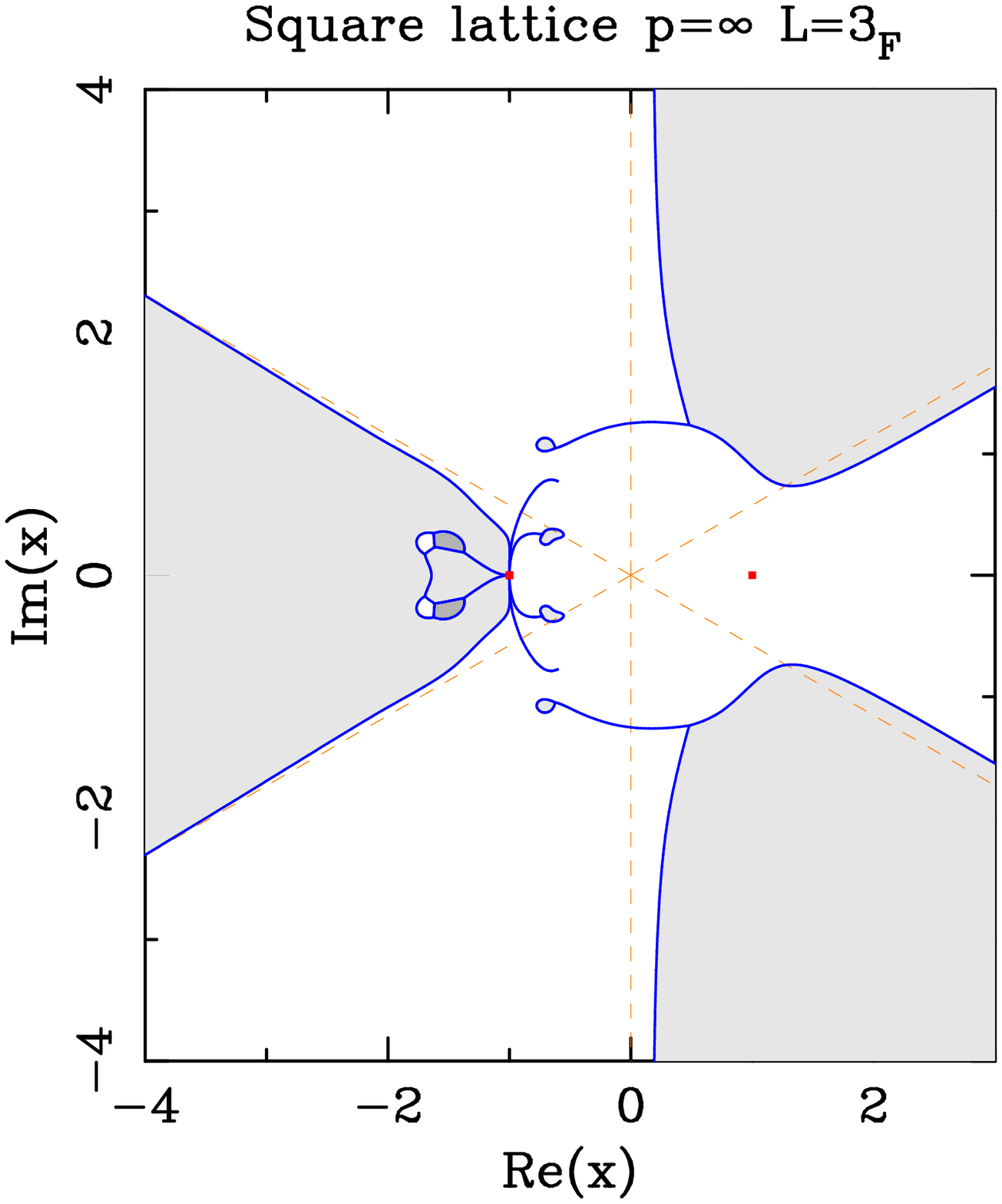} \\[2mm]
  \phantom{(((a)}(a) & \phantom{(((a)}(b)\\[5mm]
  \includegraphics[width=200pt]{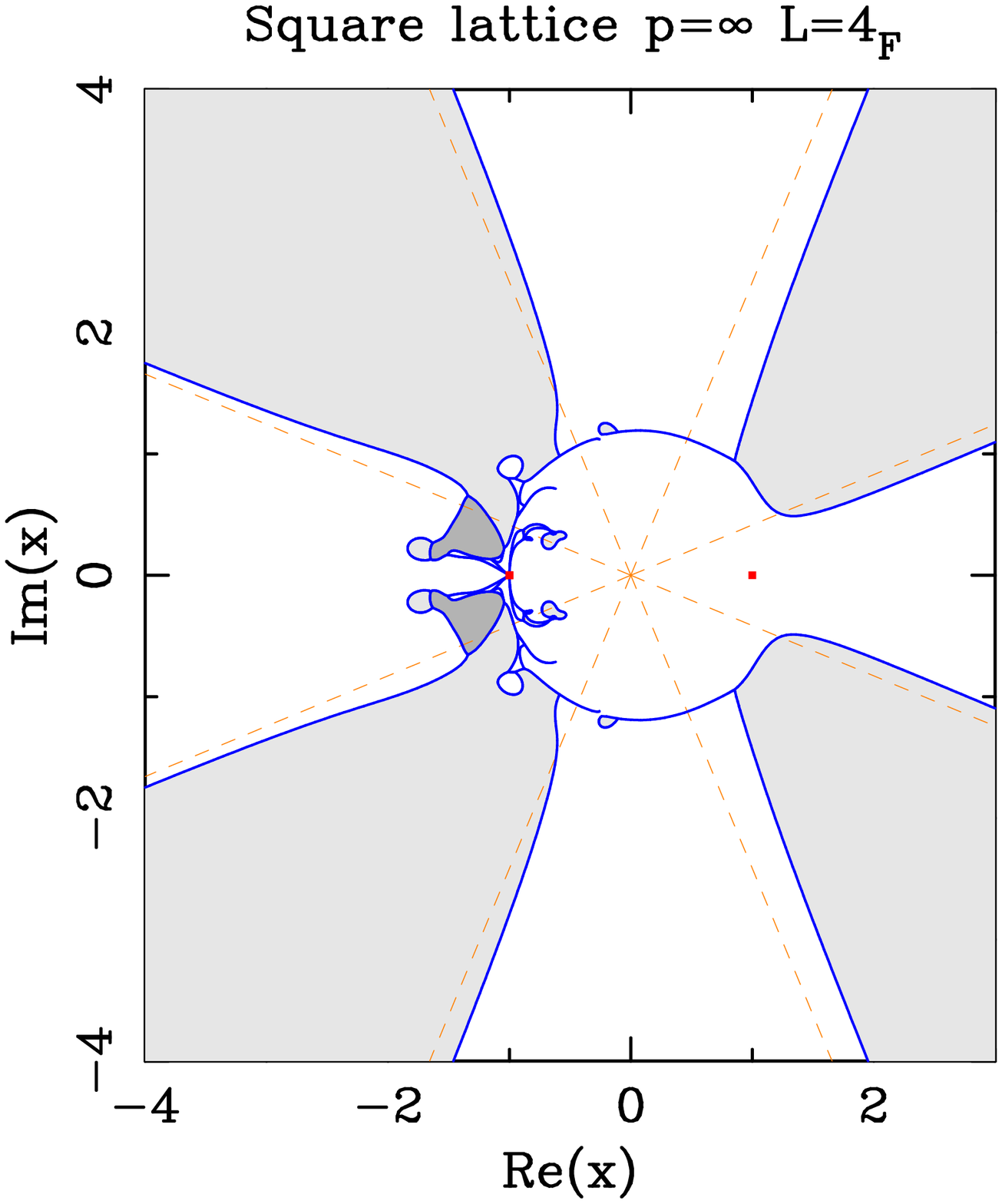} &
  \includegraphics[width=200pt]{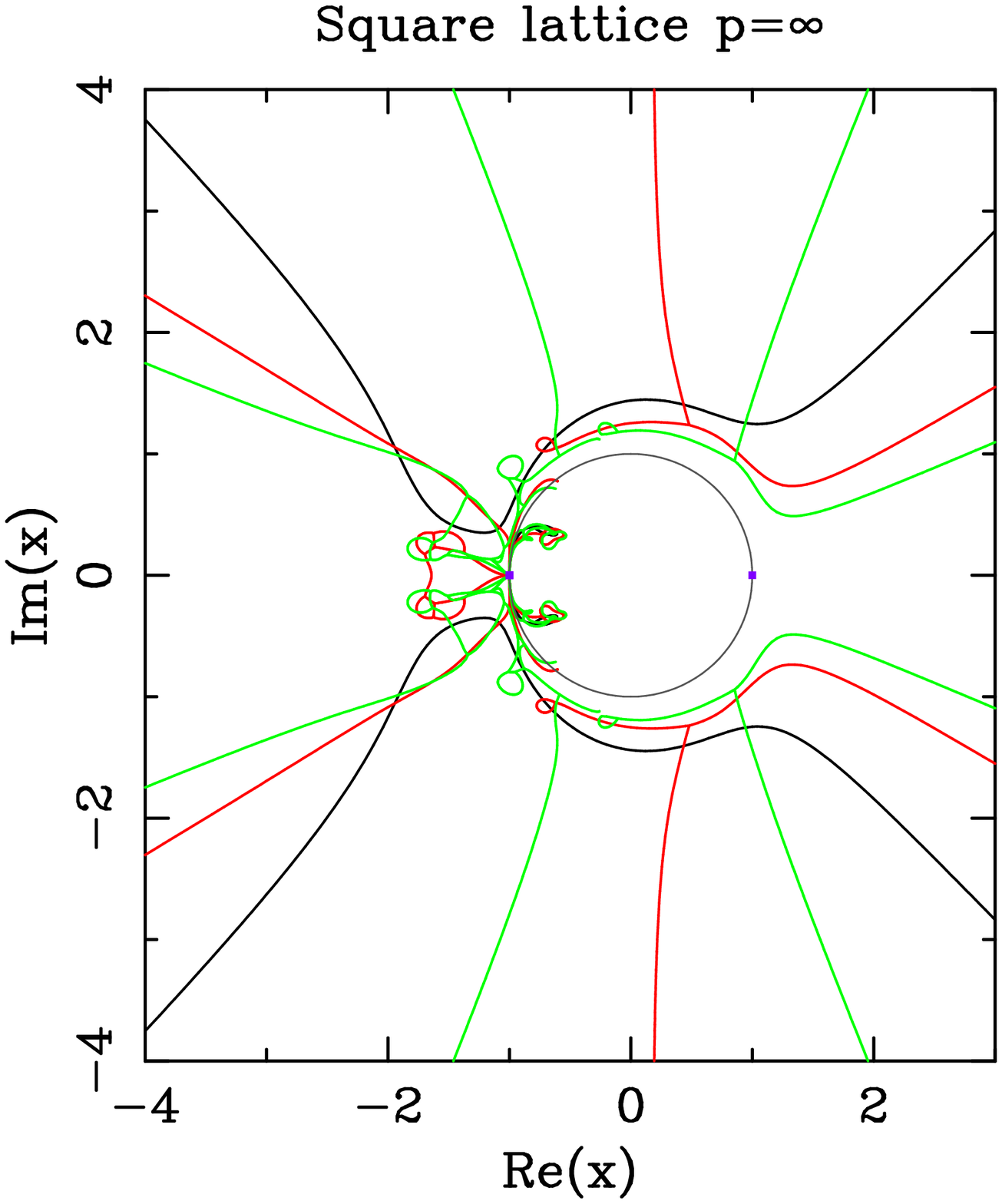} 
\\[2mm]
  \phantom{(((a)}(c) &  \phantom{(((a)}(d)
\end{tabular}
\caption{\label{Curves_sq_p=Infty}
Limiting curves for the square-lattice RSOS model with $p=\infty$ ($Q=4$)
and several widths:
$L=2$ (a), $L=3$ (b), and $L=4$ (c). 
Figure~(d) shows all these curves together: $L=2$ (black), $L=3$ (red),
$L=4$ (green).
The solid squares
$\blacksquare$ show the values where Baxter found the free energy.
In the regions displayed in light gray (resp.\  white)
the dominant eigenvalue comes from the sector $\chi_{1,3}$ 
(resp.\ $\chi_{1,1}$). In the regions displayed in a darker gray
the dominant eigenvalue comes from the sector $\chi_{1,5}$.
}
\end{figure}

\clearpage
%
%
\begin{figure}
\centering
\begin{tabular}{cc}
  \includegraphics[width=200pt]{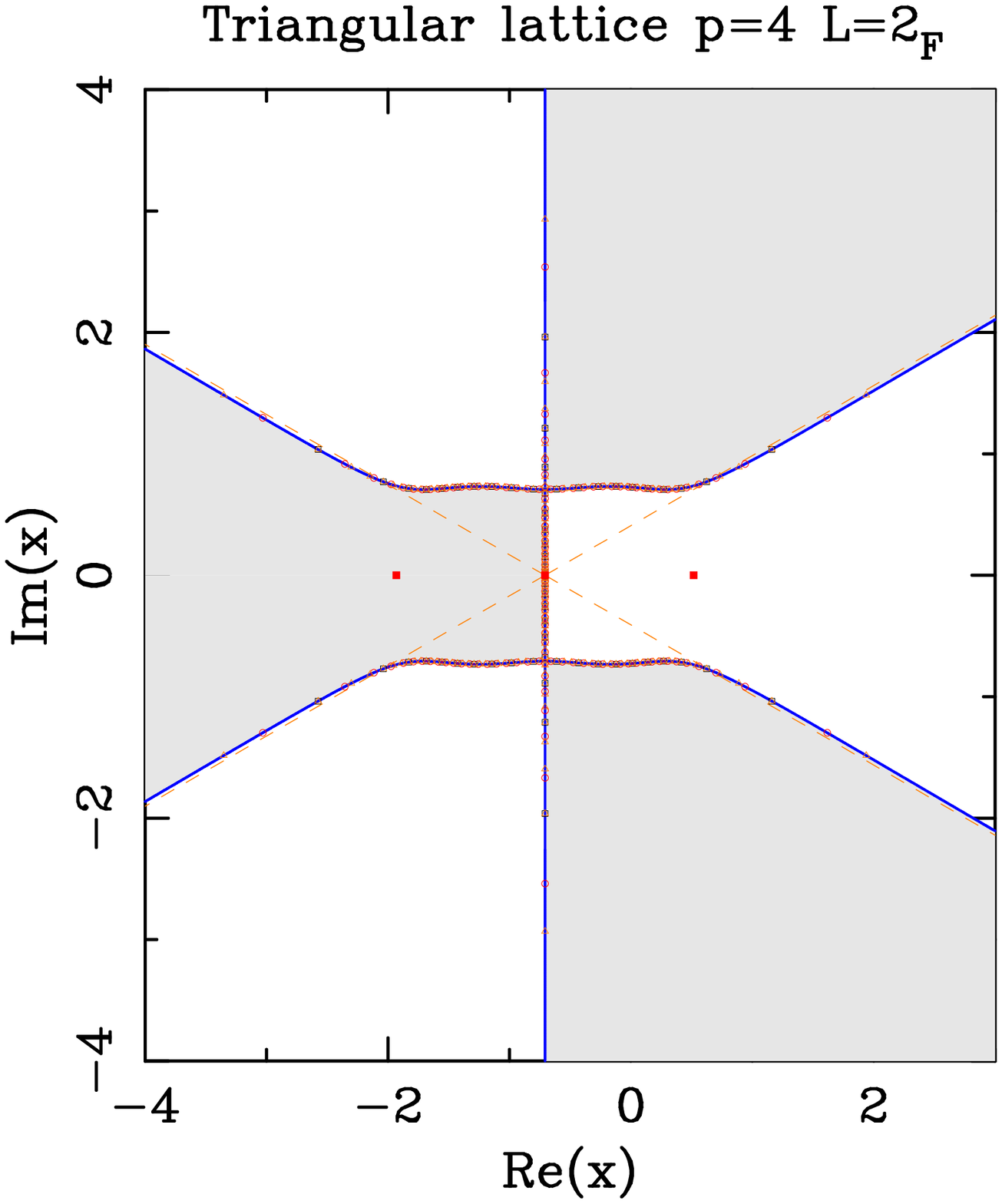} &
  \includegraphics[width=200pt]{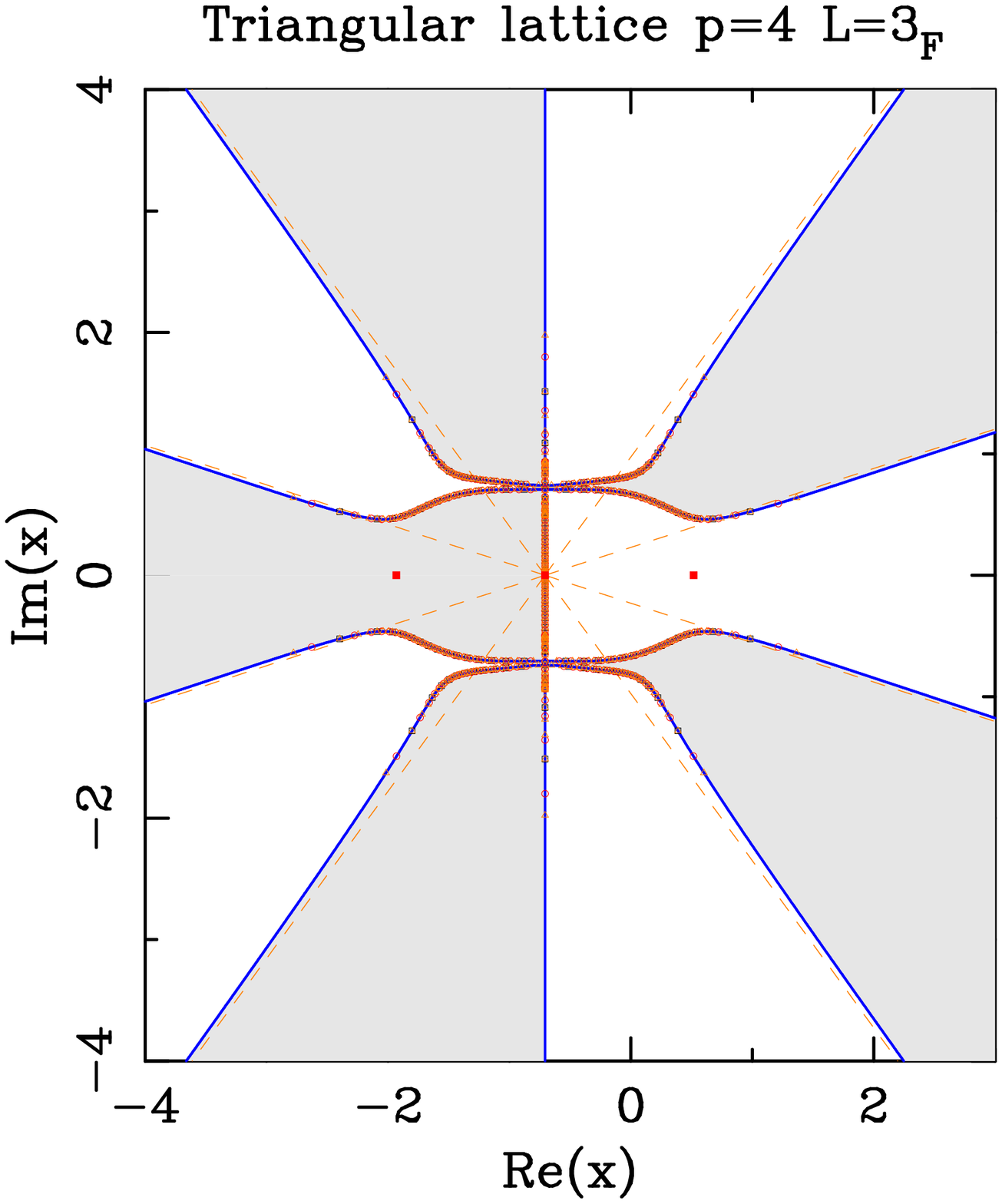} \\[2mm]
  \phantom{(((a)}(a) & \phantom{(((a)}(b)\\[5mm]
  \includegraphics[width=200pt]{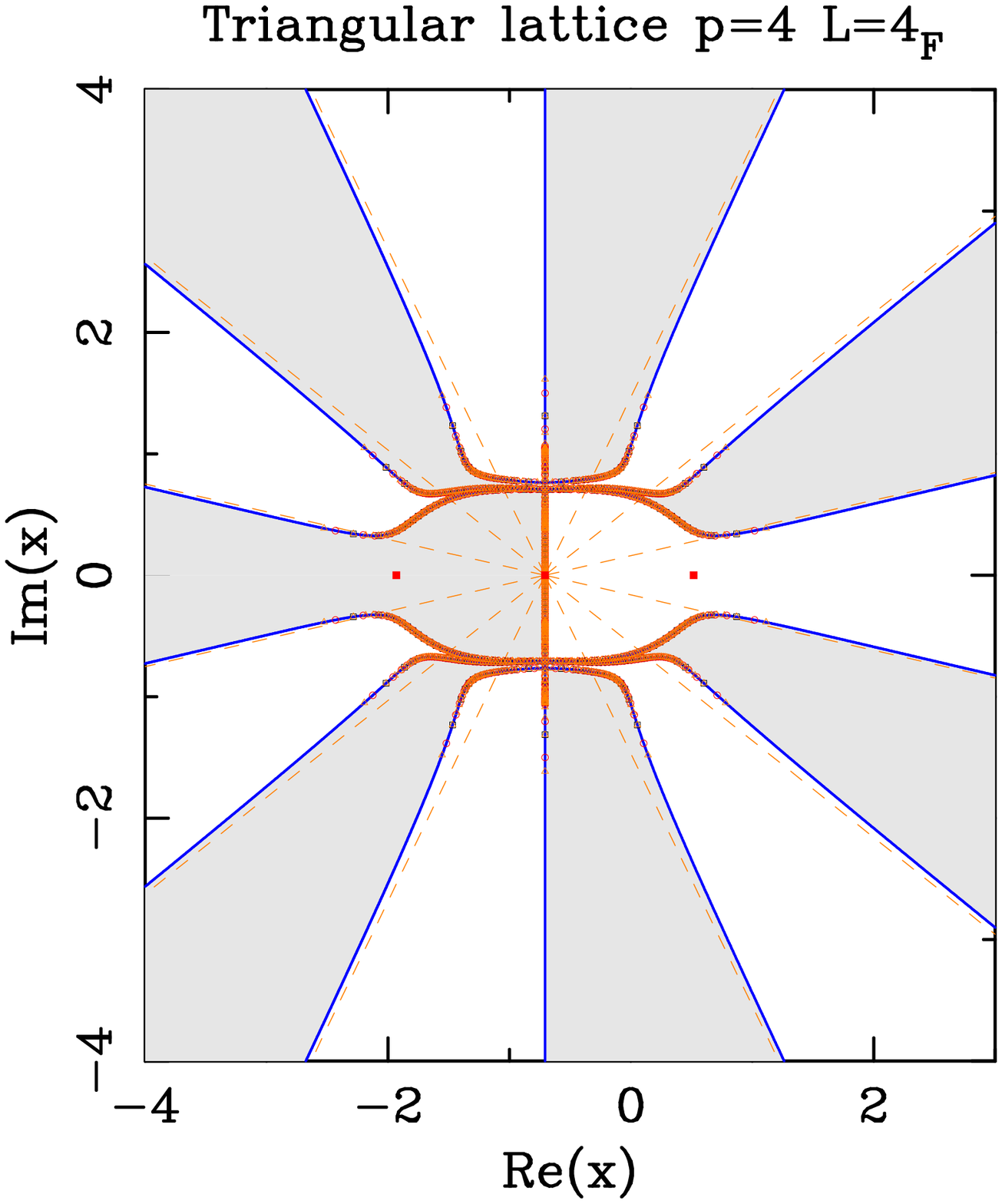} &
  \includegraphics[width=200pt]{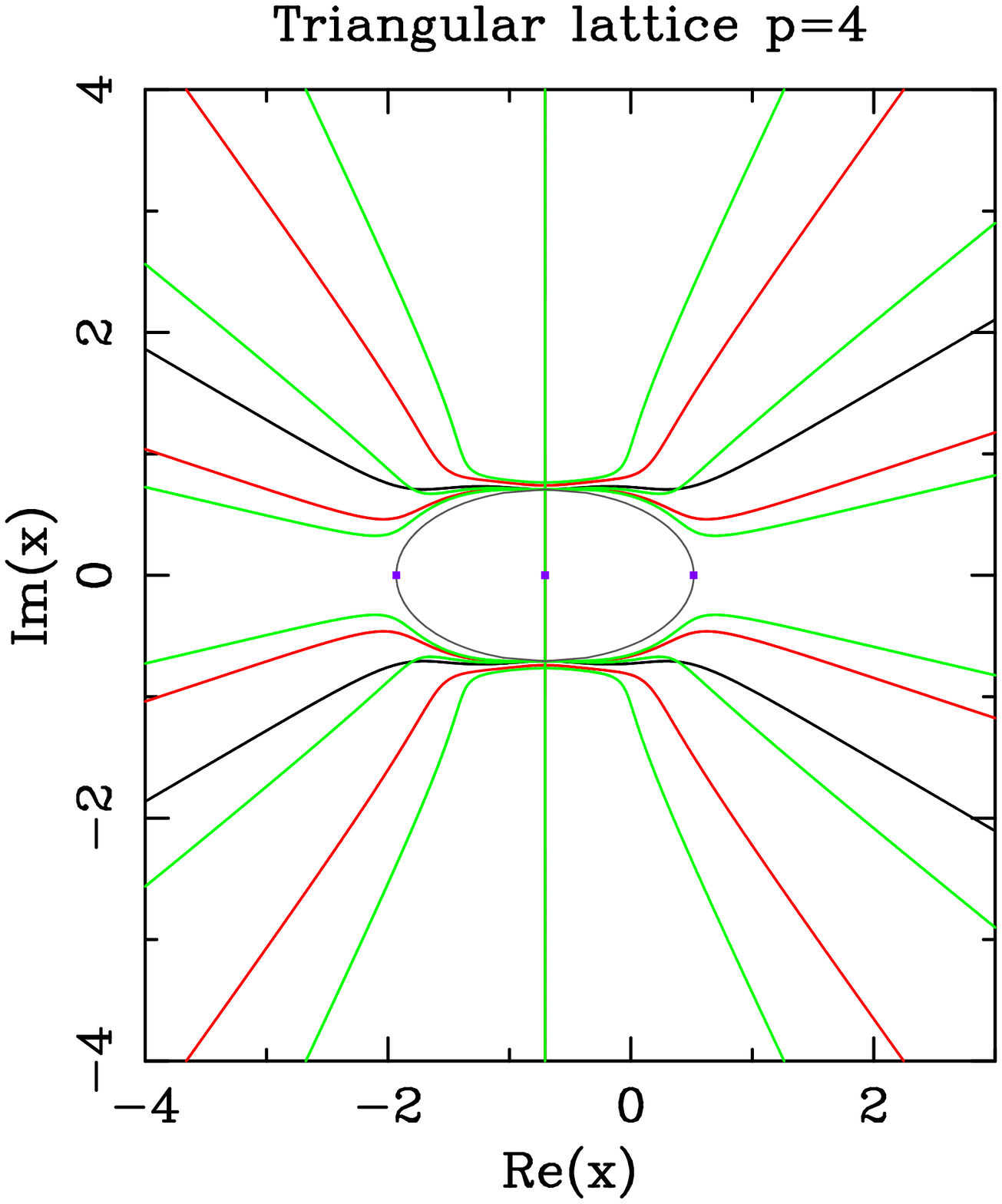} \\[2mm]
  \phantom{(((a)}(c) &  \phantom{(((a)}(d)
\end{tabular}
\caption{\label{Curves_tri_p=4}
Limiting curves for the triangular-lattice RSOS model with $p=4$ and 
several widths: $L=2$ (a), $L=3$ (b), and $L=4$ (c). For each width $L$, we
also show the partition-function zeros for finite strips of dimensions
$L_\text{F}\times (10L)_\text{P}$ (black $\square$),
$L_\text{F}\times (20L)_\text{P}$ (red $\circ$), and
$L_\text{F}\times (30L)_\text{P}$ (brown $\triangle$).
Figure~(d) shows all these limiting curves together: $L=2$ (black), $L=3$ (red),
$L=4$ (green).
The solid squares $\blacksquare$ show the values where Baxter found 
the free energy.
The symbol $\times$ in (a) marks the position of the found isolated limiting
point.
In the regions displayed in gray (resp.\  white)
the dominant eigenvalue comes from the sector $\chi_{1,3}$
(resp.\ $\chi_{1,1}$).
The gray ellipse corresponds to $(\Re x +1/\sqrt{2})^2 + 3 (\Im x)^2 = 3/2$.
This curve goes through the points $x=-e^{\pm i\,\pi/4}$. 
}
\end{figure}

%
%
\begin{figure}
\centering
\begin{tabular}{cc}
  \includegraphics[width=200pt]{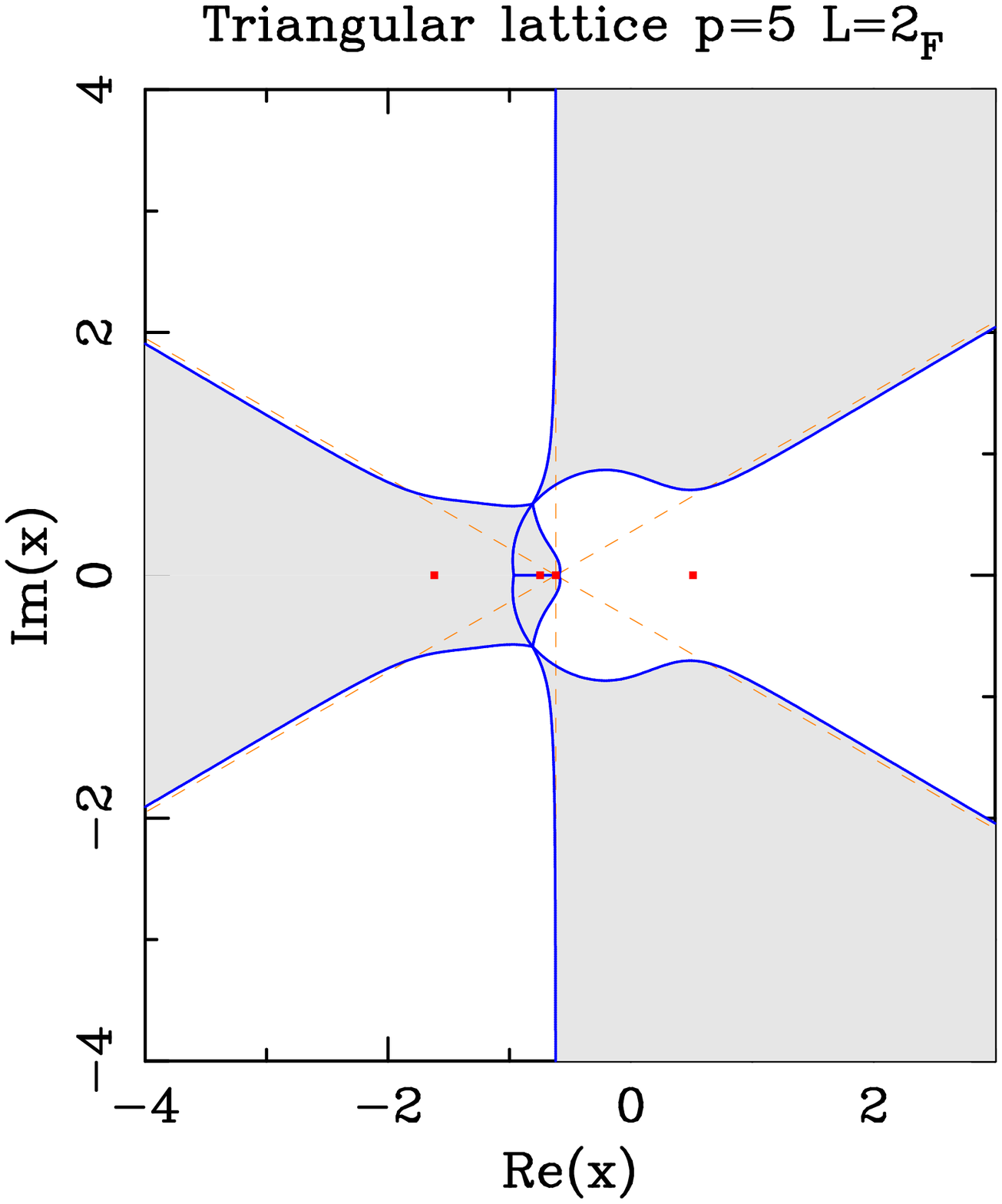} &
  \includegraphics[width=200pt]{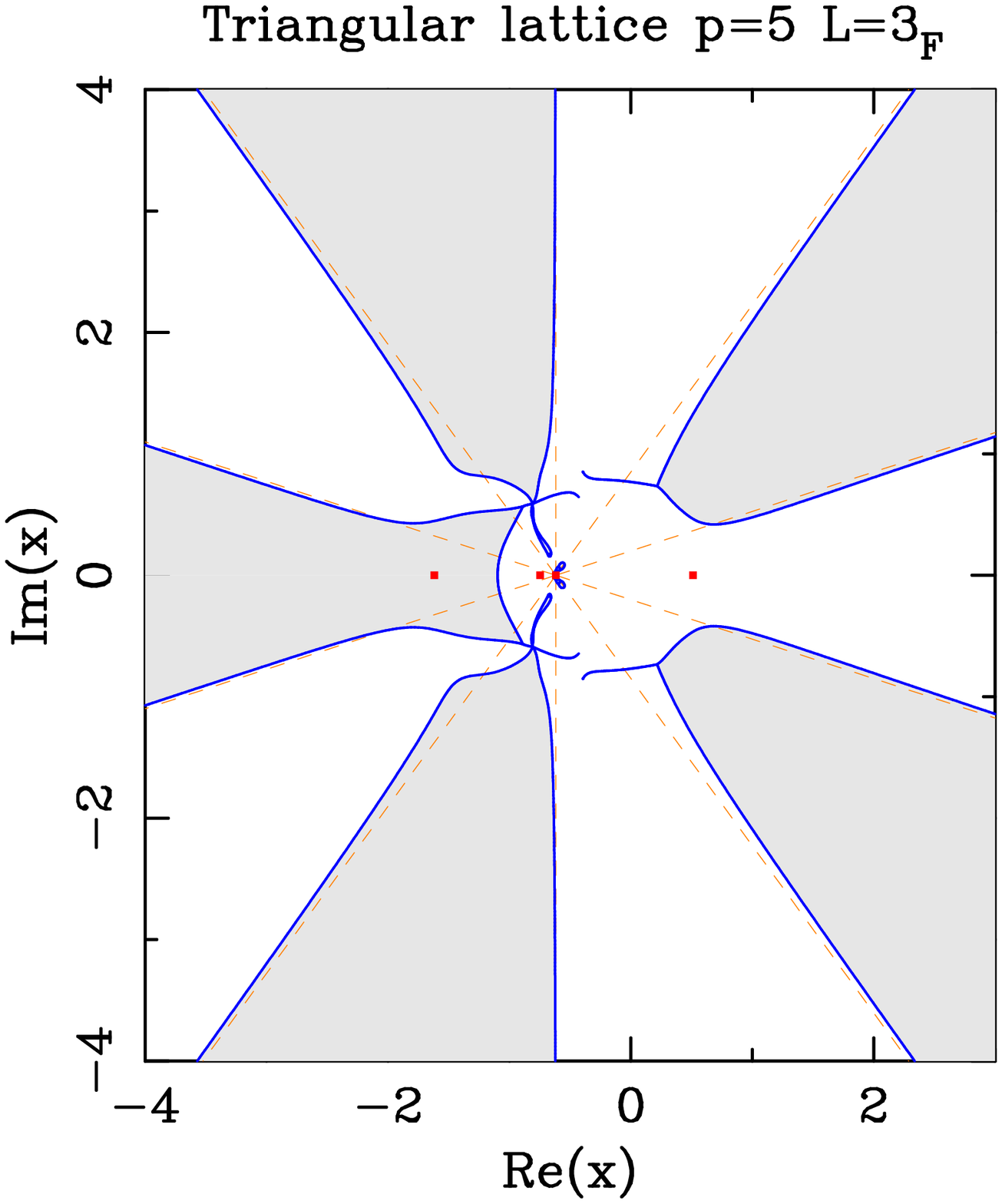} \\[2mm]
  \phantom{(((a)}(a) & \phantom{(((a)}(b)\\[5mm]
  \includegraphics[width=200pt]{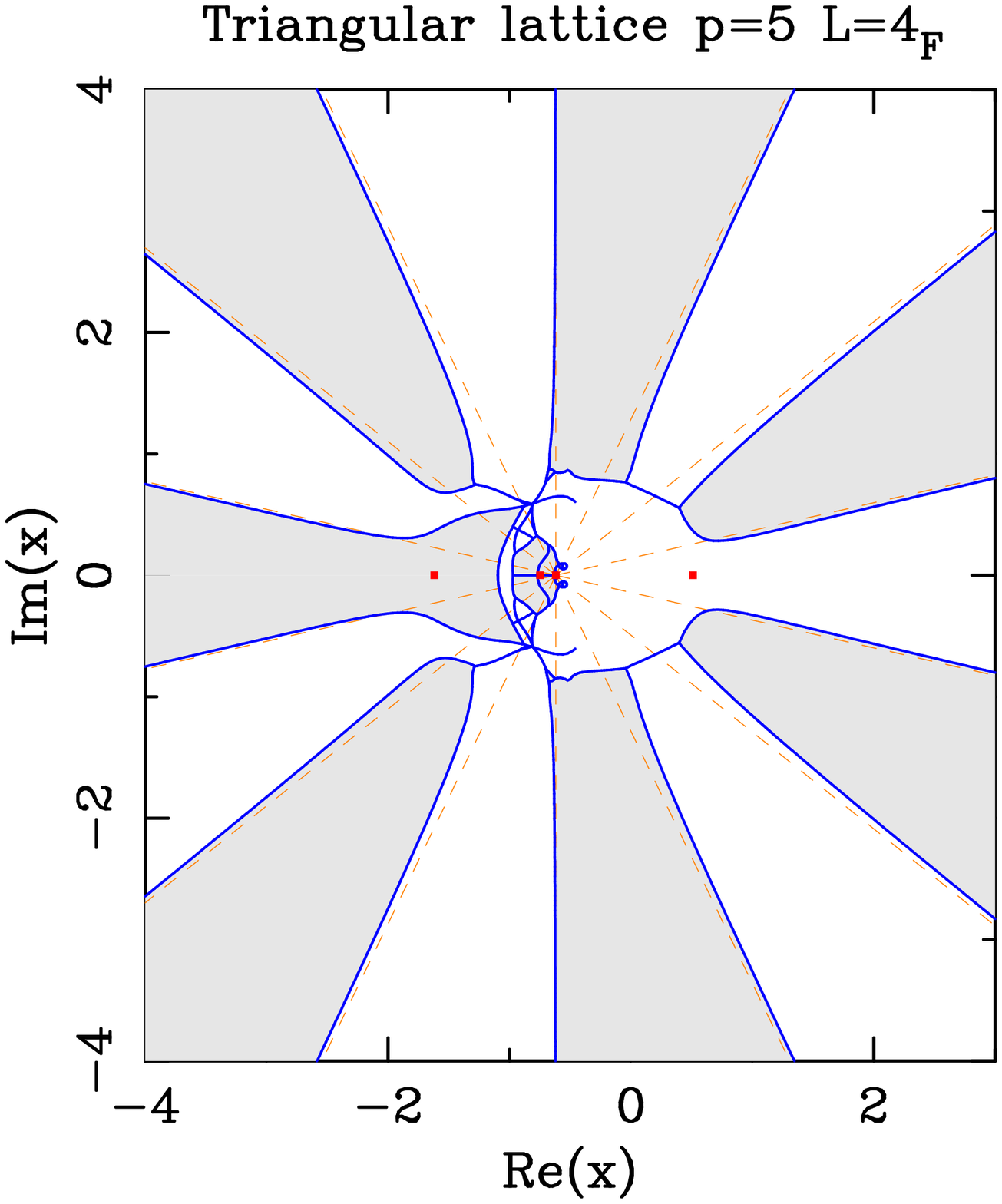} &
  \includegraphics[width=200pt]{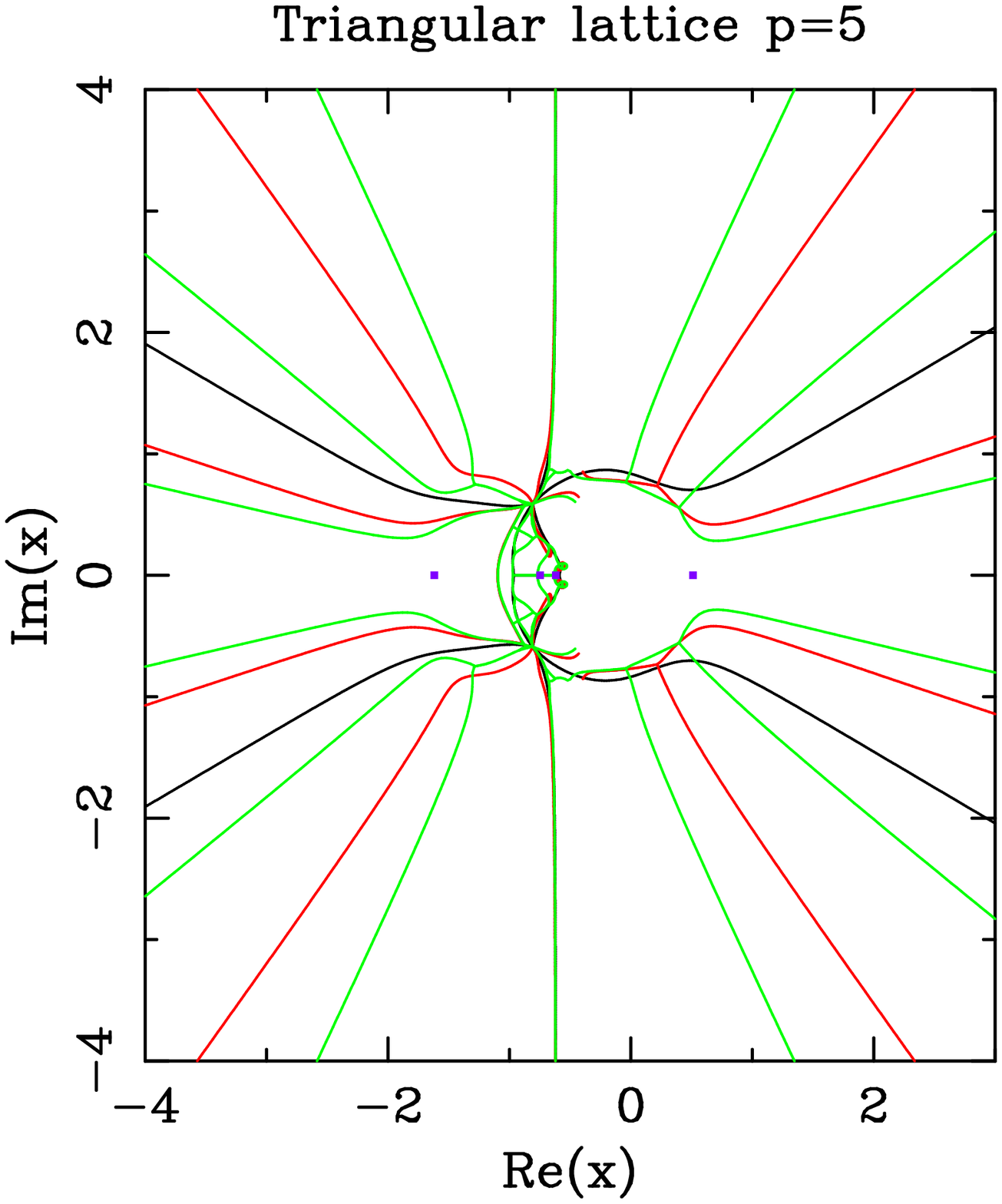} \\[2mm]
  \phantom{(((a)}(c) &  \phantom{(((a)}(d)
\end{tabular}
\caption{\label{Curves_tri_p=5}
Limiting curves for the RSOS model with $p=5$ and several widths:
$L=2$ (a), $L=3$ (b), and $L=4$ (c). 
Figure~(d) shows all these curves together: $L=2$ (black), $L=3$ (red),
$L=4$ (green).
The solid squares
$\blacksquare$ show the values where Baxter found the free energy.
In the regions displayed in light gray (resp.\  white)
the dominant eigenvalue comes from the sector $\chi_{1,3}$ 
(resp.\ $\chi_{1,1}$). 
}
\end{figure}

\clearpage
%
%
\begin{figure}
\centering
\begin{tabular}{cc}
  \includegraphics[width=200pt]{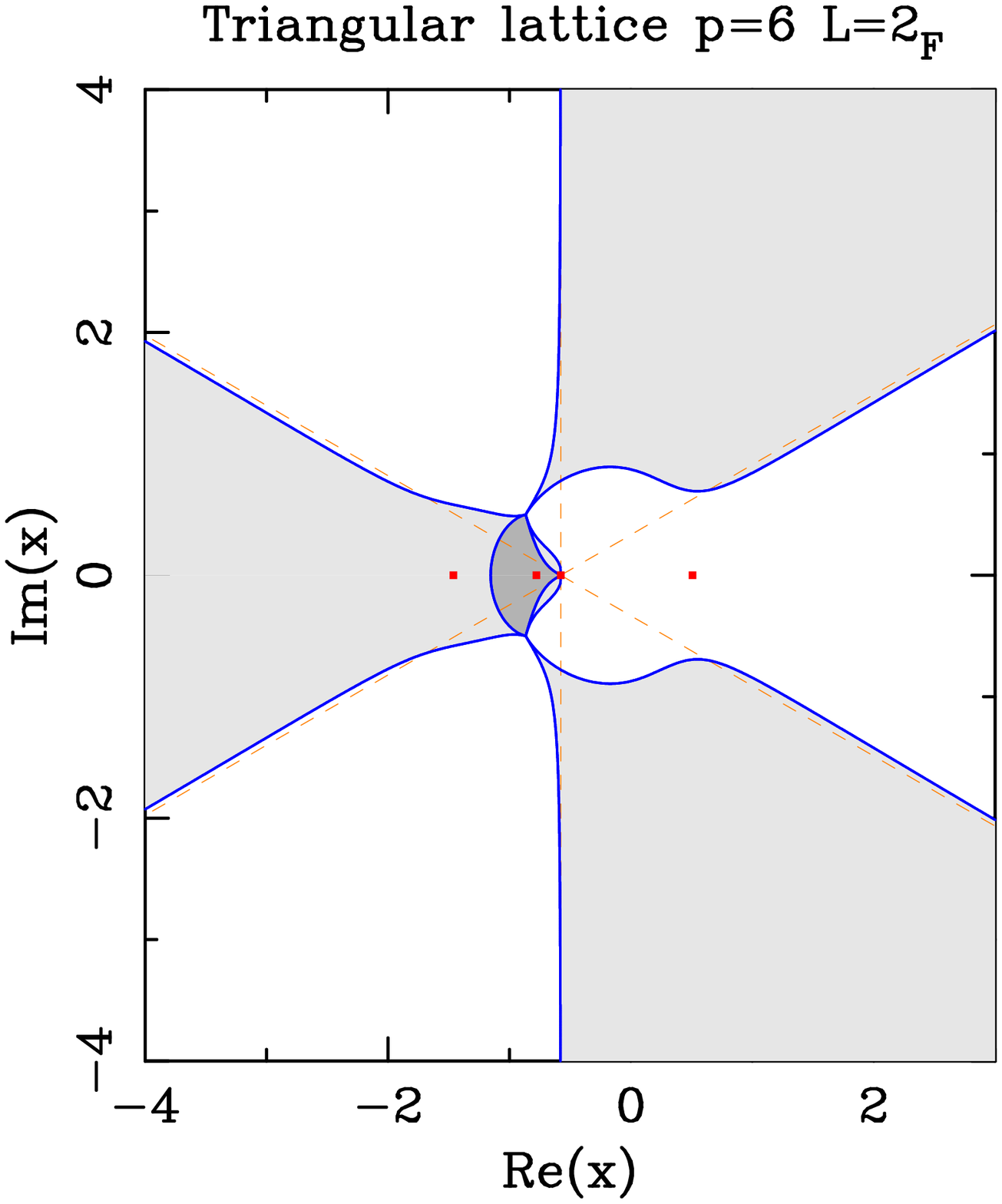} &
  \includegraphics[width=200pt]{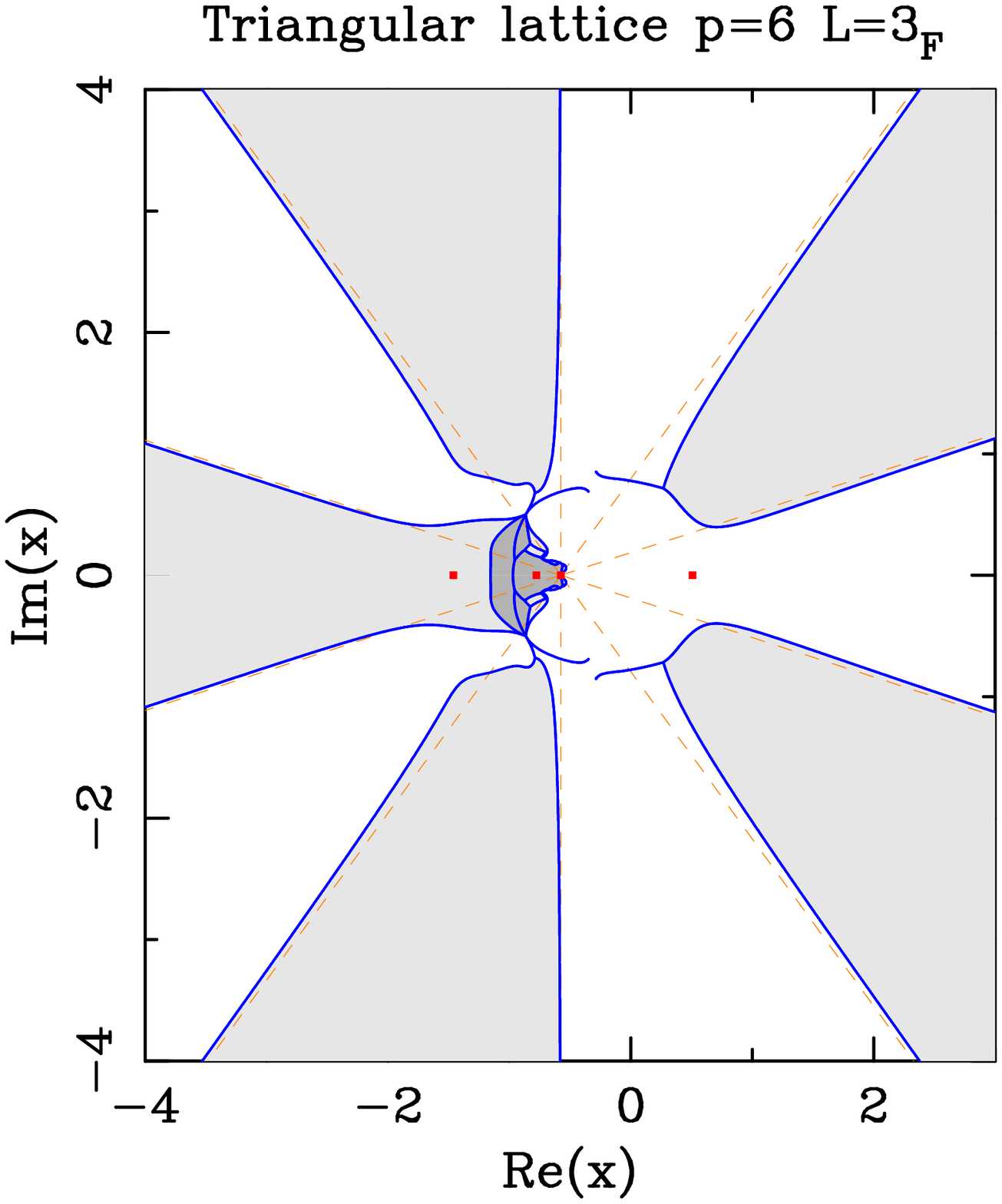} \\[2mm]
  \phantom{(((a)}(a) & \phantom{(((a)}(b)\\[5mm]
  \includegraphics[width=200pt]{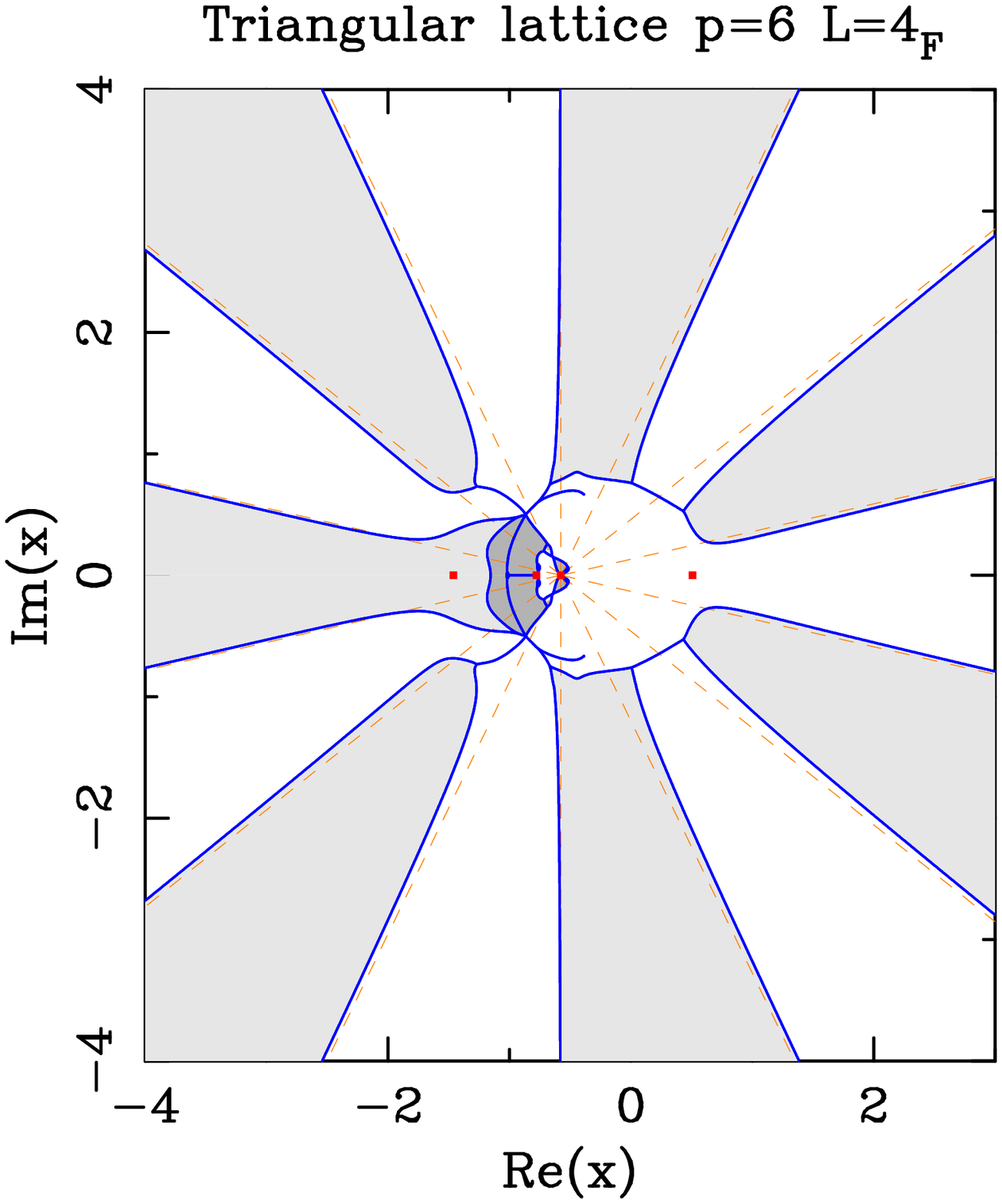} &
  \includegraphics[width=200pt]{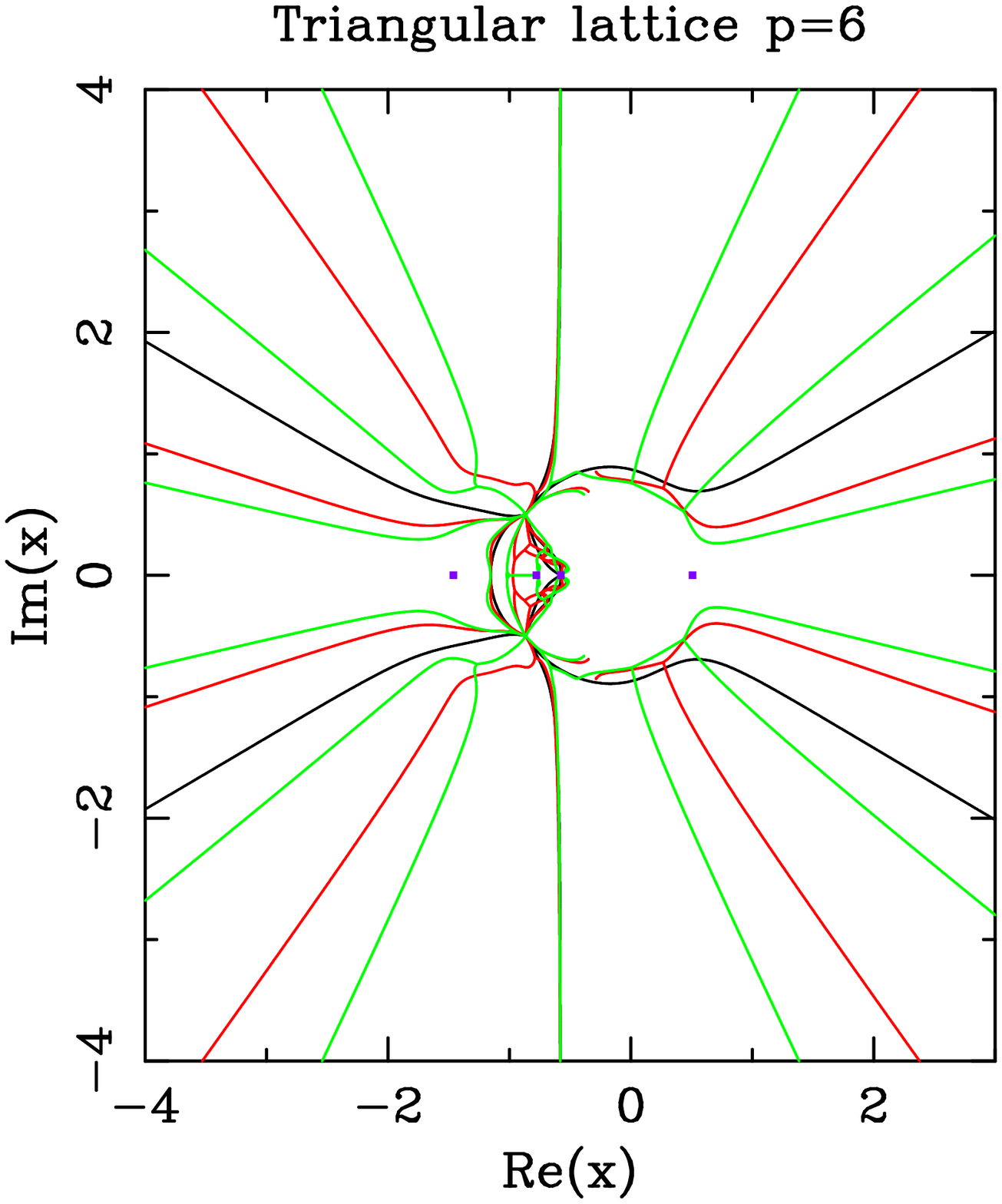} 
\\[2mm]
  \phantom{(((a)}(c) &  \phantom{(((a)}(d)
\end{tabular}
\caption{\label{Curves_tri_p=6}
Limiting curves for the triangular-lattice RSOS model with $p=6$ and several widths:
$L=2$ (a), $L=3$ (b), and $L=4$ (c). 
Figure~(d) shows all these curves together: $L=2$ (black), $L=3$ (red),
$L=4$ (green).
The solid squares
$\blacksquare$ show the values where Baxter found the free energy.
In the regions displayed in light gray (resp.\  white)
the dominant eigenvalue comes from the sector $\chi_{1,3}$ 
(resp.\ $\chi_{1,1}$). In the regions displayed in a darker gray
the dominant eigenvalue comes from the sector $\chi_{1,5}$. 
}
\end{figure}

\clearpage
%
%
\begin{figure}
\centering
\begin{tabular}{cc}
  \includegraphics[width=200pt]{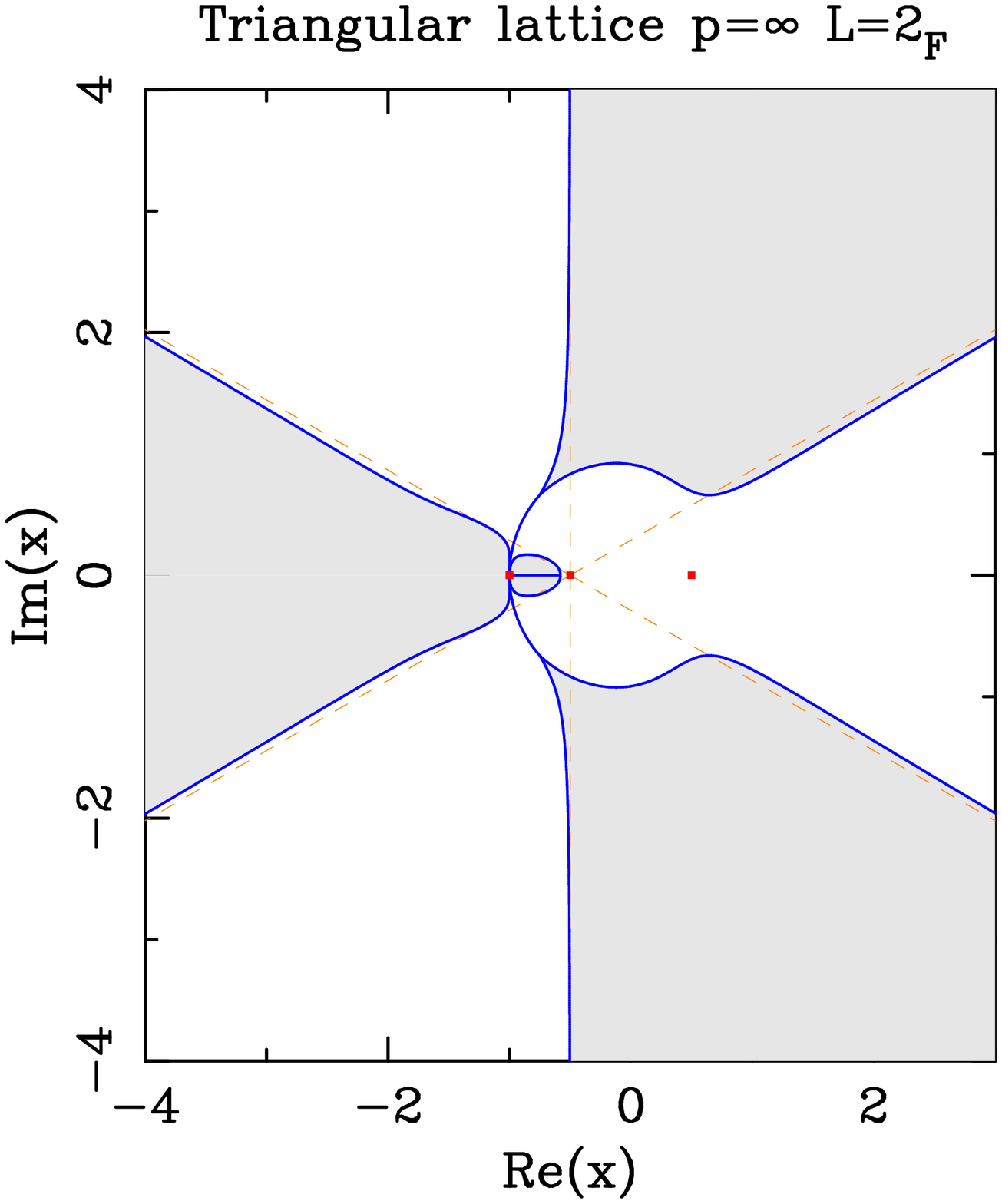} &
  \includegraphics[width=200pt]{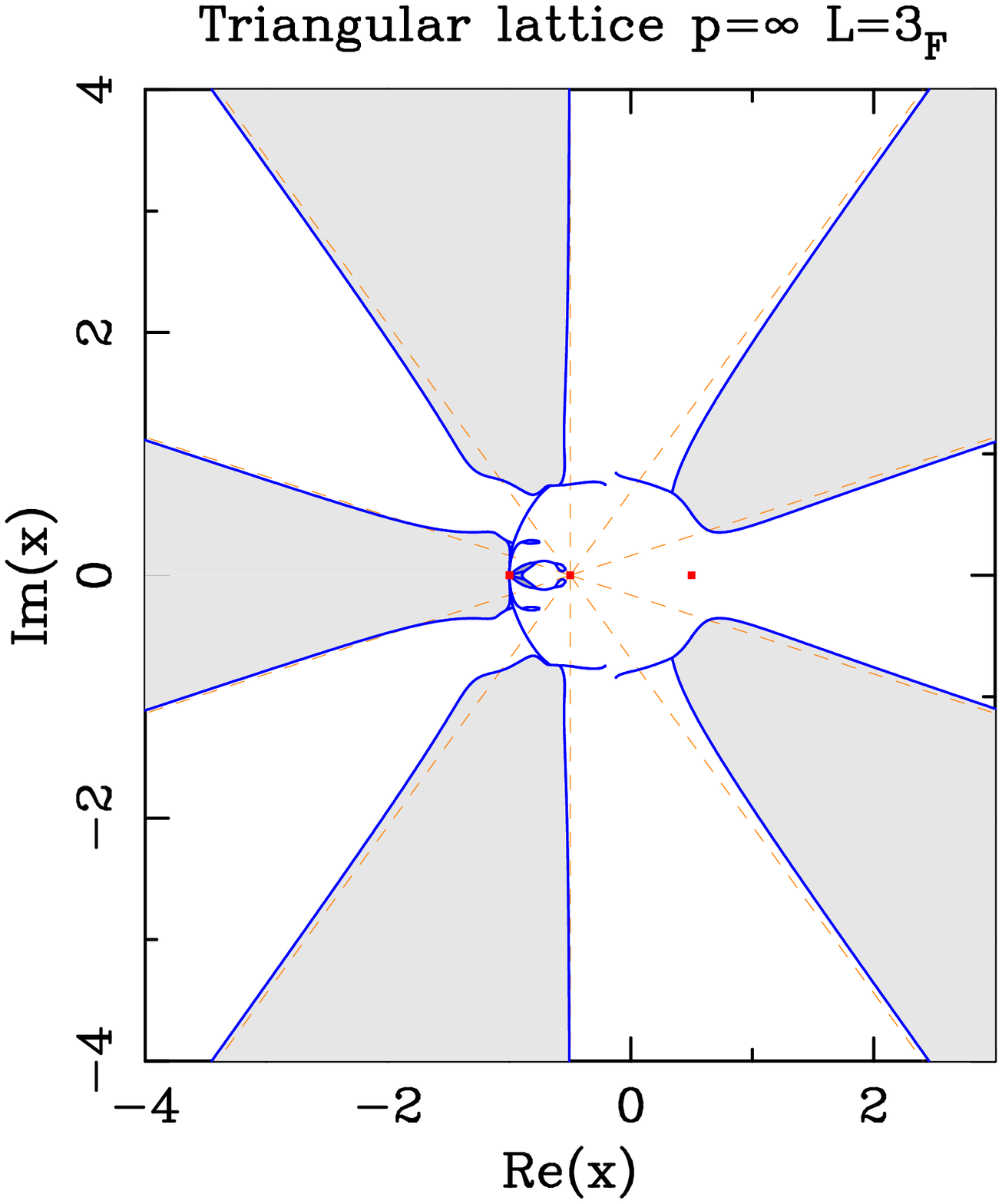} \\[2mm]
  \phantom{(((a)}(a) & \phantom{(((a)}(b)\\[5mm]
  \includegraphics[width=200pt]{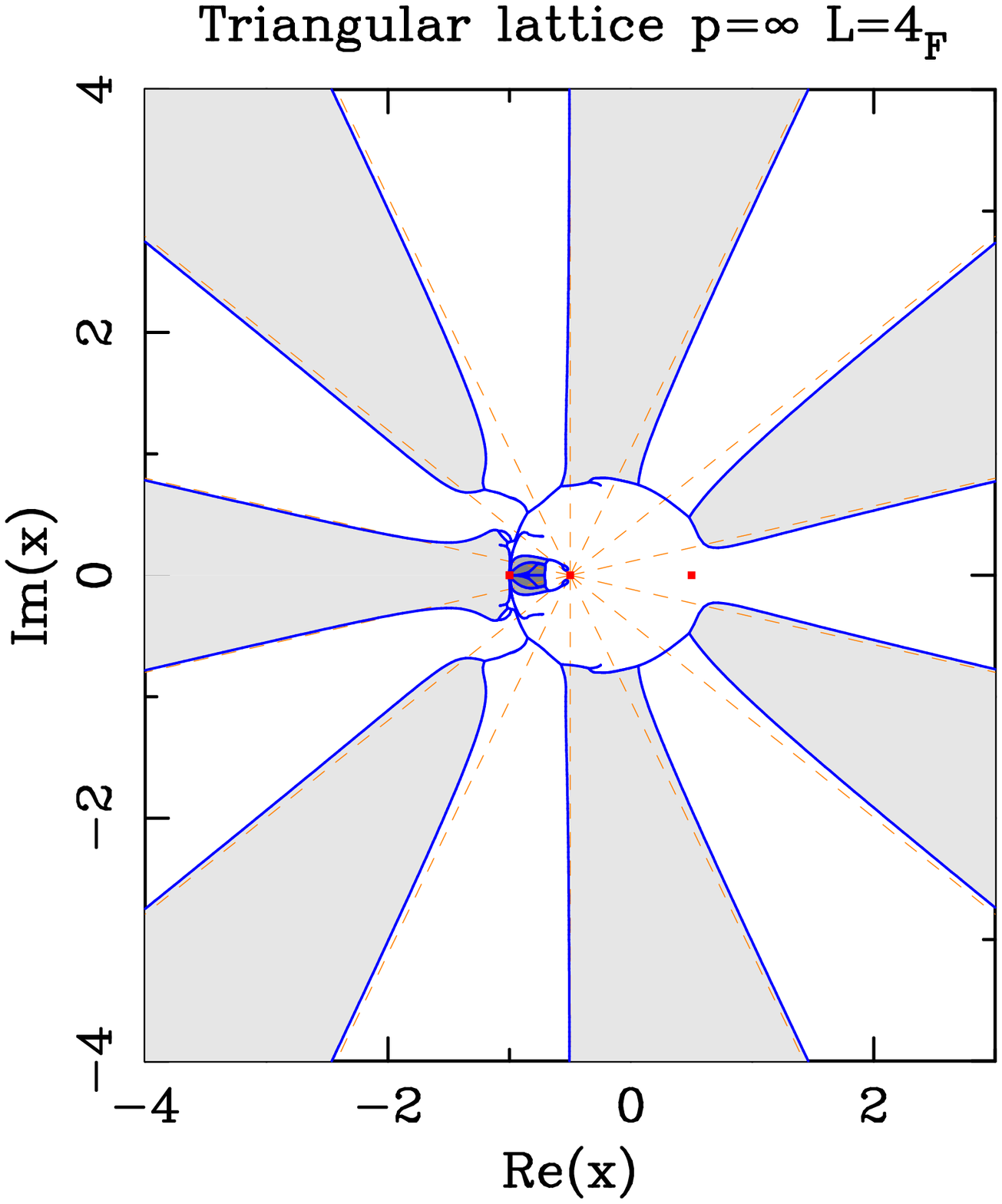} &
  \includegraphics[width=200pt]{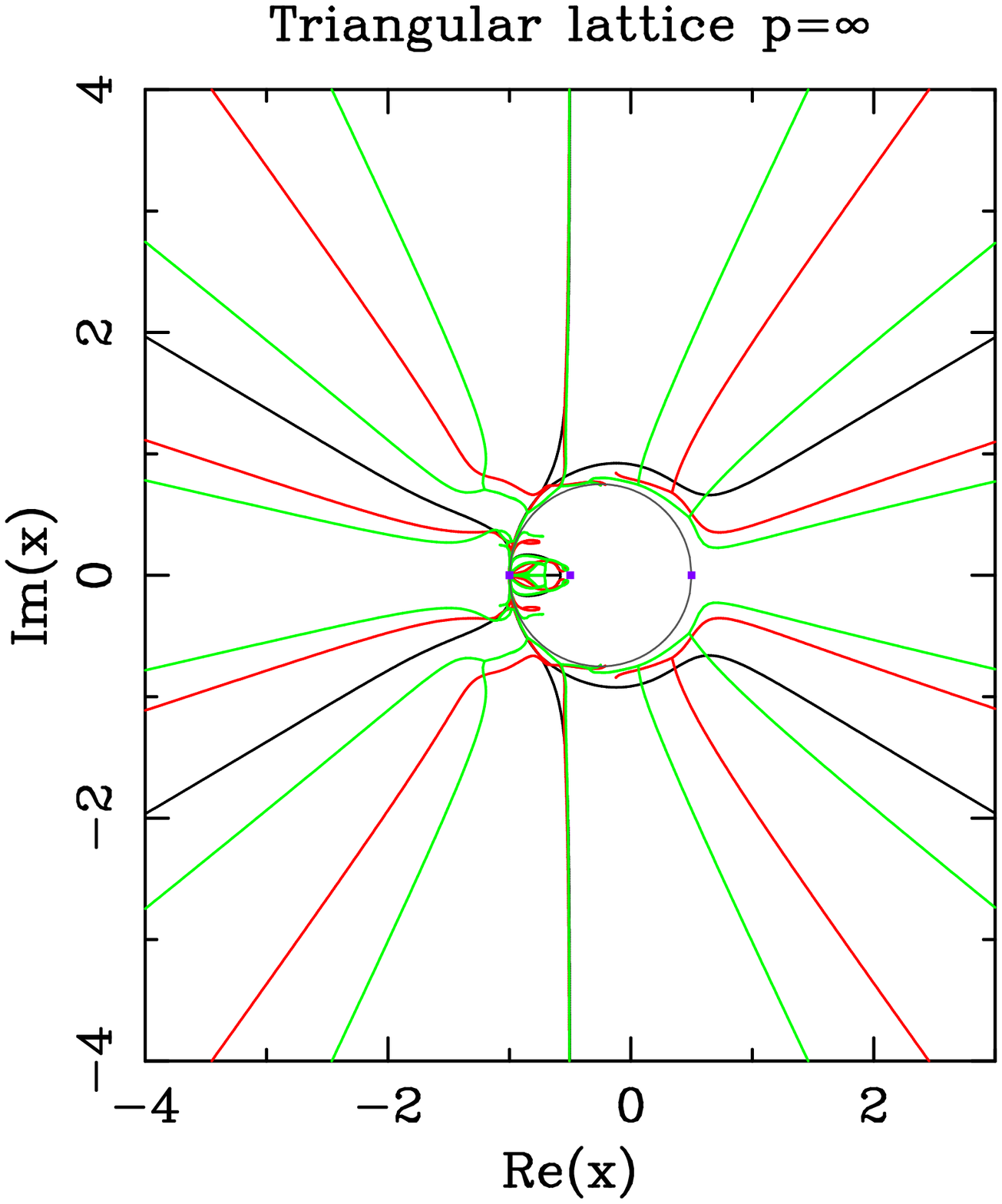} 
\\[2mm]
  \phantom{(((a)}(c) &  \phantom{(((a)}(d)
\end{tabular}
\caption{\label{Curves_tri_p=Infty}
Limiting curves for the triangular-lattice RSOS model with $p=\infty$ ($Q=4$)
and several widths:
$L=2$ (a), $L=3$ (b), and $L=4$ (c). 
Figure~(d) shows all these curves together: $L=2$ (black), $L=3$ (red),
$L=4$ (green).
The solid squares
$\blacksquare$ show the values where Baxter found the free energy.
In the regions displayed in light gray (resp.\  white)
the dominant eigenvalue comes from the sector $\chi_{1,3}$ 
(resp.\ $\chi_{1,1}$). In regions displayed in a darker gray
the dominant eigenvalue comes from the sector $\chi_{1,5}$. In (c),
an even darker gray marks the regions with a dominant eigenvalue coming 
from the sector $\chi_{1,7}$. The gray circle corresponds to
\protect\reff{circles_tri_x_p=infty}. 
}
\end{figure}

\clearpage
%
%
\begin{figure}
\centering
\begin{tabular}{cc}
  \includegraphics[width=200pt]{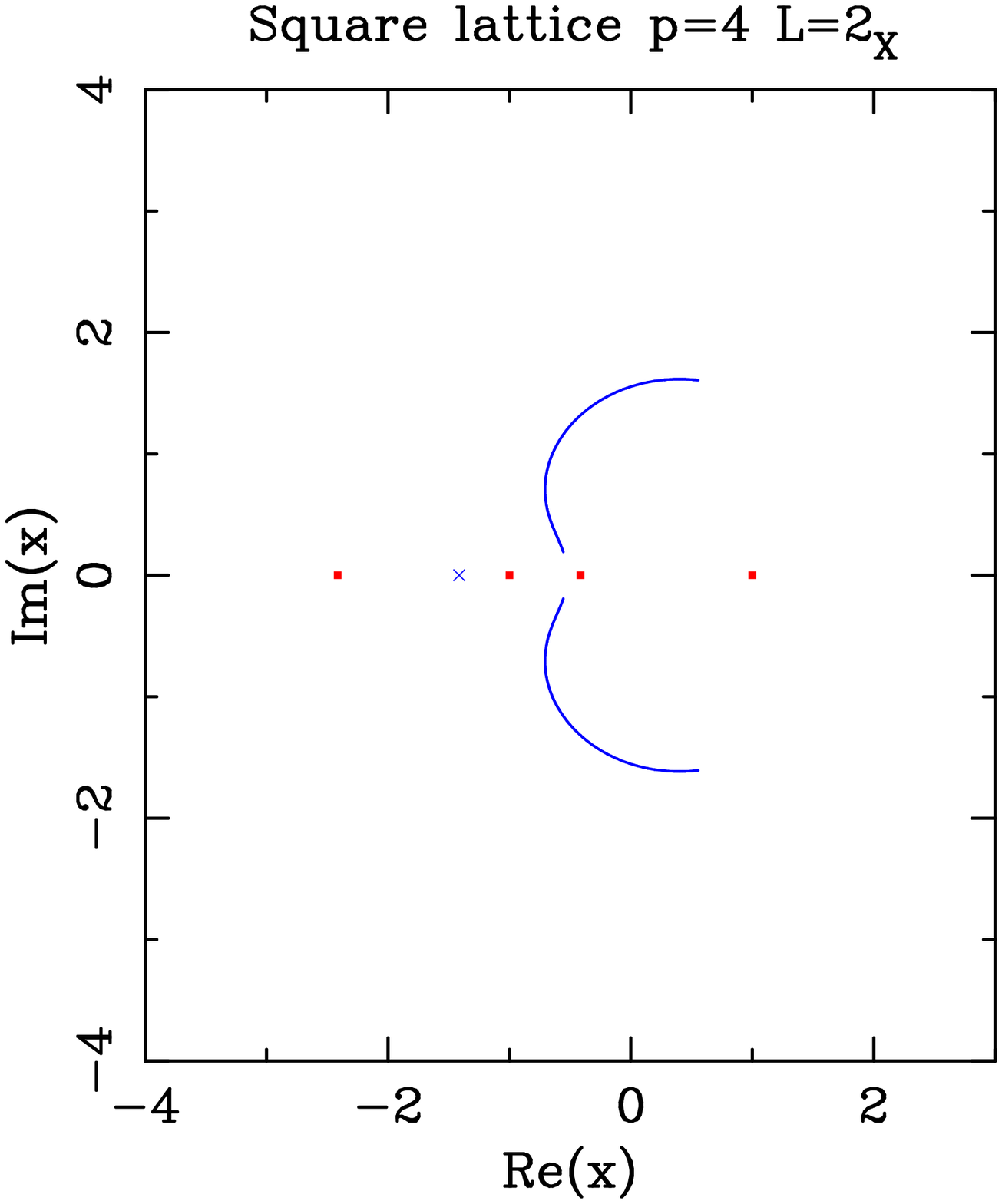} &
  \includegraphics[width=200pt]{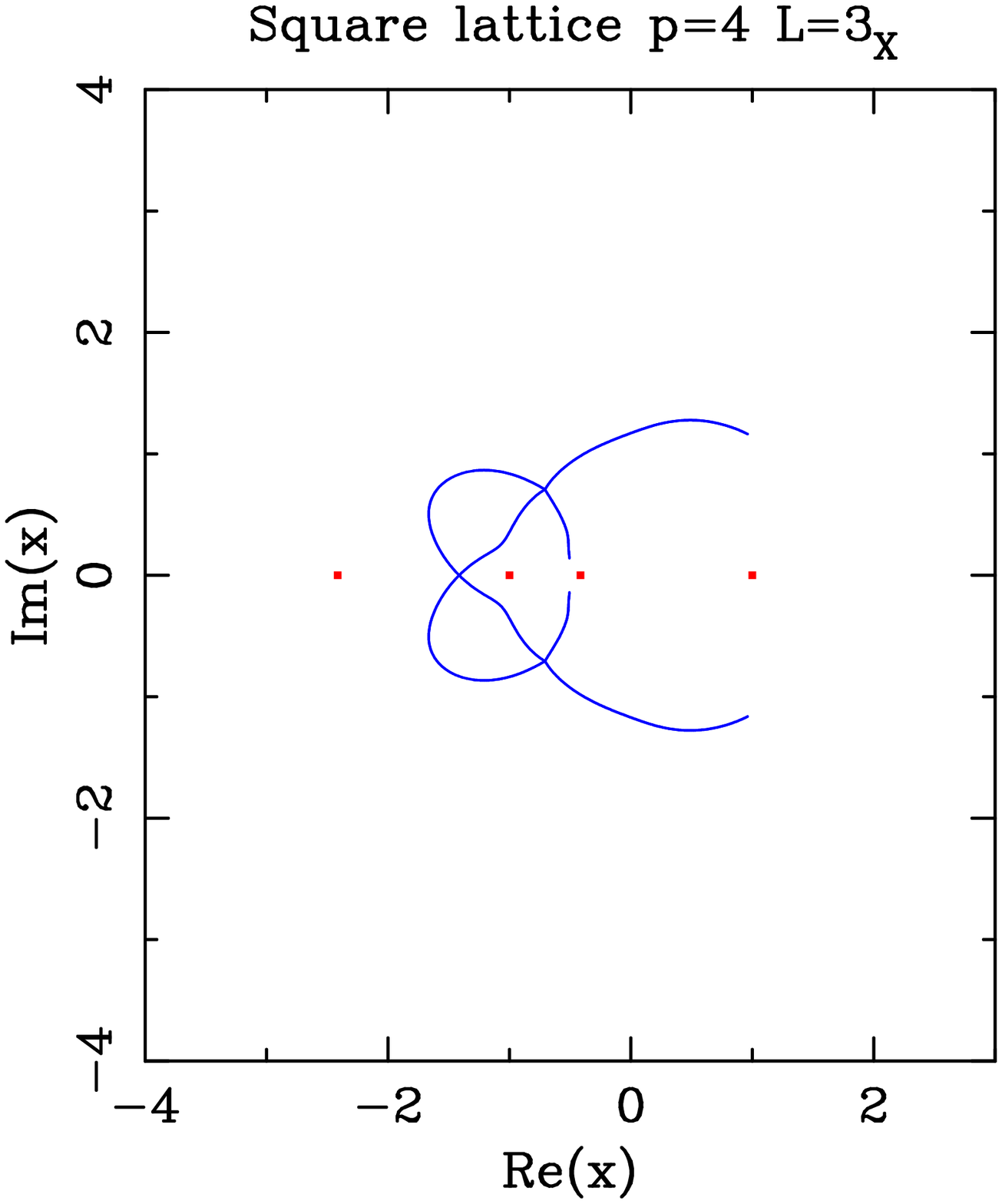} \\[2mm]
  \phantom{(((a)}(a) & \phantom{(((a)}(b)\\[5mm]
  \includegraphics[width=200pt]{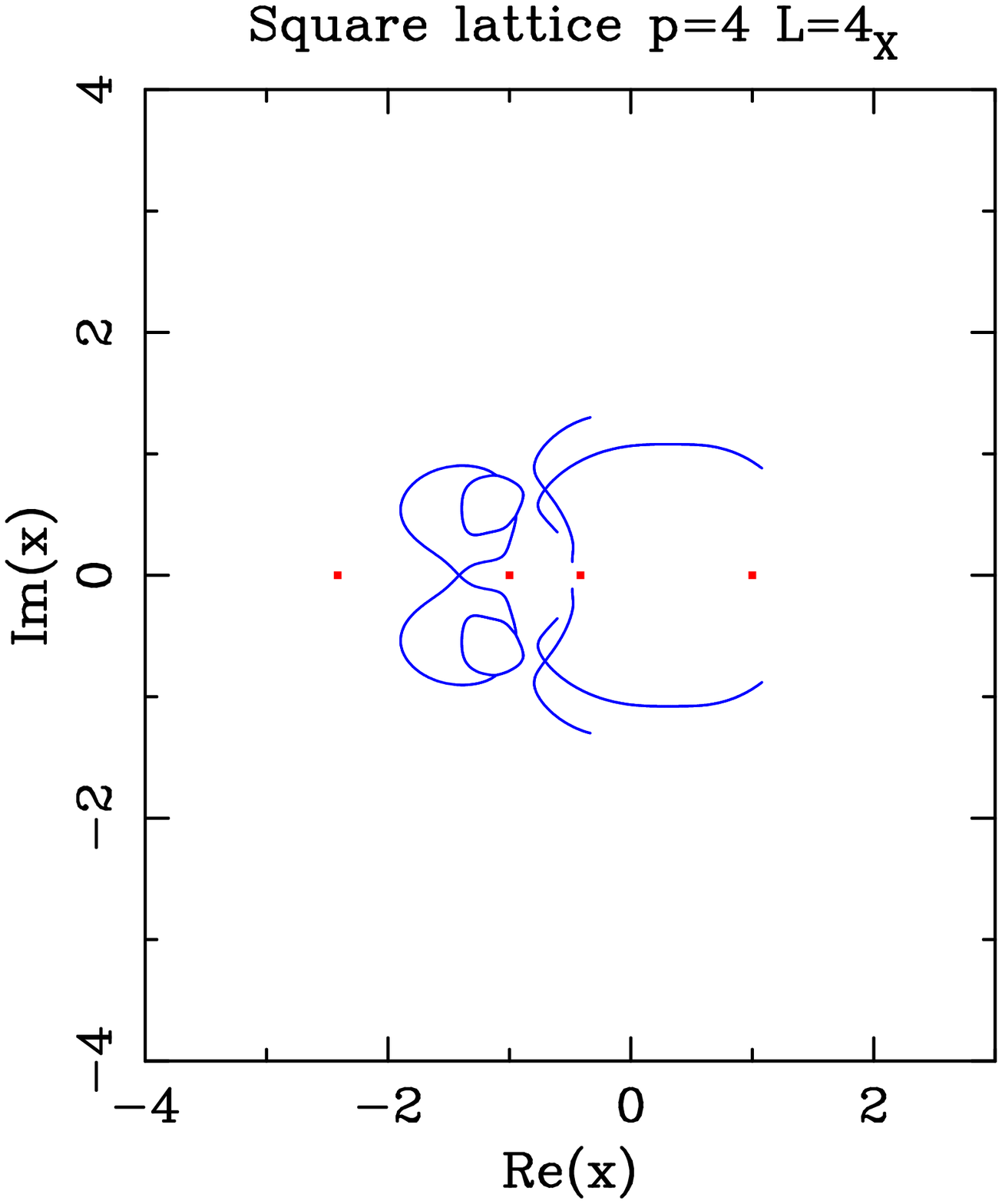} & 
  \includegraphics[width=200pt]{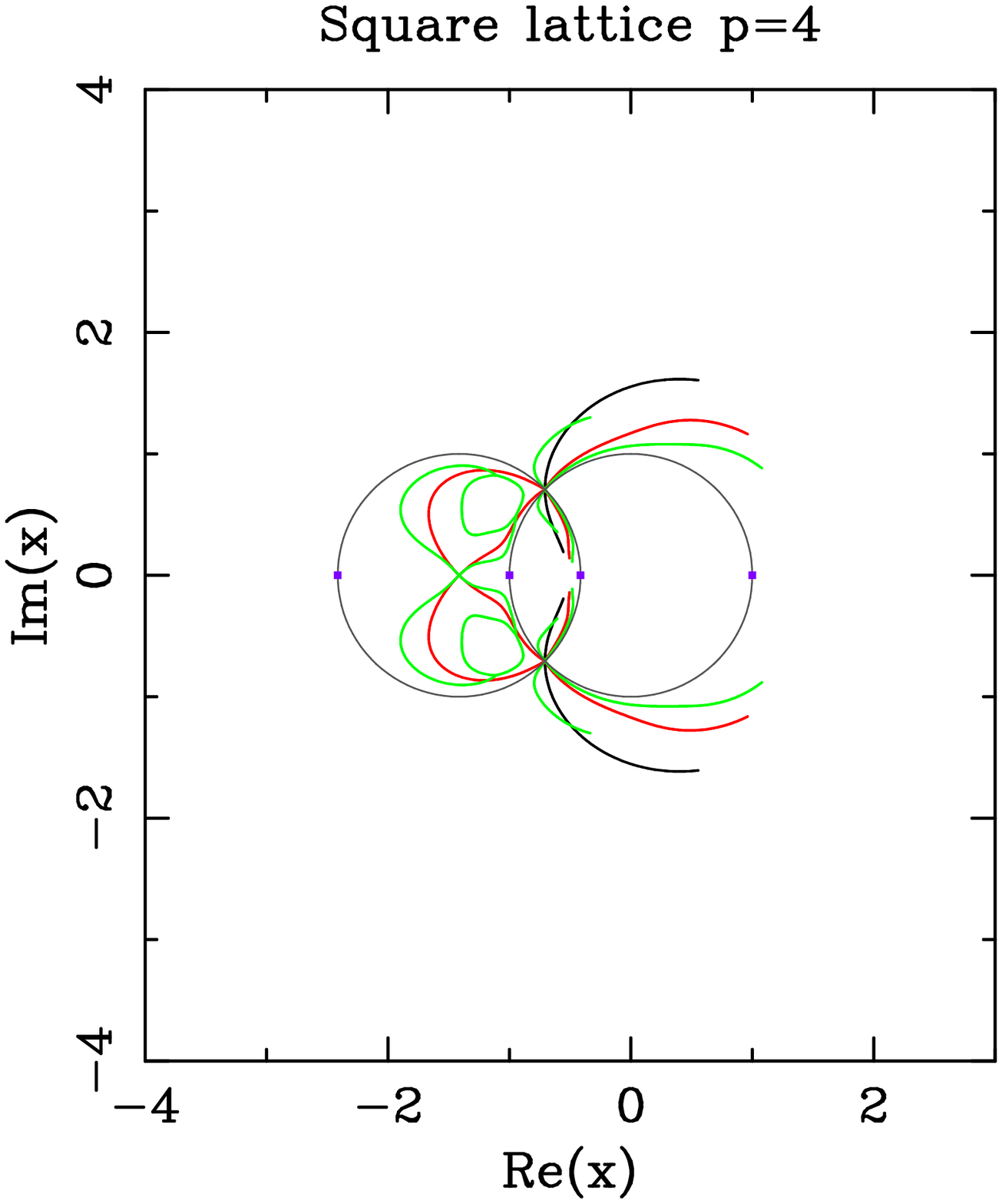} \\[2mm]
  \phantom{(((a)}(c) &  \phantom{(((a)}(d)
\end{tabular}
\caption{\label{Curves_sq_p=4_1}
Limiting curves for the square-lattice RSOS model with $p=4$ and several widths:
$L=2$ (a), $L=3$ (b), and $L=4$ (c) when only the sector $\chi_{1,1}$ is
taken into account. 
Figure~(d) shows all these curves together: $L=2$ (black), $L=3$ (red),
$L=4$ (green).
The solid squares
$\blacksquare$ show the values where Baxter found the free energy.
The dark gray circles correspond to \protect\reff{circles_sq_x}
}
\end{figure}

\clearpage
%
%
\begin{figure}
\centering
\begin{tabular}{cc}
  \includegraphics[width=200pt]{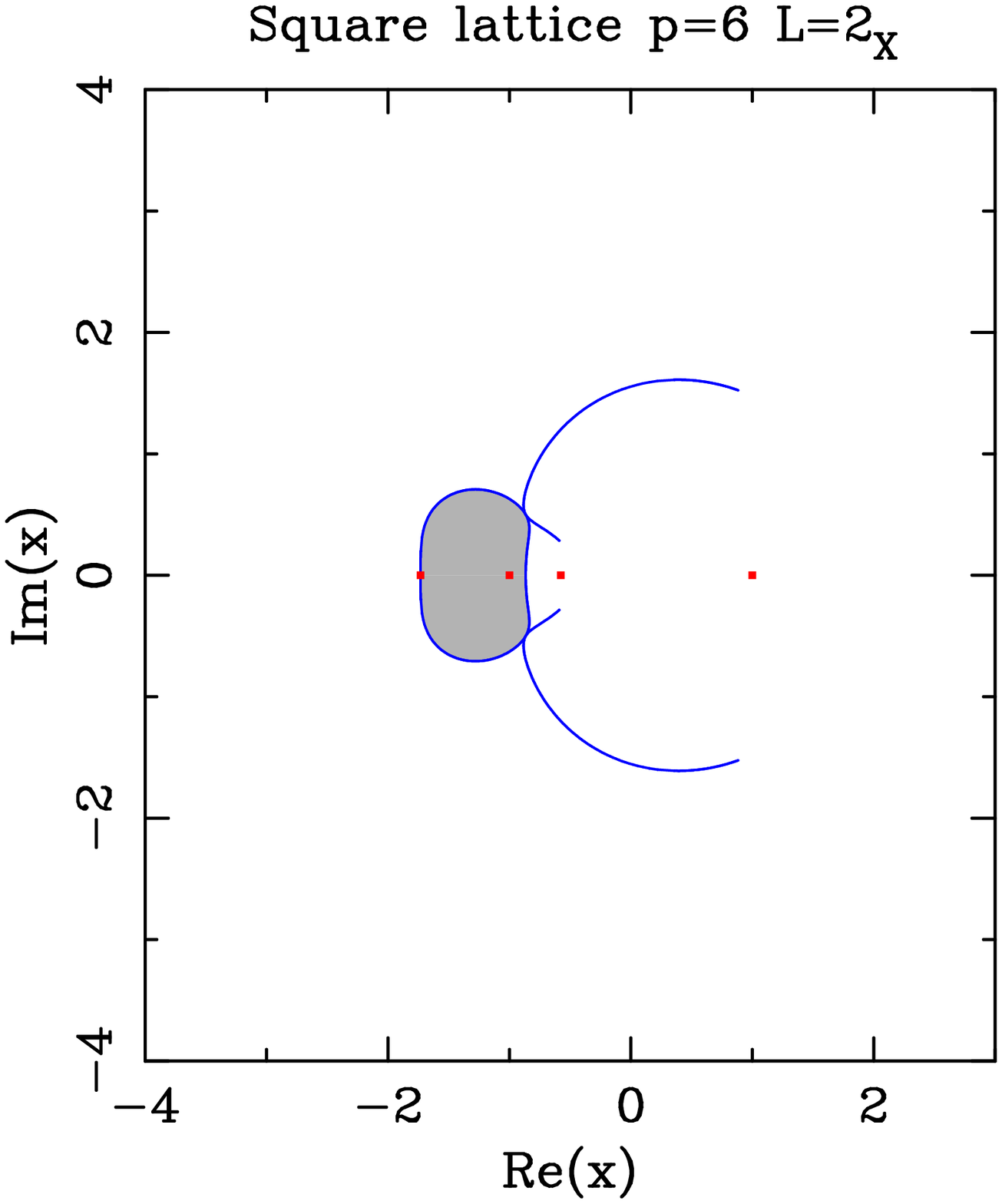} &
  \includegraphics[width=200pt]{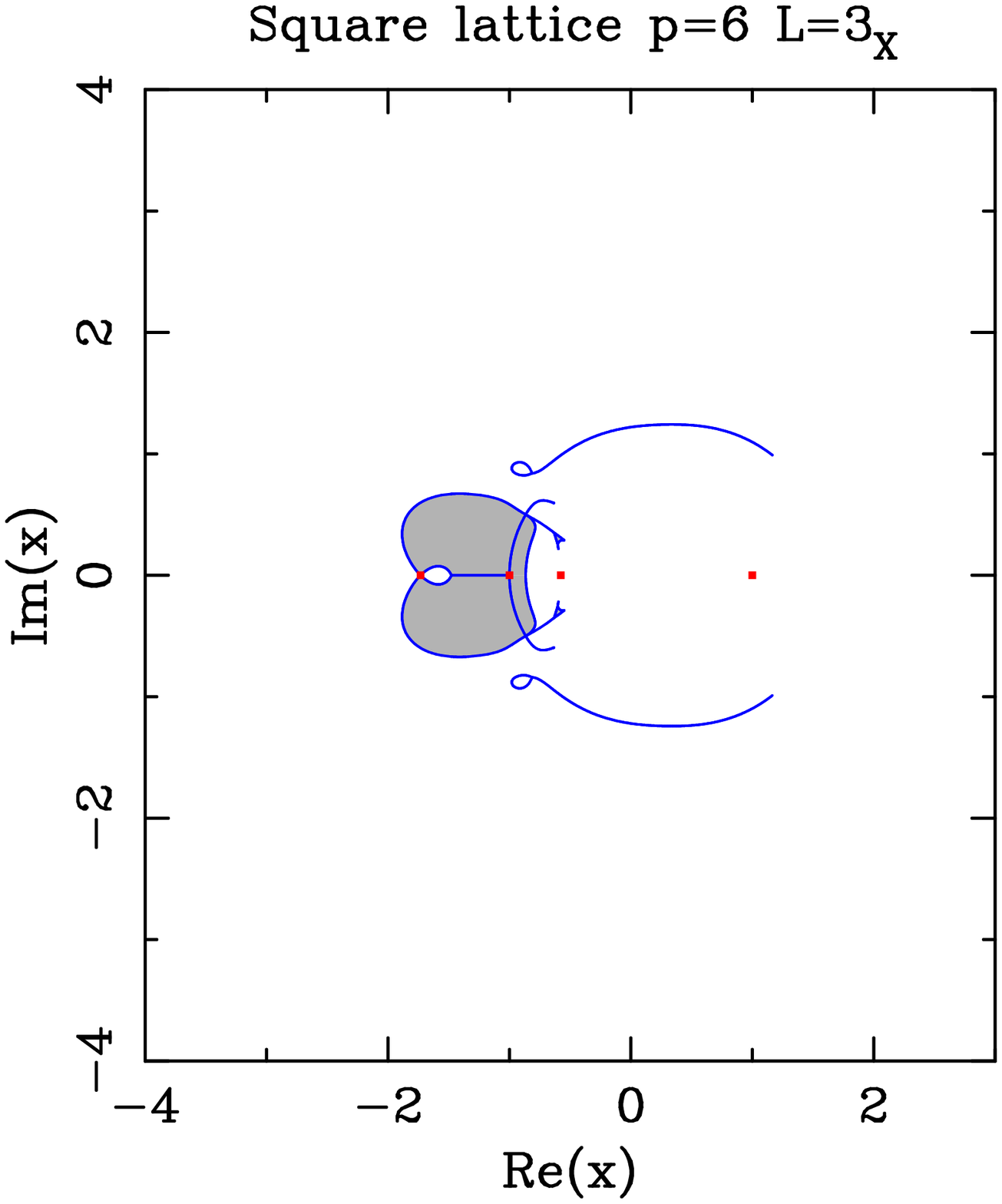} \\[2mm]
  \phantom{(((a)}(a) & \phantom{(((a)}(b)\\[5mm]
  \includegraphics[width=200pt]{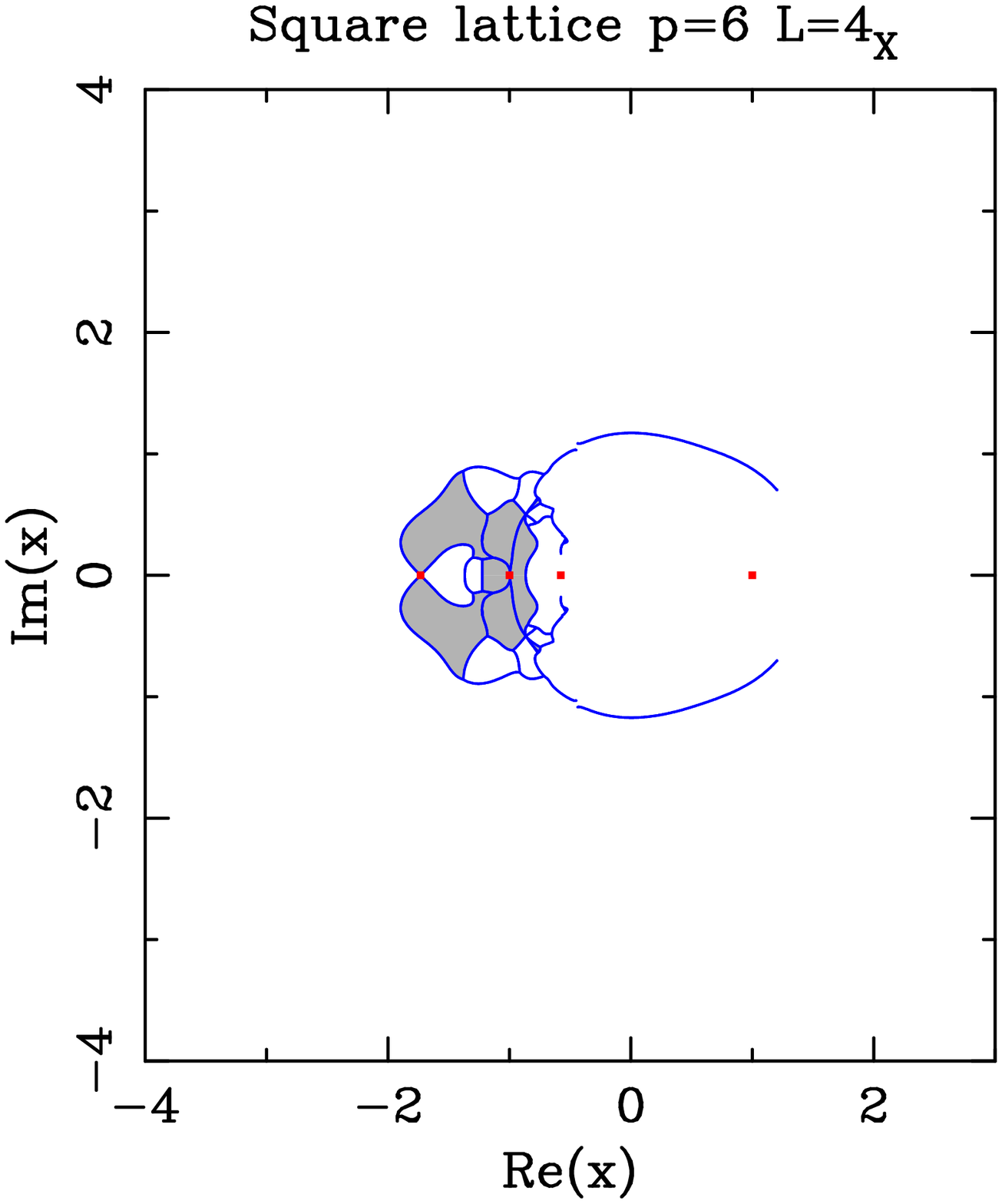} & 
  \includegraphics[width=200pt]{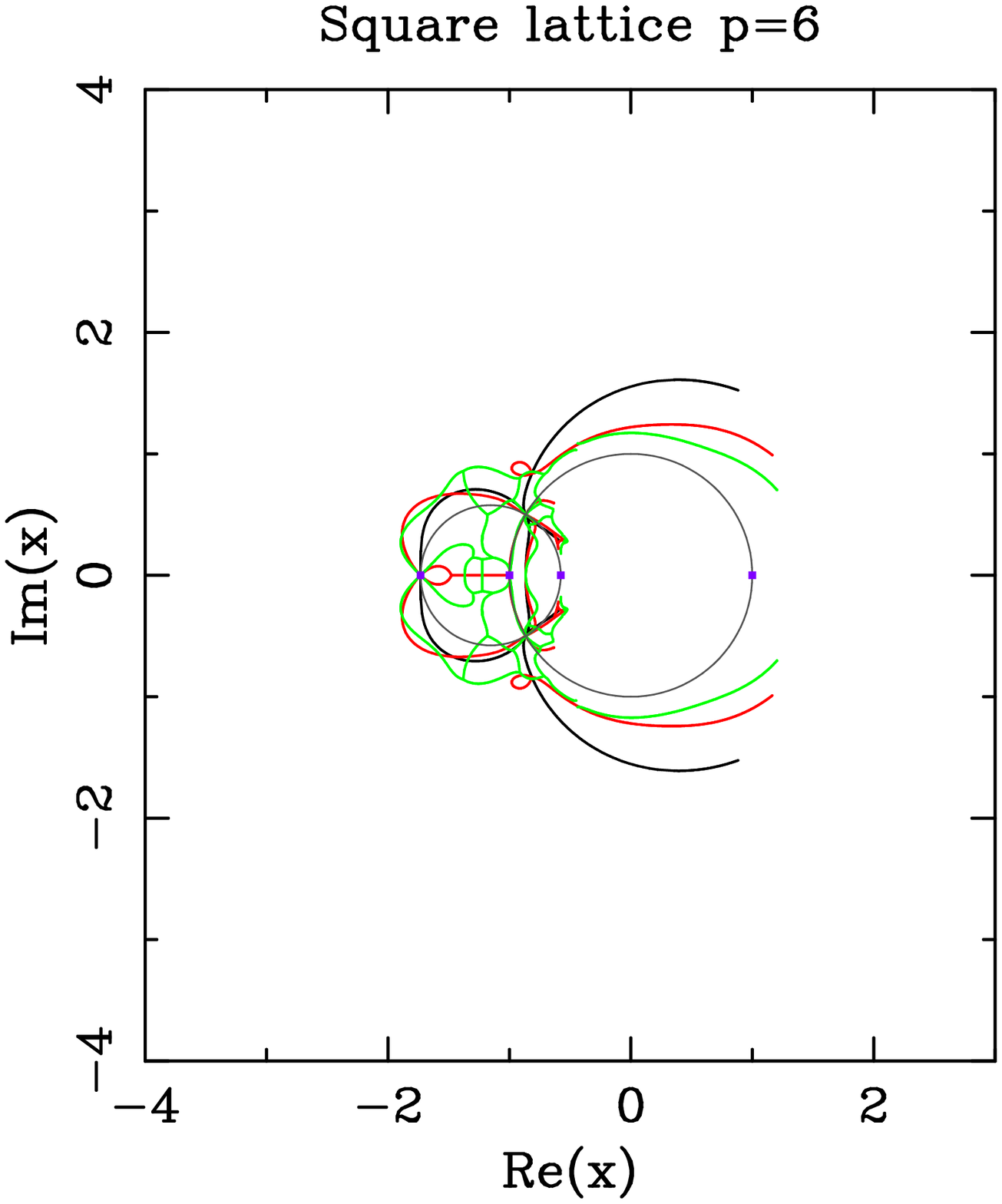} \\[2mm]
  \phantom{(((a)}(c) &  \phantom{(((a)}(d)
\end{tabular}
\caption{\label{Curves_sq_p=6_1}
Limiting curves for the square-lattice RSOS model with $p=6$ and several widths:
$L=2$ (a), $L=3$ (b), and $L=4$ (c) when only the sectors $\chi_{1,1}$ and
$\chi_{1,5}$ are taken into account. 
In the regions displayed in dark gray (resp.\  white)
the dominant eigenvalue comes from the sector $\chi_{1,5}$
(resp.\ $\chi_{1,1}$).
Figure~(d) shows all these curves together: $L=2$ (black), $L=3$ (red),
$L=4$ (green).
The solid squares
$\blacksquare$ show the values where Baxter found the free energy.
The dark gray circles correspond to \protect\reff{circles_sq_x}
}
\end{figure}

\clearpage
%
%
\begin{figure}
\centering
\begin{tabular}{cc}
  \includegraphics[width=200pt]{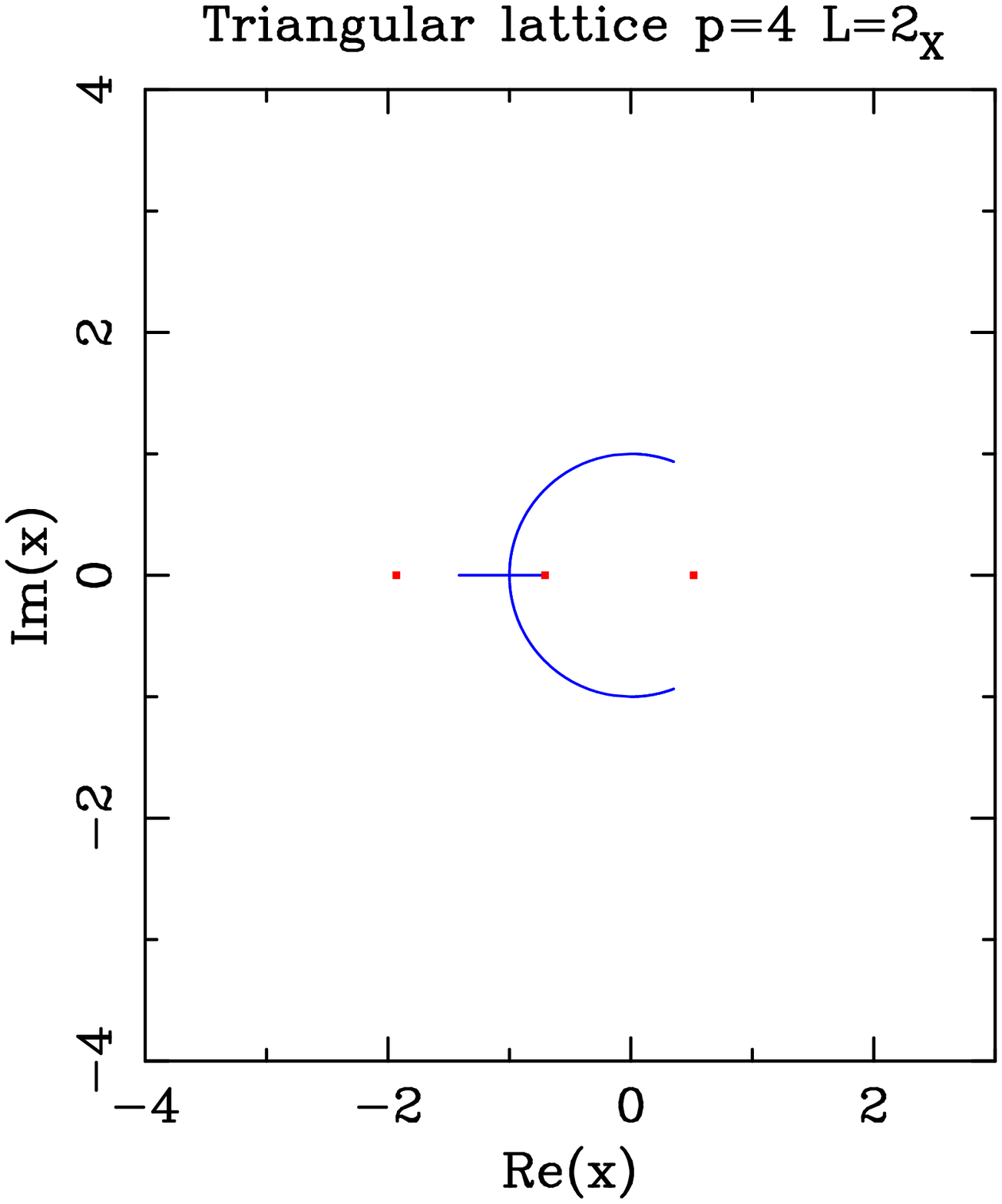} &
  \includegraphics[width=200pt]{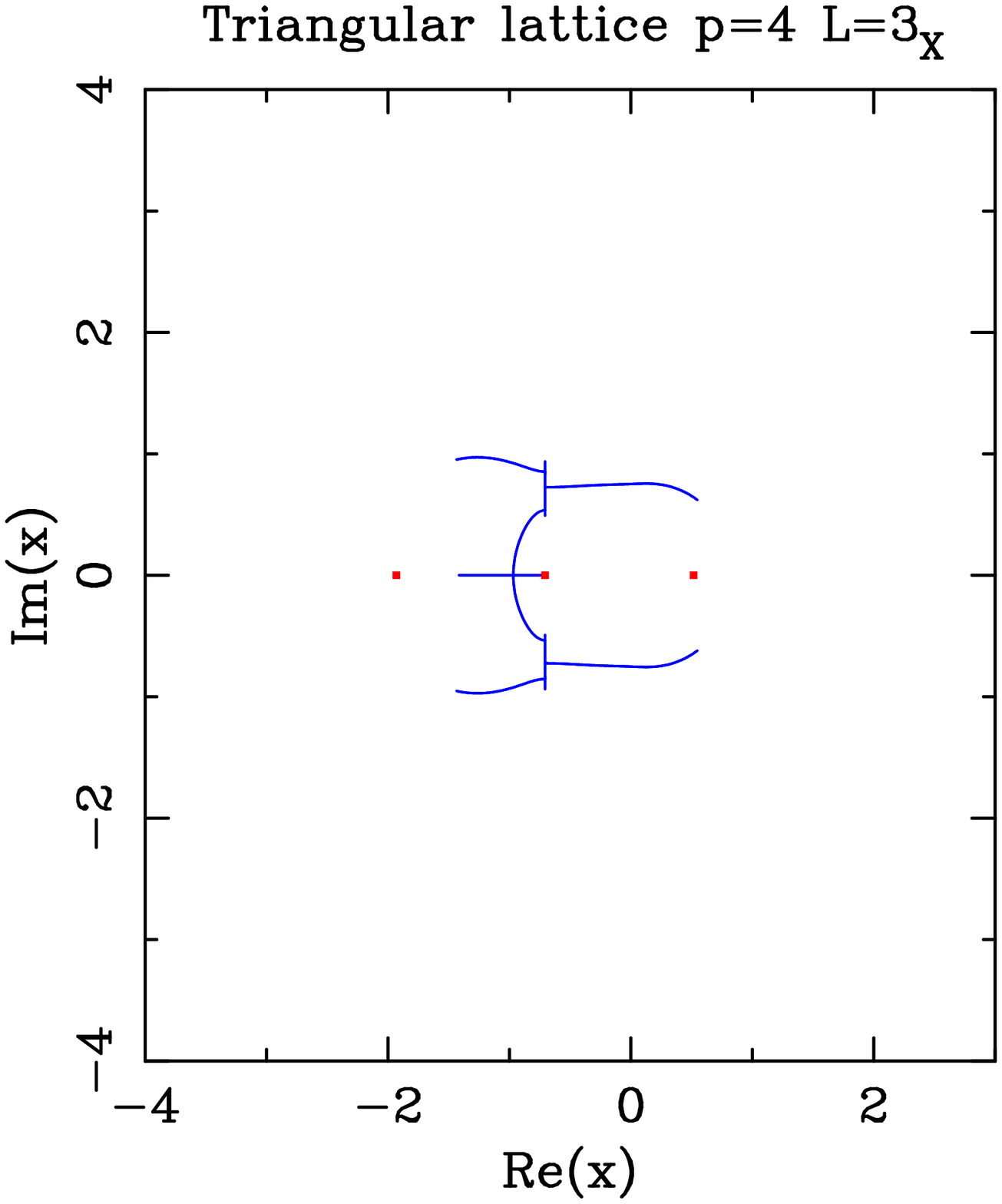}  \\[2mm]
  \phantom{(((a)}(a) & \phantom{(((a)}(b)\\[5mm]
  \includegraphics[width=200pt]{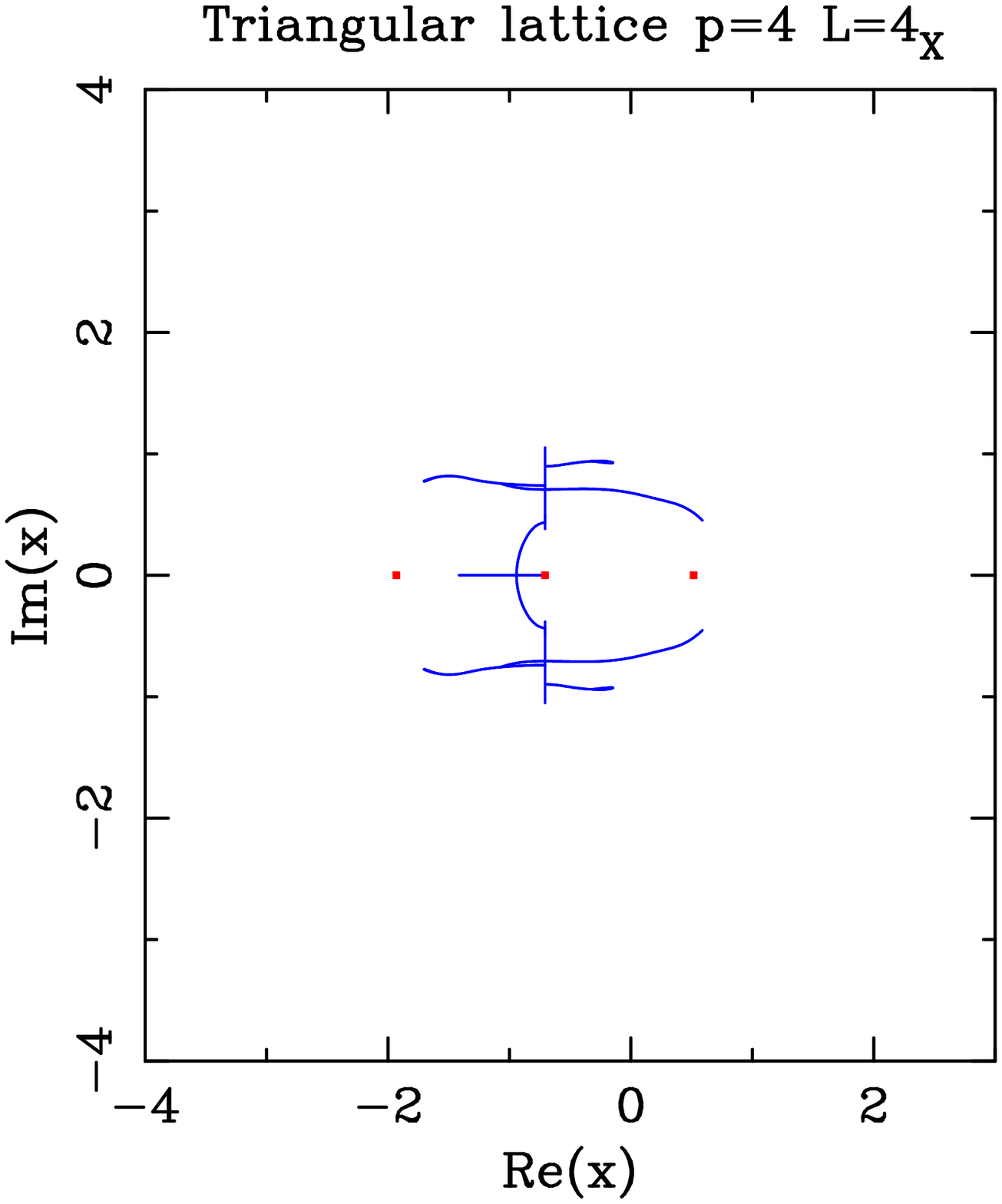} & 
  \includegraphics[width=200pt]{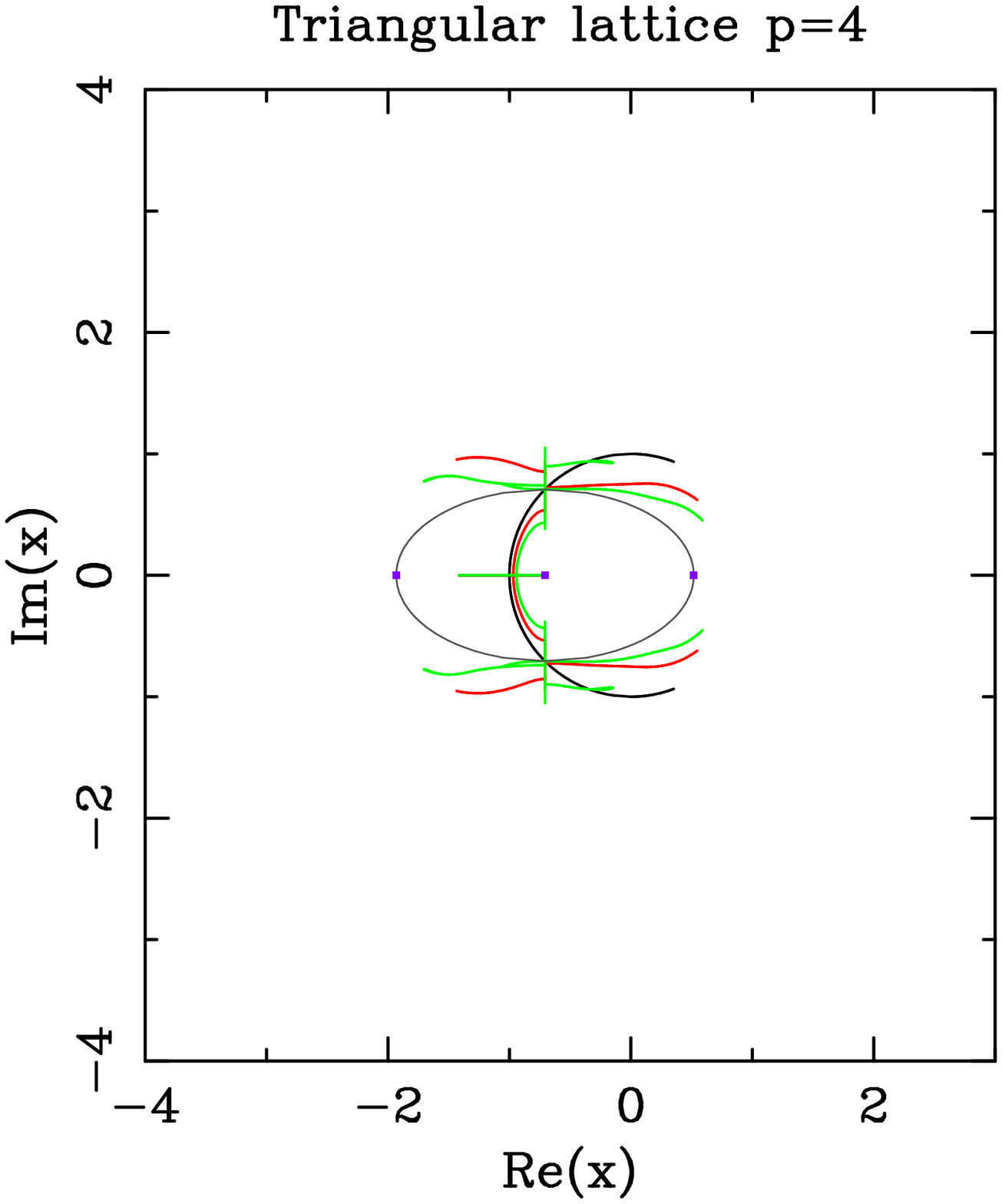} \\[2mm]
  \phantom{(((a)}(c) &  \phantom{(((a)}(d)
\end{tabular}
\caption{\label{Curves_tri_p=4_1}
Limiting curves for the triangular-lattice RSOS model with $p=4$ and 
several widths:
$L=2$ (a), $L=3$ (b), and $L=4$ (c) when only the sector $\chi_{1,1}$ is
taken into account. 
Figure~(d) shows all these curves together: $L=2$ (black), $L=3$ (red),
$L=4$ (green).
The gray ellipse corresponds to $(\Re x +1/\sqrt{2})^2 + 3 (\Im x)^2 = 3/2$.
This curve goes through the points $x=-e^{\pm i\,\pi/4}$.
}
\end{figure}

\clearpage
%
%
\begin{figure}
\centering
\begin{tabular}{cc}
  \includegraphics[width=200pt]{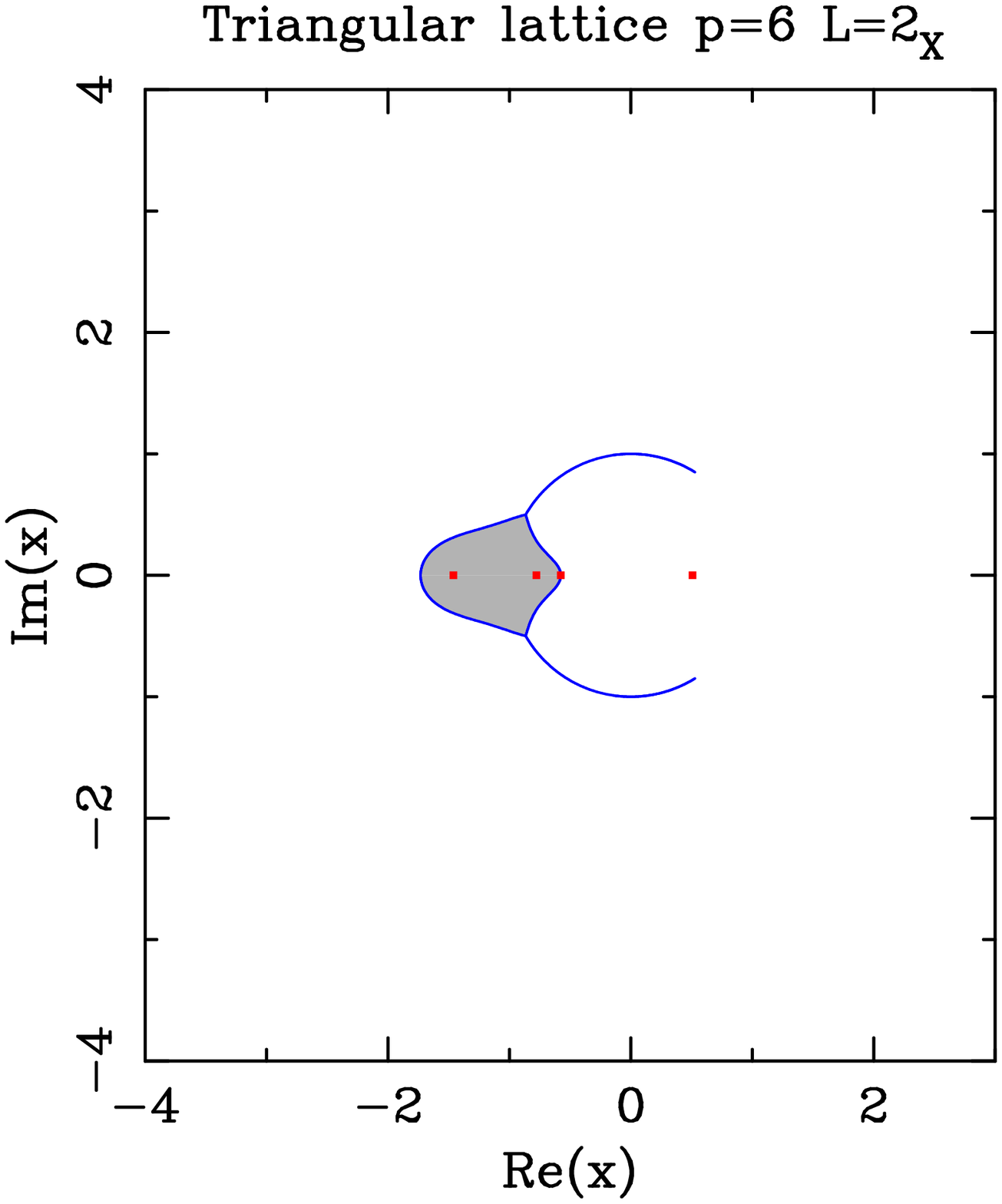} &
  \includegraphics[width=200pt]{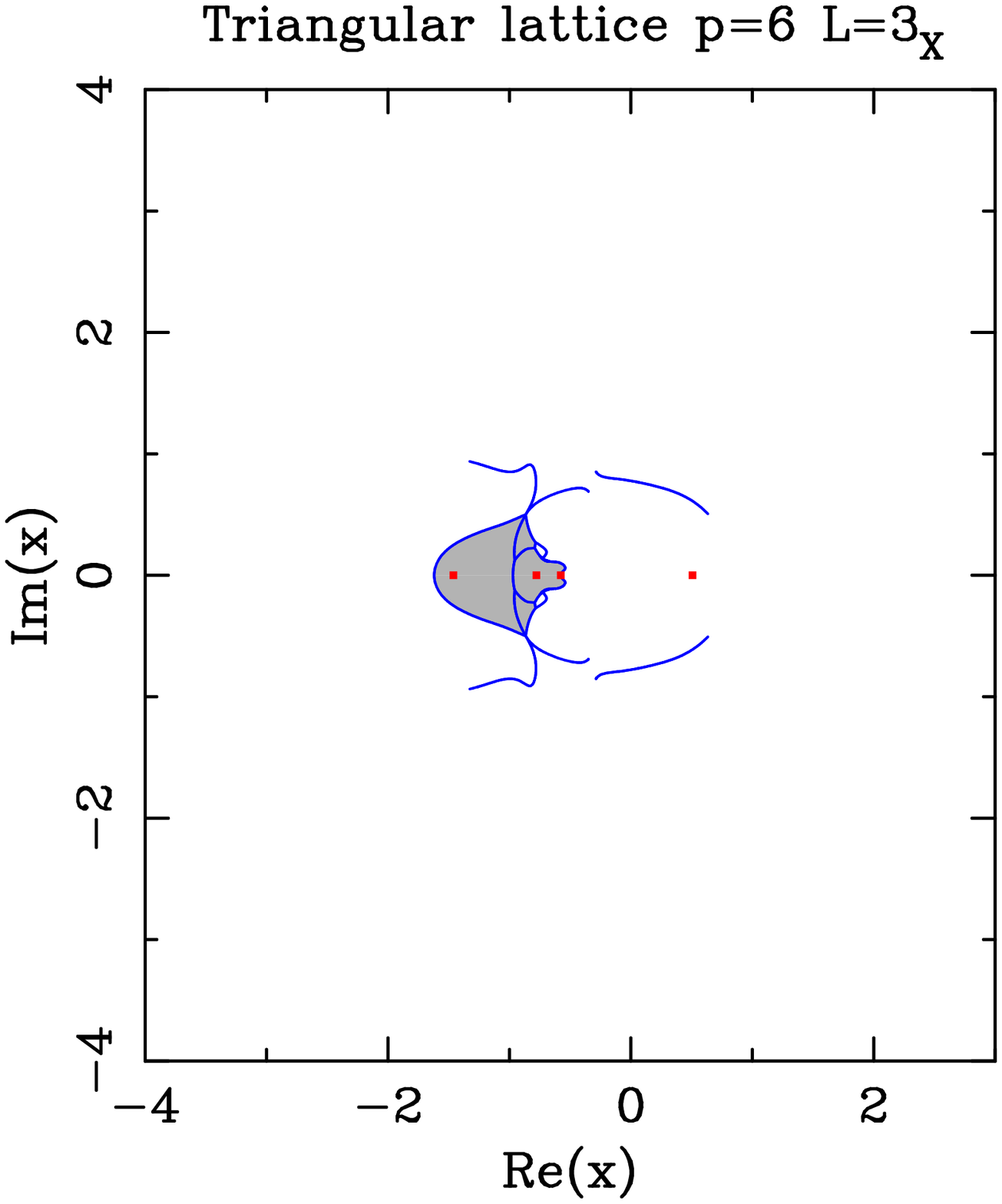} \\[2mm]
  \phantom{(((a)}(a) & \phantom{(((a)}(b)\\[5mm]
  \includegraphics[width=200pt]{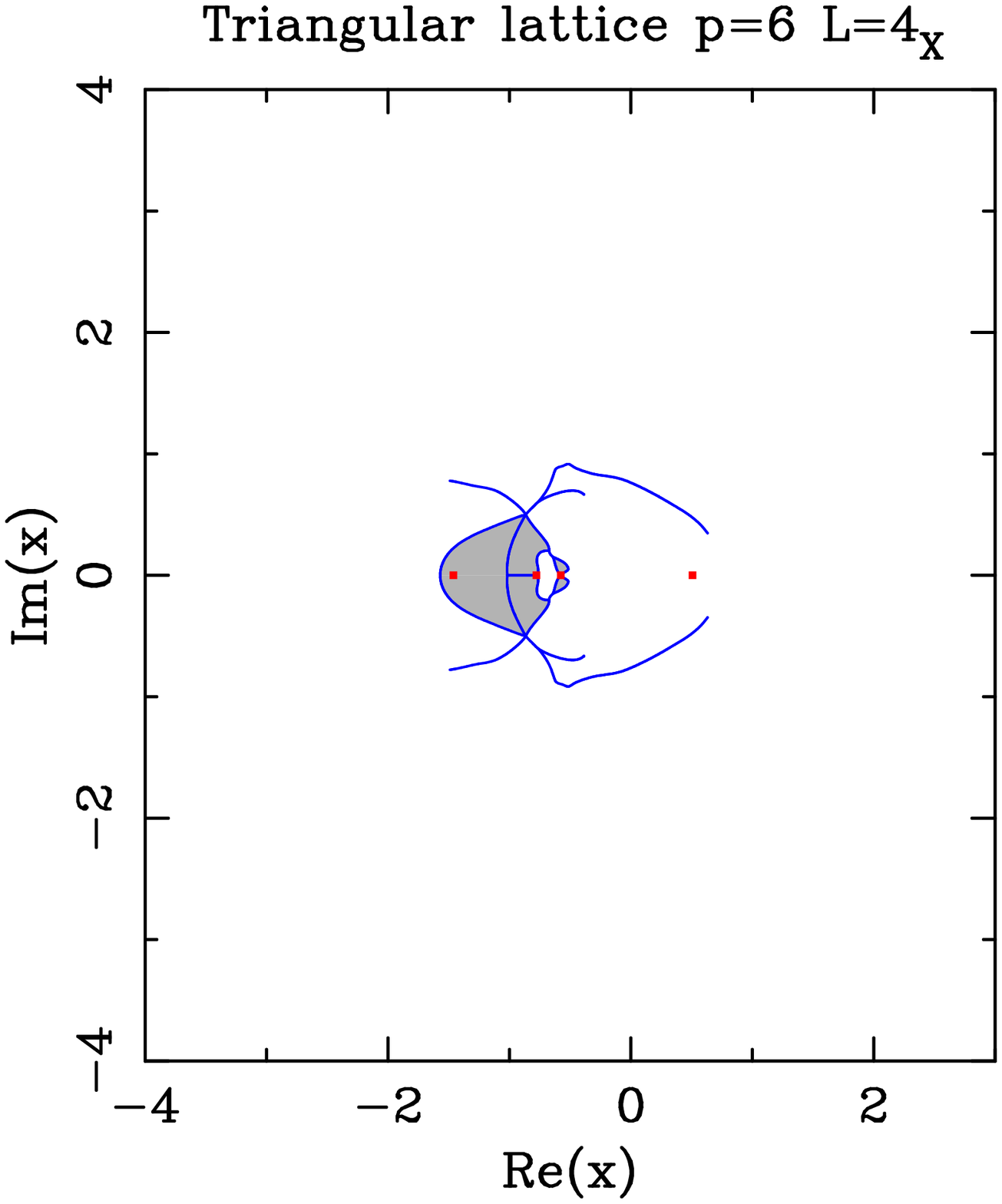} &
  \includegraphics[width=200pt]{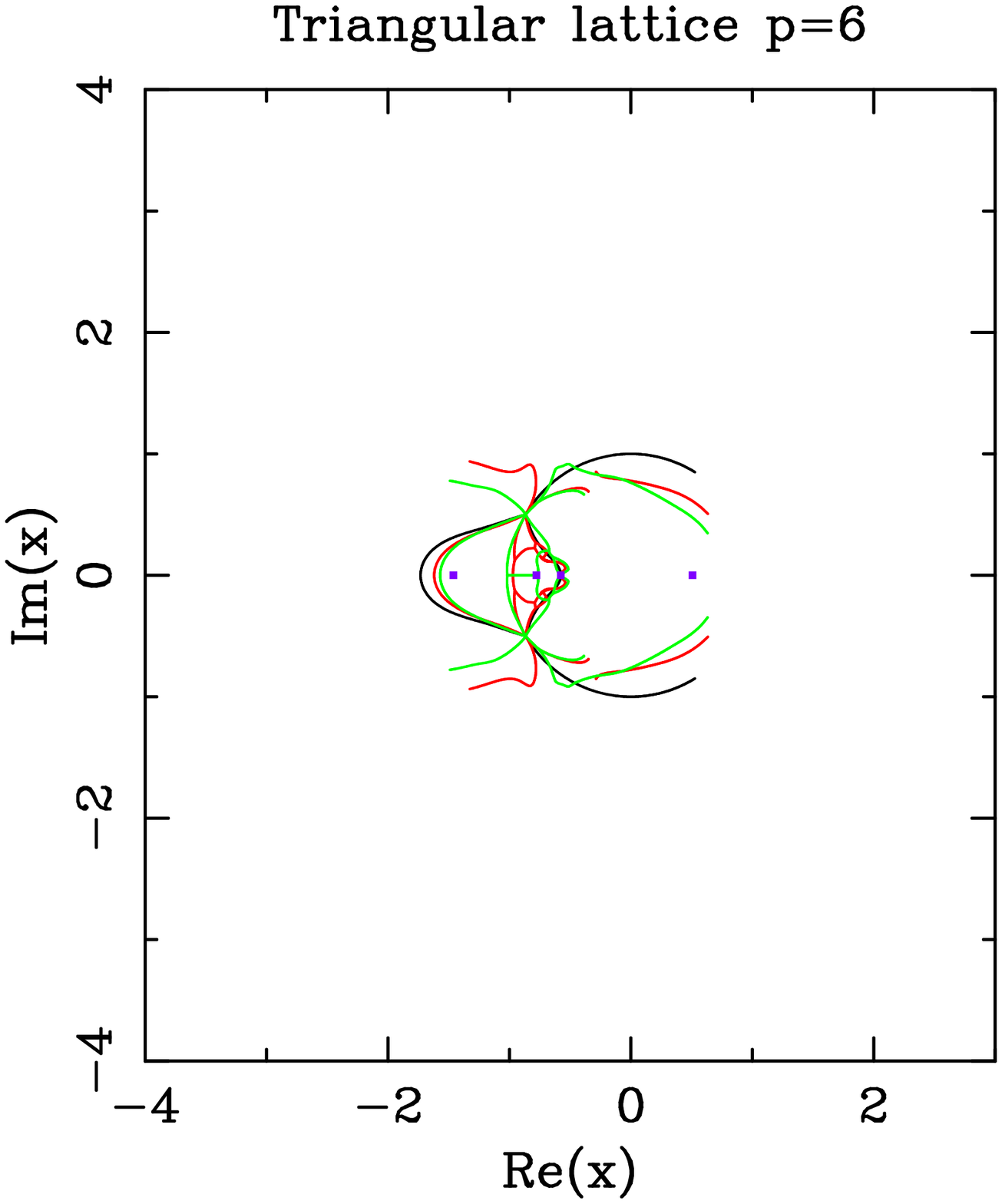} \\[2mm]
  \phantom{(((a)}(c) &  \phantom{(((a)}(d)
\end{tabular}
\caption{\label{Curves_tri_p=6_1}
Limiting curves for the triangular-lattice RSOS model with $p=6$ 
and several widths:
$L=2$ (a), $L=3$ (b), and $L=4$ (c) when only the sectors $\chi_{1,1}$ and
$\chi_{1,5}$ are taken into account.
In the regions displayed in dark gray (resp.\  white)
the dominant eigenvalue comes from the sector $\chi_{1,5}$
(resp.\ $\chi_{1,1}$).
Figure~(d) shows all these curves together: $L=2$ (black), $L=3$ (red),
$L=4$ (green).
The solid squares
$\blacksquare$ show the values where Baxter found the free energy.
}
\end{figure}

\end{document}